\documentclass[onecolumn,showpacs,showkeys,preprintnumbers,amsmath,amssymb,floatfix]{revtex4}

\usepackage{hyperref} 

\usepackage[utf8x]{inputenc} 
\usepackage{graphicx}
\usepackage{epsfig}
\usepackage[normalem]{ulem}
\usepackage{color}
\usepackage{ulem}    
\usepackage{enumerate}
\usepackage{lscape}   
\usepackage[nodisplayskipstretch]{setspace} 
\usepackage{type1cm}  
\usepackage{lettrine} 
\usepackage{dcolumn}
\usepackage{bm}

\makeatletter
\newcommand*{\defeq}{\mathrel{\rlap{%
                     \raisebox{0.42ex}{$\m@th\cdot$}}%
                     \raisebox{-0.4ex}{$\m@th\cdot$}}%
                     =}                                     
\makeatother 
\newcommand{\jump}[1]{\big[\big[{#1}\big]\big]}     
\renewcommand{\cal}[1]{\mathcal{#1}}                        
\newcommand{\calhat}[1]{\widehat{\mathcal{#1}}}             
\newcommand{\calo}[1]{\mathcal{O}\left(#1\right)}       
\newcommand{\devt}[3]{ {#1}^{(#2)}_{(#3)} }                 
\newcommand{\ab}{{\alpha\beta}}                             
\newcommand{\mE}{\mathcal{E}}                               
\newcommand{\mL}{\mathcal{L}}                               
\newcommand{\neutral}[1]{\underline{#1}}                    
\newcommand{\nX}{\underline{\mathrm{X}}}                    
\newcommand{\nY}{\underline{\mathrm{Y}}}                    
\newcommand{\nZ}{\underline{\mathrm{Z}}}                    
\newcommand{\nDrpp}{\underline{\Delta r}'^*_p}              
\newcommand{\nDrp}{\underline{\Delta r}^*_p}                
\newcommand{\ntau}{\neutral{\tau}}                          
\newcommand{\tildelim}[1]{\underset{#1}{\widetilde{\lim}}}  

\renewcommand{\bar}[1]{\overline{#1}}
\newcommand{\scirc}{{\scalebox{0.5}{$\circ$}}}
\newcommand{\scircc}{{\scalebox{0.5}{$\circ\circ$}}}
\newcommand{\dott}[1]{\overset{\scirc}{#1}\hspace{0.0pt}}       
\newcommand{\ddott}[1]{\overset{\scircc}{#1}\hspace{0.0pt}}     
\newcommand{\ds}{\displaystyle}
\newcommand{\dottb}[1]{{\stackrel{\ _\circ}{#1}\hspace{0.0pt}\ }} 

\newcommand{\beq}{\begin{equation}}
\newcommand{\eeq}{\end{equation}}
\newcommand{\ba}{\begin{aligned}}
\newcommand{\ea}{\end{aligned}}
\newcommand{\bea}{\begin{equation}\begin{aligned}}
\newcommand{\eea}{\end{aligned}\end{equation}}
\newcommand{\ub}{\underbrace}
\newcommand{\nLun}[1]{\|{#1}\|_{\text{L}^1}}
\definecolor{grey1}{rgb}{0.153,0.157,0.133}          
\definecolor{grey2}{rgb}{0.074,0.074,0.066}          

\newcommand{\nid}{\noindent}

\definecolor{airforceblue}{rgb}{0.36, 0.54, 0.66} 

\makeatletter
\def\l@subsubsection#1#2{}
\makeatother

\begin{document}
\preprint{Preprint}

\title[Self-force]{Indirect (source-free) integration method. II. Self-force consistent radial fall}

\author
{
Patxi Ritter\textsuperscript{1,2,3}
, 
Sofiane Aoudia\textsuperscript{4,5}
, 
Alessandro D.A.M. Spallicci\textsuperscript{1,6}\footnote{Corresponding author: spallicci@cnrs-orleans.fr, http://lpc2e.cnrs-orleans.fr/$\sim$spallicci/},  
St\'ephane Cordier\textsuperscript{2,7}
}

\affiliation{
\textsuperscript{1}Universit\'e d'Orl\'eans\\ 
Observatoire des Sciences de l'Univers en r\'egion Centre, UMS 3116 \\
Centre Nationale de la Recherche Scientifique\\
Laboratoire de Physique et Chimie de l'Environnement et de l'Espace, UMR 7328\\
\mbox {3A Av. de la Recherche Scientifique, 45071 Orl\'eans, France}\\
\textsuperscript{2}Universit\'e d'Orl\'eans\\
Centre Nationale de la Recherche Scientifique\\
\mbox {Math\'ematiques - Analyse, Probabilit\'es, Mod\`elisation - Orl\'eans, UMR 7349}\\
Rue de Chartres, 45067 Orl\'eans, France\\
\textsuperscript{3}Univerzita Karlova\\
Matematicko-fyzik\'{a}ln\'{i} fakulta, \'{U}stav teoretick\'{e} fyziky\\
V Hole\u{s}ovi\u{c}k\'{a}ch 2, 180 00 Praha 8, \u{C}esk\'{a} Republika\\
\textsuperscript{4}Laboratoire de Physique Th\'{e}orique - Facult\'{e} des Sciences Exactes \\ 
Universit\'{e} de Bejaia, 06000 Bejaia, Algeria\\
\textsuperscript{5}Max Planck Institut f\"{u}r Gravitationphysik, A. Einstein\\
1 Am M\"uhlenberg, 14476 Golm, Deutschland\\
\textsuperscript{6}Chaire Fran\c{c}aise, Universidade do Estado do Rio de Janeiro\\
\mbox{Instituto de F\'isica, Departamento de F\'isica Te{\'o}rica}\\
\mbox{Rua S\~{a}o Francisco Xavier 524, Maracan\~{a}, 20550-900 Rio de Janeiro, Brasil}\\
\textsuperscript{7}Universit\'{e} Joseph Fourier\\
Agence pour les Math\'{e}matiques en Interaction\\ 
Centre Nationale de la Recherche Scientifique\\
Laboratoire Jean Kuntzmann, UMR 5224\\ 
\mbox{Campus de Saint Martin d'H\'{e}res, Tour IRMA, 51 rue des Math\'{e}matiques, 38041 Grenoble, France} 
}

\date{14 September 2015}

\begin{abstract}

We apply our method of indirect integration, described in Part I, at fourth order, to the radial fall affected by the self-force.  
The Mode-Sum regularisation is performed in the Regge-Wheeler gauge using the equivalence with the harmonic gauge for this orbit. 
We consider also the motion subjected to a self-consistent and iterative correction determined by the self-force through osculating stretches of geodesics. 
The convergence of the results confirms the validity of the integration method. This work complements and justifies the analysis and the results appeared in Int. J. Geom. Meth. Mod. Phys., 11, 1450090 (2014). 
\end{abstract}



\maketitle
\vspace{-0.65 cm}
\hspace{1.55 cm}{\footnotesize Mathematics Subject Classification 2010: 35Q75, 35L05, 65M70, 70F05, 83C10, 83C35, 83C57}

\section{Introduction and motivations}
 
We apply the indirect method of Part I to the motion of a particle perturbed by the back-action, that is the influence of the emitted radiation and of the mass $m_0$ on its own worldline, thanks to the interaction with the field of the other mass $M$. 

The problem of the back-action for massive point particles moving in a strong field with any velocity has been tackled by concurring approaches all yielding the same result, exclusively defined in the harmonic (H) gauge. Result derived in 1997 by Mino, Sasaki and Tanaka \cite{misata97}, Quinn and Wald \cite{quwa97}, around an expansion of the mass ratio ${ m_0/M}$. 
The main achievement has been the identification of the regular and singular perturbation components, and their playing or not-playing role in the motion, respectively. The conclusive equation has been baptised MiSaTaQuWa from the first two initials of its discoverers. 
Later, Detweiler and Whiting \cite{dewh03} have shown an alternative approach, not any longer based on the computation of the tails, but derived from the Dirac solution \cite{di38}. It is customary to call self-force (SF) the expression resulting from MiSaTaQuWa and DeWh approaches, and to switch between the former (the SF externally breaking the background geodesic as non-null right hand-side term) and the latter (the particle following a geodesic of the total metric, background plus perturbations) intepretations of the same phenomenon.  

A full introduction to the SF is to be found in \cite{blspwh11}, while for a first acquaintance the reader might be satisfied by the arguments exposed in \cite{spriao14}. 

In the MiSaTaQuWa conception, the gravitational waves are partly radiated to infinity (the instantaneous, also named direct, component), and partly scattered back by the black hole potential, thus forming back-waves (the tail part) which impinge on the particle and give origin to the SF. 
Alternatively, the same phenomenon is described by an interaction particle-black hole generating on one hand a field which behaves as outgoing radiation in the wave-zone, and thereby extracts energy from the particle; on the other hand, in the near zone, the field acts on the particle and determines the SF which impedes the particle to move on the geodesic of the background metric.
From these works, it emerges the splitting between the instantaneous and tail components of the perturbations, the latter  acting on the motion. Unfortunately the tail component can't be computed directly, if not as a difference between the total and the instantaneous components. Instead, the DeWh approach reproduces  the Dirac definition. It consists of half of the difference between the retarded and the  advanced fields, to which is added an {\it ad hoc} field including the contributions from the past light inner cone, while avoiding non-causal future contributions.     

The SF computation is not an easy task because the field perturbation is divergent at the position of the particle, and it is therefore  necessary to use a suitable procedure of regularisation. The latter deals with the divergences coming from the infinitesimal size of the particle.
In the Regge-Wheeler (RW) gauge \cite{rewh57},  we benefit of the wave-equation and of the gauge invariance of its wave-function. Regrettably, the singularity in the perturbed metric has a complicated structure which has made impossible so far to find a suitable regularisation scheme.  Nevertheless,  current investigations attempt to identify gauge transformations between the RW and H gauges, {\it e.g.}, Hopper and Evans \cite{hoev10,hoev13}, or use numerical integration approaches that deal {\it de facto} only with the homogeneous form of the Regge-Wheeler-Zerilli (RWZ) equation \cite{rewh57,ze70c}. 

In the H gauge, a regularisation recipe in spherical harmonics and named Mode-Sum, was conceived by Barack and Ori \cite{baor00,baor01}.  Such a procedure is to be carried out partially or totally in the H gauge. There is an exception though for a purely radial orbit. In this case, there is a regular connection between the RW and H gauges; thus the quantification of the SF may be carried out entirely in the RW gauge. Further, the outcome is invariant for these two gauges and all regularly related gauges \cite{baor01}. Herein, we thus proceed with a detailed computation  {\it in toto} in the RW gauge, announced by Barack and Lousto in \cite{balo02} but never appeared.  

In the '70s, Zerilli computed the gravitational radiation emitted during the radial fall \cite{ze70a,ze70b,ze70c} into a Schwarzschild-Droste (SD) black hole \cite{sc16, dro16a, dro16b}. 
Many studies followed later on. The first was from Davis {\it et al.} \cite{daruprpr71}, who considered, in the frequency domain, the radiation emitted by a particle initially at rest in free fall from infinity. Later, Ferrari and Ruffini \cite{ferrariruffini1981} resume, still in the frequency domain, the same system, but conferring an initial speed to the particle from infinity. The first to solve the problem of the fall of the particle still initially at rest but for a finite distance from the black hole were Lousto and Price in a series of papers \cite{lopr97a,lopr97b,lopr98}, where they detail and give a numerical technique to deal with the point source in the time domain. Martel and Poisson \cite{mapo02} resume the same problem by proposing a  family of parametrised initial conditions, all of them being solutions of the Hamiltonian constraint; further they study the influence of these initial conditions on the wave-forms and energy spectra.

Thirty years later, back-action - without orbital evolution - was partially analysed only in two works \cite{lo00,balo02}, and with contrasting predictions (in the former Lousto suggests that back-action is repulsive for most modes, conversely to the latter where Barack and Lousto attribute always an attractive feature). We have largely commented these papers in \cite{spri14}.
Needless to say, the time shortness of the fall forbids any important accumulation of back-action effects but, from the epistemological point of view, radial fall for gravitation remains the most classical problem of all, and raising the most delicate technical questions.
Early gravitational SF computation were carried out in the H gauge by Barack and Sago for circular \cite{basa07,basa09} and eccentric orbits \cite{basa10}.  

In the context of the Extreme Mass Ratio Inspiral (EMRI) gravitational wave sources, the gravitational SF heavily impacts the wave-forms. It has been suggested to evolve the most relativistic orbits through the iterative application of the SF on the particle worldline, {\it i.e.}, the self-consistent approach by Gralla and Wald \cite{grwa08,grwa11}. We implement it for the least adiabatic orbit of all, that is radial infall, using our integration method.  The strict self-consistency would imply that the applied SF at some instant is what arises from the actual field at that same instant. 
So far this has been done only for a scalar charged particle around an SD black hole by Diener {\it et al.} \cite{divewade12},
 and never for a massive particle. 
For quasi-circular and inspiral orbits, dealt by Warburton {\it et al.} \cite{waakbagasa12}, Lackeos and Burko \cite{labu12}, the applied SF is what would have resulted if the particle
were moving along the geodesic that only instantaneously matches the true orbit. 
Herein, we adopt the latter acception. 

We thus study how the back-action affects the motion, the radiated energy and the wave-forms of a particle without and with the self-consistent approach. According to the different inclinations of the reader, his interest may raise from one or more of the following considerations.
\vskip4pt
{\nid Technical assessments and advancements} 
\begin{itemize}
\item{The feeble differences between a radial orbit for which the self-force corrections are neglected or conversely taken into account (without and with orbital evolution) allow to test our numerical integration scheme, as we expect similar results, while possibly appreciating any difference, among these three cases.}
\item{The inclusion of back-action effects demands a sophisticated algorithm of at least fourth order, since considering third time derivatives of the wave-function.} 
\item{The contrasting results in \cite{lo00,balo02} need a resolution. We have recovered the results in \cite{balo02} (self-force is attractive in H and RW gauges) and proved wrong those in \cite{lo00} (claiming that the self-force is mainly repulsive and divergent at the horizon), see the full discussion in \cite{spri14}. Anyway, the work in \cite{balo02} does not consider the impact of the self-force on the trajectory, which we deal with herein.}
\item{Radial infall is the least adiabatic orbit of all types. Imposing the identity between the radiated energy and the lost orbital energy for computing the corrections on the motion would be most unjustified as shown by Quinn and Wald \cite{quwa99}. Indeed, it is just for non-adiabatic orbits, that is required applying a continuous correction on the trajectory due to the SF effects, {\it i.e.} the self-consistent method \cite{grwa08,grwa11}. Thus, radial infall imposes such an application, though it is not rewarding due to the feebleness of the SF effects themselves.}
\item {Given  the limitations of numerical relativity in evolving circular and elliptic orbits for small mass ratio binaries, the comparison of results for head-on collisions from numerical and perturbation methods is of interest.}
\item{In particle physics, when referring to the transplanckian regime and black hole production, back-action has a pivotal role in head-on collisions according to Gal'tsov {\it et al.} \cite{gakospto10a,gakospto10b}.}
\item{The regular transformation between H and RW gauges for radial trajectories allow to carry out the Mode-Sum regularisation entirely in the RW gauge.}
\end{itemize}
When endeavouring towards astrophysical scenarios, we recall that 
\begin{itemize}
\item{It has been estimated by Amaro-Seoane, Sopuerta {\it et al.} \cite{assofr13,brasso14} that a relevant number of EMRIs will consist of direct plunges when the supermassive black hole (SMBH) is not rotating.}
\item{Radial trajectories are comparable to portions of highly eccentric orbits producing an EMRB (Extreme Mass Ratio Burst) following Berg and Gair \cite{bega13a,bega13b,bega13c}.}
\item{The last stages of EMRI plunges were analysed by Keden, Gair and Kamionkowski \cite{kegaka05} for discriminating supermassive black holes from boson stars, supposedly horizonless objects, and  by Macedo {\it et al.} \cite{mapacacr13} for signatures of dark matter.}
\item{The concept of maximal velocity in radial fall is discussed in high energy astrophysics for jets and tidal disruption by Chicone and  Mashhoon \cite{chma05,ma05}, Kojima and Takami \cite{kota06}.}  
\end{itemize}
\nid{General motivations} 
\begin{itemize}
\item{Radial fall is the most classic problem in physics instantiated by the stone of Aristot\'el\=es, the tower of Galilei, the apple of Newton, and the cabin of Einstein. The solutions represent the level of understanding of gravitation at a given epoch, and have thereby an epistemological relevance.}
\item{It is a worthwhile problem \`a la Feynman: {\it The worthwhile problems are the ones you can really solve or help solve, the ones you can really contribute something to. No problem is too small or too trivial if we can really do something about it} \cite{feynmanwiki}.} 
\end{itemize}

The paper is structured as follows. Section II, after a brief review of the SF, is largely devoted to the computation in the RW gauge of the regularisation parameters through the Mode-Sum method. Section III deals with some numerical issues, the performance and validation of the code. In Sect. IV, we deal with the impact of the SF on the motion of the particle without and with the self-consistent evolution for the radial fall through osculating orbits. 
The appendixes deal with the Riemann-Hurwitz regularisation \cite{ri59,hu82}, the numerical extraction of the field at the particle position and display the jump conditions for the radial orbit.  

Geometric units ($G = c = 1$) are used, unless stated otherwise. The metric signature is $(-, +, +, +)$. The particle position  on the perturbed metric is noted by $r_p(\tau)$ while on the background metric by $R$.  

\section{Gravitational SF}\label{section:GSF:RW}

\subsection{Foreword}

The SF equation, defined in the H gauge, is given by \cite{misata97, quwa97, dewh03}

\beq
F^\alpha_{\rm self} = - \frac{1}{2} m_0
(g^{\alpha\beta} + u^\alpha u^\beta) 
\left (2h_{\mu\beta ;\nu}^{\text{R(H)}}- h_{\mu\nu ;\beta}^{\text{ R(H)}}\right) u^\mu u^\nu~, 
\label{gweq}
\eeq
where $\text R$ stands for the regular part of the perturbations $h_{\mu\nu}$, either tail (MiSaTaQuWa) or radiative (DeWh). The two contributions are not equivalent, but the final results are. The other quantities are the background metric $g_{\mu\nu}$ and the 
four-velocity $u^\alpha$.   
The SF is obtained by subtracting the singular part from the retarded force

\beq
F^{\alpha(\text{H})}_\text{self}=F^{\alpha(\text{H})}_\text{ret}-F^{\alpha(\text{H})}_\text{S}~.
\label{ret-s}
\eeq

The retarded force is computed from the retarded field

\begin{align}
&F^{\alpha(\text{H})}_{\text{ret}}=F^{\alpha}\left[h^{\text{ret}(\text{H})}_\ab\right] = 
m_0k^{\alpha\beta\gamma\delta}\nabla_\delta\bar{h}^{\text{ret}(\text{H})}_{\beta\gamma} =
-\frac{1}2m_0\left(g^{\ab}+u^\alpha u^\beta\right)\left(2\nabla_\delta h^{\text{ret}(\text{H})}_{\beta\gamma}-\nabla_\alpha h^{\text{ret}(\text{H})}_{\gamma\delta}\right)u^\gamma u^\delta ~,
\label{rappel:Fret}
\end{align}
where $\bar{h}_\ab=h_\ab-1/2 g_\ab h$, and $k^{\alpha\beta\gamma\delta}$ is given by 

\beq
\begin{aligned}
k^{\ab\gamma\delta}=&\frac{1}2g^{\alpha\delta}u^\beta u^\gamma - g^{\ab}u^\gamma u^\delta - \frac{1}2u^\alpha u^\beta u^\gamma u^\delta + 
\frac{1}4u^\alpha g^{\beta\gamma}u^\delta + \frac{1}4g^{\alpha\delta}g^{\beta\gamma}~.
\label{k}
\end{aligned}
\eeq

As shown in \cite{baor01}, for a transformation to any gauge (G) 
 
\beq
h^{\text{ret}(\text{G})}_\ab=h^{\text{ret}(\text{H})}_\ab+\delta h^{(\text{H}\to\text{G})}_\ab~,
\label{chgt:jauge}
\eeq

the SF changes as

\begin{align}
F^{\alpha(\text{G})}_\text{self}= & F^{\alpha(\text{H})}_\text{self}+\delta F^{\alpha(\text{H}\to\text{G})} = 
 F^{\alpha(\text{H})}_\text{ret}-F^{\alpha(\text{H})}_\text{S}+\delta F^{\alpha(\text{H}\to\text{G})} = 
 F^{\alpha(\text{G})}_\text{ret}-F^{\alpha(\text{H})}_\text{S}~.
\label{transform:Fself}
\end{align}

Thus, in an arbitrary gauge G, the singular term - to be extracted from the retarded force - is always expressed in the H gauge and not in the G one, as it might be supposed. In the H gauge, the isotropy of the singularity around the particle eases the computation of $F^{\alpha(\text{H})}_\text{S}$, while guaranteeing its inconsequential role on the motion. Instead, in other gauges we are confronted with the lack of isotropy \cite{quwa97}.
We recall the expression of the Mode-Sum decomposition in the H gauge \cite{baor00}

\begin{align}
F^{\alpha(\text{H})}_\text{self} & = \sum_{\ell=0}^\infty F^{\alpha\ell(\text{H})}_{\text{ret}} - F^{\alpha(\text{H})}_\text{S}= \sum_{\ell = 0}^\infty F^{\alpha\ell(\text{H})}_{\text{ret}} - 
\sum_{\ell=0}^\infty \Big[ A^{\alpha(\text{H})} L + B^{\alpha(\text{H})} + C^{\alpha(\text{H})} L^{-1}\Big] - D^\alpha~,
\label{mode:sum:H}
\end{align}
where $L=\ell+1/2$, and $\ell$ is the mode index. Inserting the Mode-Sum expression of $F^{\alpha(\text{H})}_\text{S}$ from  Eq. (\ref{mode:sum:H})  into Eq. (\ref{transform:Fself}), and decomposing $F^{\alpha(\text{G})}_\text{ret}$ in ${\ell}$ modes, we get \cite{baor01}

\beq
F^{\alpha(\text{G})}_\text{self}=\sum_{\ell=0}^\infty \Big[F^{\alpha\ell(\text{G})}_{\text{ret}}-
A^{\alpha(\text{H})} L-B^{\alpha(\text{H})}-C^{\alpha(\text{H})} L^{-1}\Big]-D^\alpha~.
\label{mode:sum:G:2}
\eeq

In Eq. (\ref{mode:sum:G:2}), the regularisation parameters are computed in the H gauge, but we can go a step further and totally dismiss the H gauge. For a restricted class of gauges for which the transformation gauge vector is regular, the Mode-Sum technique can be used to regularise the retarded force directly in those gauges \cite{baor01}. We then have

\beq
F^{\alpha(\text{G})}_\text{self} = \sum_{\ell=0}^\infty \Big[F^{\alpha\ell(\text{G})}_{\text{ret}}-A^{\alpha(\text{G})} L-B^{\alpha(\text{G})}-C^{\alpha(\text{G})} L^{-1}]-D^{\alpha(\text{G})}~.
\label{mode:sum:Gideal}
\eeq

In \cite{baor01} several orbits were examined. It was concluded that the RW gauge is regularly connected to the H gauge only for purely radial orbits. Further, it has been shown that the components of the transformation gauge vector are not only regular at the position of the particle but they can  be made vanishing.  That is to say, for radial orbits, the regularisation parameters share the same expression in the RW and H gauges. The SF is thus gauge invariant for RW, H and all other gauges interrelated via a regular transformation gauge vector. 

We thus derive the regularisation parameters entirely in the RW gauge, and confirm their identity with those in the H gauge found by Barack {\it et al.} \cite{baetal02}. 
The RW gauge has the distinct advantage of giving easy access to the components of the perturbation tensor (instead strongly coupled in the H gauge) via the RWZ wave-functions. 

We deal from here onwards with radial infall. This implies that i) the odd modes vanish, and the source term for even modes is simplified; ii) there { are not} $m$ modes; iii) the perturbation $K$ vanishes for a fixed $\theta$; iv) for symmetry, the terms  $F^t_\text{self}$ and $F^r_\text{self}$ don't vanish, conversely to $F^\theta_\text{self}=F^\phi_\text{self}=0$.

\subsection{{ Computation of the regularisation parameters in the RW gauge}}\label{regmsp}

\begin{figure}[h!]
    \centering
    \includegraphics[width=0.3\linewidth]{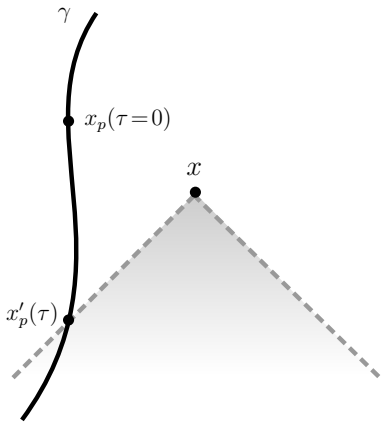}
    \caption{We define $x'_p(\tau)=(t'_p,r'_p)$ any point of the $\gamma$ world-line followed by the particle, and $x_p=(t,r_p)=x'_p(\tau=0)$ the point where we finally evaluate the SF, $\tau$ being the proper time; $x=(t,r)$ indicates a point taken in the neighbourhood of $x_p$ where the field $\psi(x)$ is evaluated, before taking the limit $x \rightarrow x_p$. }
    \label{fig01}
\end{figure}

The value $x'_p(\tau)=(t'_p,r'_p)$ represents any point of the $\gamma$ world-line followed by the particle, while $x_p=(t,r_p)=x'_p(\tau=0)$ the point where the SF is evaluated, Fig. (\ref{fig01}), $\tau$ being the proper time. Further, $x=(t,r)$ indicates a point taken in the neighbourhood of $x_p$ where the field $\psi(x)$ is evaluated, before taking the limit $x \rightarrow x_p$.

We define the Green function $G\left(x,x_p(\tau)\right)$ as
\beq
\psi(x)=\int_{-\infty}^{0^+}G\big(x,x'_p(\tau)\big)d\tau~.
\label{link:psi:G}
\eeq

When associated to the RWZ equation \cite{rewh57,ze70c}, we get 
\beq
\Big[-\partial_t^2+\partial_{r^*}^2-V(r)\Big]G = \widehat{\cal{G}}(r)\delta(r-r'_p)\delta(t-t'_p)+
 \widehat{\cal{F}}(r)\partial_r\delta(r-r'_p)\delta(t-t'_p)~, 
\label{rwz:G}
\eeq

where the even potential is 

\beq
V^{\ell}_e(r) = 2 f \frac{\lambda^2(\lambda+1)r^3+3\lambda^2Mr^2+9\lambda M^2r+9M^3}{r^3(\lambda r+3M)^2 }~,
\label{veven}
\eeq
where $\lambda=(\ell-1)(\ell+2)/2$, and the source term coefficients are

\beq
\begin{aligned}
\widehat{\cal{F}}(r)&=-\frac{\kappa rf^2(r)}{4(\lambda+1)(\lambda r+3M)}~,\\
\widehat{\cal{G}}(r)&=\frac{\kappa rf}{2(\lambda+1)(\lambda r+3M)}\left[\frac{r(\lambda+1)-M}{2r}-\frac{3M \mE^2}{\lambda r+3M}\right]~,
\end{aligned}
\eeq
for $\kappa=8\pi m_0Y^{\ell0}=4m_0\sqrt{2\pi L}$, and $\mE=f(r_p)u^t$.  

According to the properties on distributions, Appendix (B) in Part I, Eq. (\ref{rwz:G}) is rewritten using an alternative version  of the Green function, named $\widehat{G}$ and defined such that 

\beq
G(x,x'_p)=\left[\calhat{Q}(r'_p)-\calhat{F}(r'_p)\partial/\partial r'_p\right]\widehat{G}(x,x'_p)~,
\label{green:reduced}
\eeq
where $\calhat{Q}(r'_p)\defeq\left[\calhat{G}(r)-d\calhat{F}(r)/dr\right]_{r=r'_p}$. Using $\delta(r-r'_p)=f(r'_p)^{-1}\delta(r^*-r'^*_p)$ and $\mathcal{Z}\defeq-\partial^2_t+\partial^2_{r^*}-V(r)$, Eq. (\ref{rwz:G}) becomes

 \bea
 \mathcal{Z}G(x,x'_p)&=\left[\calhat{G}(r)-\frac{d\calhat{F}(r)}{dr}\right]_{r=r'_p}\delta(r-r_p')\delta(t-t_p')-\calhat{F}(r'_p)\frac{\partial}{\partial r'_p}\delta(r-r_p')\delta(t-t_p') \\
 &=\left[\calhat{Q}(r'_p)-\calhat{F}(r'_p)\frac{\partial}{\partial r'_p}\right]\delta(r-r_p')\delta(t-t_p')
 ={X}\delta(r-r_p')\delta(t-t_p')
 \eea
 \beq
\mathcal{Z}{X}\widehat{G}={X}\mathcal{Z}\widehat{G}={X}\delta(r-r_p')\delta(t-t_p')~,
 \eeq

\beq
\Big[-\partial_t^2+\partial_{r^*}^2-V(r)\Big]\widehat{G}=f(r'_p)^{-1}\delta(r^*-r'^*_p)\delta(t-t'_p)~.
\label{rwz:reduced:tr}
\eeq

Considering the causal structure of the Green function, it is now useful to introduce the Eddington-Finkelstein coordinates $(u,v)$ \cite{ed24,fi58}. In these new variables, $v= t + r^*$ ingoing and $u= t - r^*$ outgoing, the expression of the wave-operator is simply given by $\partial^2_{r^*}-\partial^2_{t} = -4 \partial_{uv}$. In the same way, casting  $\delta(r^*-r'^*_p)\delta(t-t'_p)$ in $(u,v)$ { variables},  requires { to deal with the  product of a function $\mu(t,r^*)$ with $\delta$}. Under the integral definition of the latter

\beq
\int_{\mu(\mathbb{R}^2)}\delta(x)\phi(x)dx=\int_{\mathbb{R}^2}\delta(\mu(x))\phi(\mu(x))\left|J_\mu\right|dx~,
\eeq
where $\phi\in\cal{D}(\mathbb{R}^2)$ and $\left|J_\mu\right|$ is the determinant of the Jacobian matrix associated to $\mu$ $\left|J_\mu\right|=2$. Then, in $(u,v)$ coordinates, Eq. (\ref{rwz:reduced:tr}) turns into

\beq
\Big[4\partial_{uv}+V(r)\Big]\widehat{G}=2f(r'_p)^{-1}\delta(u-u'_p)\delta(v-v'_p)~.
\label{rwz:reduced:uv}
\eeq

The Mode-Sum regularisation in the RW gauge, and thus the determination of the SF, will be achieved by the undertaking of two pursuits (i) the analytic computation of the regularisation parameter; (ii) the numerical computation of the $\ell$-modes of the retarded force by solving the RWZ equation.

For the analytic venture, the regularisation of the SF by the Mode-Sum technique requires the evaluation of the divergency, {\it i.e.} the singular part $F^{\alpha}_\text{S}$ of the retarded solution. The singular part is fitted by a $1/L$ power series, of which coefficients are the regularisation parameters. The computation of the latter is based on a local analysis, {\it i.e.} at the neighbourhood of the particle, of the wave-function $\psi$, or more exactly of its associated Green's function.  The technique consists of a perturbative expansion of the Green function modes in a small spacetime region around the particle for great values of $\ell$.  

To accomplish the local analysis, we write the Green function $G$ in a reduced form that takes into account the causal nature of its support, {\it i.e.}, its non-zero value in the future light cone of $x'_p$. Then, we expand the reduced Green function in powers of $1/L$, where each term of the series, namely $G_n$, will also be locally expanded for $x\to x_p$, such that the coefficients of the expansion $G_n$will be function of $x_p$ and $x'_p$. 

The next step considers integrating $G$ with respect to the proper time $\tau$. To this end, we have to express the $G_n$ coefficients  explicitly in terms of $\tau$. This is achieved by a Taylor series of $G_n\left(x'_p(\tau)\right)$ around $\tau=0$.   Finally, we integrate $G$ and $\partial_r G$ to get the asymptotic behaviour of $\psi^\ell$ and $\partial_r\psi^\ell$ when $\ell\to\infty$. The computation of the other derivatives 
 $\partial_t^n\partial_r^mG$, is performed by borrowing from Part I the relationships on partial derivatives, which were used for the jump conditions to get $\partial_t^n\partial_r^m\psi^{\ell\to\infty}$ quantities for which $n+m\leq3$.

We will gain access to the behaviour of the wave-function and its derivatives versus $\ell$, thereby testing the convergence of our numerical code for very high modes. We will then compute the $\ell$-modes of the perturbations $H^\ell_{1,2}$, { K} and of the retarded force $F^{\alpha\ell}[h^{{\text ret}\ell}_\ab]$, for large values of $\ell$, thereby accomplishing the other (numerical) venture. 

The reader may skip this very technical discussion and get directly to the results expressed by Eqs. (\ref{F:sing:RP}). Otherwise, we assume the reader be well acquainted with the Mode-Sum by Barack \cite{barack2000,barack2001} and the coming after literature. 

\subsubsection{Computation strategy and detailed description}

For the asymptotic behaviour of $\partial_t^n\partial_r^m\psi^\ell$ when $\ell\to\infty$ for $n+m\leq 3$, that is up the third derivative of the wave-function, we apply our strategy through the following steps 

\begin{itemize}
\item {\it a. Reduced Green's function.}
\item {\it b. Expansion of the reduced Green function in powers of ${1/L}$ around ${x=x_p}$.}
\item {\it c. Reconstruction of the Green function ${G(x,x'_p)}$.}
\item {\it d. Expansion around ${x'_p=x_p}$.}
\item {\it e. Computation of $\psi^{\ell\to\infty}(x_p)$, $\partial_r\psi^{\ell\to\infty}(x_p)$, and $\partial_t^n\partial_r^m\psi^{\ell\to\infty}(x_p)$.}
\end{itemize}


\paragraph{Reduced Green's function.}
The causal structure allows to rewrite $\widehat{G}$ in a reduced form $g(x,x'_p)$. Indeed, $\widehat{G}(x,x'_p)$ has support in the future light cone of $x'_p$, so $\widehat{G}(u<u'_p,v)=\widehat{G}(u,v<v'_p)=0$. Thus

\beq
\widehat{G}(x,x'_p)=2f(r'_p)^{-1}g(x,x'_p)\cal{H}(u-u'_p)\cal{H}(v-v'_p)~,
\label{green:causal}
\eeq
where $\cal{H}(u-u'_p)\cal{H}(v-v'_p)$ are Heaviside or step distributions, which confine the support of $\widehat{G}(x,x'_p)$  to  the area made by all points { $x$} belonging to the future light cone of $x'_p$. By inserting Eq. (\ref{green:causal}) into Eq. (\ref{rwz:reduced:uv}), we express the wave-operator applied to $\widehat{G}$

\beq
\partial_{uv}\Big[g(x,x'_p)\cal{H}(u-u'_p)\cal{H}(v-v'_p)\Big]
=\partial_u\Big[\partial_vg\cal{H}_{u'_p}\cal{H}_{v'_p}+g\cal{H}_{u'_p}\delta_{v'_p}\Big]\\
=\partial_{uv}g\cal{H}_{u'_p}\cal{H}_{v'_p}+\partial_vg\delta_{u'_p}\cal{H}_{v'_p}+\partial_ug\cal{H}_{u'_p}\delta_{v'_p}+g\delta_{u'_p}\delta_{v'_p}~.
\eeq

Equation (\ref{rwz:reduced:uv}) involves four distinct types of quantities
\begin{itemize}
  \item (i) $(\cdots)\times\cal{H}_{u'_p}\cal{H}_{v'_p}$,
  \item (ii) $(\cdots)\times\delta_{u'_p}\cal{H}_{v'_p}$, 
  \item (iii) $(\cdots)\times\cal{H}_{u'_p}\delta_{v'_p}$, 
  \item (iv) $(\cdots)\times\delta_{u'_p}\delta_{v'_p}$.
\end{itemize} 

The action of each term relies upon the behaviour along the characteristic lines $u=u'_p$ and $v=v'_p$.
\begin{itemize}
\item {\it (i)}. For $u>u'_p$ and $v>v'_p$, only the term (i) has a contribution; $g$ satisfies the homogeneous equation associated to Eq. (\ref{rwz:reduced:uv}).
\item {\it (ii)}. If $u=u'_p$ is constant, only the term (ii) has a contribution;  then $\partial_v g(u'_p,v)=0$.
\item {\it (ii)}. If $v=v'_p$ is constant, only the term (iii) has a contribution; then $\partial_u g(u,v'_p)=0$.
\item {\it (iv)}. On the world line $(u,v)=(u'_p,v'_p)$, the coefficient of term (iv) must be equal to the coefficient of the source term of Eq. (\ref{rwz:reduced:uv}); then $g(u'_p,v'_p)=1$.
\end{itemize}
According to {\it (i)}, we have
\beq
4\partial_{uv}g+V(r)g=0~,\quad \forall\ u>u'_p\text{ and }v>v'_p~,
\label{rwz:causal}
\eeq
while according to {\it (ii)}, {\it (iii)} and {\it (iv)}, we have

\beq
g(u=u'_p,v)=g(u,v=v'_p)=1~.
\label{rwz:causal:CI}
\eeq

\begin{figure}[h!]
\centering
\includegraphics[width=0.5\linewidth]{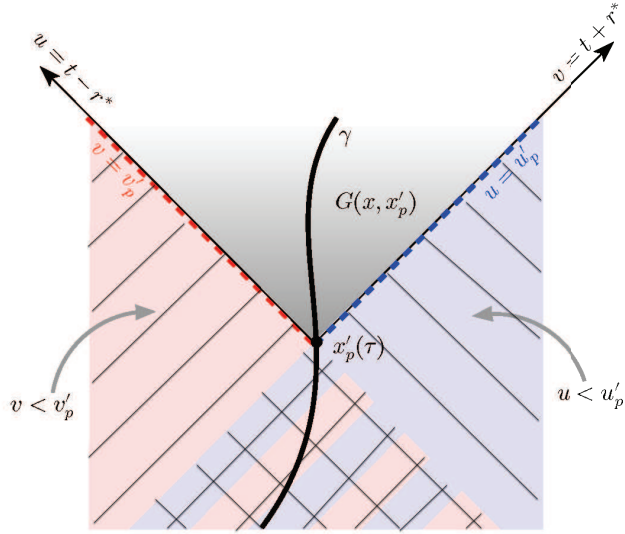}
\caption{The reduced Green function $\widehat{G}(x,x'_p)$ has support in the future light cone of $x'_p$ (dark grey area); thus $\widehat{G}(u<u'_p,v)=0$ (blue area with top-right oblique lines) and $\widehat{G}(u,v<v'_p)=0$ (pink area with top-left oblique lines).}
\label{fig02}
\end{figure}

Equation (\ref{rwz:causal:CI}) is in fact the initial condition to be associated with Eq. (\ref{rwz:causal}); it ensures the uniqueness of the solution. Figure (\ref{fig02}) shows the support of the reduced Green function. 

\paragraph{Expansion of the reduced Green function in powers of ${1/L}$ around ${x=x_p}$.}

%

We are now looking for a solution $g$ of Eqs. (\ref{rwz:causal},\ref{rwz:causal:CI}) near the evaluation point $x=x_p$, that is $r=r_p$while considering large values of $\ell$. Thus, we Taylor expand the quantities around $r=r_p$, and express them as power series in $1/L$. For dealing with both very small quantities such as the spatial separation $r-r_p$ and large quantities proportional to $L$, we introduce new variables of the product form $L\times\textit{small spatial separation}$; these variables are called "neutral" by Barack \cite{barack2000}. We first consider this procedure for the potential. In the neighbourhood of $r=r_p$, or similarly around $r^*=r^*_p$, we have

\bea
V(r)=&V(r_p)+\left.\frac{dV}{dr^*}\right|_{r_p^*}\left(r^*-r_p^*\right)+\frac{1}2\left.\frac{d^2V}{dr^{*2}}\right|_{r_p^*}\left(r^*-r_p^*\right)^2
+\mathcal{O}\left((r^*-r_p^*)^3\right) \\
=&V^{(0)}(r_p)+V^{(1)}(r_p)\Delta r_p^*+V^{(2)}(r_p)\Delta r_p^{*2}+\mathcal{O}(\Delta r_p^{*3})~,
\eea
with
\bea
&V^{(0)}(r_p)=V(r_p)~,\\
&V^{(1)}(r_p)=f(r_p)\left.\frac{dV}{dr}\right|_{r_p}~,\\
&V^{(2)}(r_p)=\frac{1}2\left[f(r_p)\left.\frac{df}{dr}\right|_{r_p}\left.\frac{dV}{dr}\right|_{r_p}+f^2(r_p)\left.\frac{d^2V}{dr^2}\right|_{r_p}\right]~,
\eea
and $\Delta r_p^*=r^*-r^*_p$. The asymptotic behaviour of $V^{(0)}$, $V^{(1)}$ and $V^{(2)}$ for $1/L\to0$ is
\beq
V^{(0)}(r_p)=\frac{f}{r_p^2}\left[L^2-\left(\frac{6M}{r_p}+\frac{1}4\right)\right]+\mathcal{O}\left(L^{-1}\right)
=\devt{V}{0}{2}L^2+\devt{V}{0}{0}+\mathcal{O}\left(L^{-1}\right)~,
\eeq
\beq
V^{(1)}(r_p)=\frac{f}{r_p^2}\left[\frac{6M-2r_p}{r_p^2}L^2-\left(\frac{96M^2-33Mr-r_p^2}{2r_p^3}\right)\right]
+\mathcal{O}\left(L^{-1}\right)
=\devt{V}{1}{2}L^2+\devt{V}{1}{0}+\mathcal{O}\left(L^{-1}\right)~,
\eeq
\beq
V^{(2)}(r_p)=\frac{f}{r_p^2}\left[\frac{60M^2-40Mr_p+6r_p^2}{2r_p^4}L^2
-\frac{1152M^3-810M^2r_p+124Mr_p^2+3r_p^3}{4r_p^5}\right]
+\mathcal{O}\left(L^{-1}\right)\\
=\devt{V}{2}{2}L^2+\devt{V}{2}{0}+\mathcal{O}\left(L^{-1}\right)~.
\eeq

\begin{figure}[h!]
\centering
\includegraphics[width=0.5\linewidth]{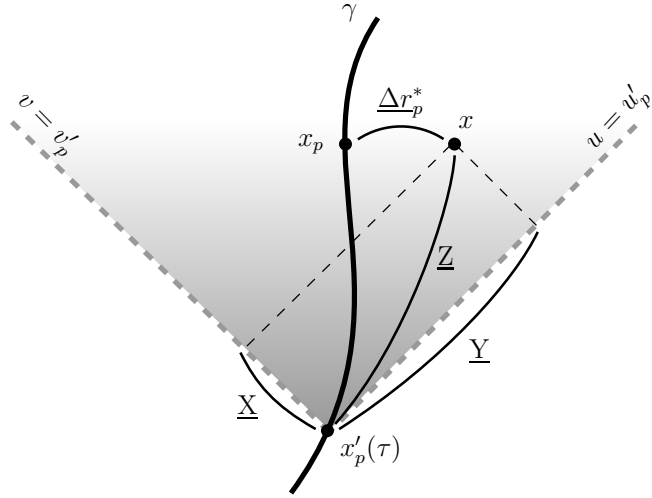}
\caption{Geometric representation of the neutral variables used in the local analysis of $g(x,x'_p)$. The grey area shows the support of the Green function corresponding to the set of points belonging to $x$, the chronological future of $x'_p$.}
\label{fig03}
\end{figure}

In this notation, $\devt{V}{k}{n}$ refers to the $n$-th Taylor coefficient in $1/L$ of the $k$-th coefficient in $\Delta r^*_p$. 
Figure (\ref{fig03}) shows the geometric representation of the neutral variables.

The expanded potential becomes
\beq
V(r_p)=V^{(0)}(r_p)+ \ub{\devt{V}{1}{2}(r_p)L^2\Delta r_p^*}_{\mathcal{O}(L)}+\ub{\devt{V}{1}{0}(r_p)\Delta r_p^*}_{\mathcal{O}(L^{-1})}+\ub{\devt{V}{2}{2}(r_p)L^2\Delta r_p^{*2}}_{\mathcal{O}(1)}+ 
\ub{\devt{V}{2}{0}(r_p)\Delta r_p^{*2}}_{\mathcal{O}(L^{-2})}+\calo{\Delta r_p^{*3}}~,
\label{dev:V}
\eeq
where we labelled the order of each term. The terms such as $L^n\Delta r_p^{*n}$ $n\in\mathbb{N}$ are of $0{^\text{th}}$ { order}, and do not catch the behaviour of $V$ with respect to $L$ because $\cal{O}(L^{-1})\sim\cal{O}(\Delta r_p^*)$. We then choose to introduce the neutral variables (underlined) for which the product form can be appraised as constant. We define
\beq
\neutral{\Delta }r_p^*\defeq L\Delta r_p^*~.
\label{neutral:Deltars}
\eeq

Truncating the expansion in Eq. (\ref{dev:V}) at $\calo{L^{-1}}$, we obtain
\beq
V(r_p)=\frac{f(r_p)}{r_p^2}\Big[L^2+\left(\nu_1\neutral{\Delta}r_p^{*}\right)L+\left(\nu_2+\nu_3\neutral{\Delta}r_p^{*2}\right)\Big]+\calo{L^{-1}}~,
\eeq
with
\bea
&\nu_1=\frac{2}{r_p}\left(\frac{3M}{r_p}-1\right)~,\\
&\nu_2=-\frac{6M}{r_p}-\frac{1}4~,\\
&\nu_3=\frac{1}{r_p^2}\left(3-\frac{20M}{r_p}+\frac{30M^2}{r_p^2}\right)~.
\eea

Similarly to Eq. (\ref{neutral:Deltars}), we introduce $\nX$ and $\nY$ as two neutral variables such that
\beq
\nX\defeq\frac{1}2\rho(r_p)L(u-u'_p)\quad\text{and}\quad \nY\defeq\frac{1}2\rho(r_p)L(v-v'_p)~,
\eeq
where the pre-factor $\rho(r_p)\defeq f(r_p)^{1/2}/r_p$ simplifies Eq. (\ref{rwz:causal}) after the change of variables 
\beq
\partial_{uv}g=(1/4)\rho(r_p)^2L^2\partial_{\nX\nY}g
\eeq
is made. In addition, we introduce another neutral variable $\nZ$
\beq
\nZ\defeq2\sqrt{\nX\nY} = \rho(r_p)L\sqrt{(u-u'_p)(v-v'_p)} = \nZ=(L/r_p)s~,
\eeq
where $s$ is the geodesic distance between the point $x$ and the point $x'_p$. Indeed, in the SD metric, we have $ds^2=-fdudv$, and therefore $s=\int_{x'_p}^{x}\left|g_\ab dx^\alpha dx^\beta\right|^{1/2}\approx\sqrt{f(r_p)(u-u'_p)(v-v'_p)}$. We define also the variable 
$\neutral{\Delta r}'^*_p$ as
\beq
\neutral{\Delta r}'^*_p\defeq\rho(r_p)L(r'^*_p-r^*_p)~.
\eeq

The equation to be solved is now

\beq
\partial_{\nX\nY}g+
\Big[1+\left(\nu_1\neutral{\Delta}r_p^{*}\right)L^{-1}+\left(\nu_2+\nu_3\neutral{\Delta}r_p^{*2}\right)L^{-2}+\calo{L^{-3}}\Big]g=0~,
\eeq
where the reduced Green function $g$ is to be expressed as a power series of $1/L$,  { whose}   coefficients   are function of $\neutral{\Delta r}^*_p$, $\neutral{\Delta r}'^*_p$ and $\nZ$ only
\beq
g=\sum^{\infty}_{k=0}L^{-k}g_k(\neutral{\Delta r}^*_p,\neutral{\Delta r}'^*_p,\nZ)~.
\eeq

However, from a practical point of view,  to get the desired accuracy, it is sufficient to truncate the sum at $k=2$. Equation (\ref{rwz:causal}) becomes

\beq
\sum^{2}_{k=0}L^{-k}\partial_{\nX\nY}g_k+\Big[1+(f_1\Delta_r)L^{-1}+(f_2+f_3\Delta_r^2)L^{-2}+
\mathcal{O}(L^{-3})\Big]\sum^{2}_{k=0}L^{-k}g_k=0~.
\eeq

Now, by identifying powers of $L$, we will have a hierarchical system of equations supplemented by the initial conditions, Eq. (\ref{rwz:causal:CI}), of the form

\beq
\begin{aligned}
&\partial_{\nX\nY}g_k+g_k=S_k~,\\
&g_k(u=u'_p,v)=g_k(u,v=v'_p)=\delta_{k0}~.
\end{aligned}
\label{rwz:causal2}
\eeq

Concretely, we have
\begin{align}
    &\text{order 0 :}\quad \partial_{\nX\nY}g_0+g_0=0~,
    \label{rwz:causal:ordre0}\\
    &\text{order 1 :}\quad \partial_{\nX\nY}g_1+g_1=-\nu_1\neutral{\Delta r}^*_pg_0~,
    \label{rwz:causal:ordre1}\\
    &\text{order 2 :}\quad \partial_{\nX\nY}g_2+g_2=-\nu_1\neutral{\Delta r}^*_pg_1-(\nu_2+\nu_3\neutral{\Delta r}^{*2}_p)g_0~.
    \label{rwz:causal:ordre2}
\end{align}

Through a a change of variable $\nZ=2\sqrt{\nX\nY}$, the left hand side changes into

\beq
\frac{\partial^2}{\partial\nX\partial\nY}g_k+g_k=\nZ^2\frac{\partial^2g_k}{\partial \nZ^2}+\nZ\frac{\partial g_k}{\partial \nZ}+\nZ^2g_k~.
\eeq
Thus, Eq. (\ref{rwz:causal:ordre0}) implies to solve a Bessel equation of order 0
\beq
\nZ^2\frac{\partial^2g_0}{\partial \nZ^2}+\nZ\frac{\partial g_0}{\partial \nZ}+(\nZ^2-0^2){ g_0}=0~,
\eeq
which solution of is a Bessel function of the first kind of order 0

\beq
g_0=J_0(\nZ)~.
\eeq

Equations (\ref{rwz:causal:ordre1},\ref{rwz:causal:ordre2}) are also Bessel equations with source terms. By working on the relationships between neutral variables $\neutral{\Delta r}^*_p$, $\neutral{\Delta r}'^*_p$, $\nX$ and $\nY$, we can rewrite the source terms $S_k$ solely as function of $\nZ$ and of the difference $\nY-\nX=\neutral{\Delta r}^*_p-\neutral{\Delta r}'^*_p$. Implementing the relationships in Tab. \ref{TAB:bessel}, the solutions of the Eqs. (\ref{rwz:causal:ordre1},\ref{rwz:causal:ordre2}) are built compatibly with the initial conditions of Eq. (\ref{rwz:causal2})

\beq
g_1=-\frac{1}4\nu_1\nZ J_1(\nZ)(\neutral{\Delta r}^*_p+\neutral{\Delta r}'^*_p)~,
\eeq
\beq
g_2=-\frac{1}6\nZ J_1(\nZ)\Big[\nu_3\left(\neutral{\Delta r}^{*2}_p+\neutral{\Delta r}^*_p\neutral{\Delta r}'^*_p+\neutral{\Delta r}'^{*2}_p\right)+\nu_2\Big]
+\frac{1}{96}\nZ^2 J_2(\nZ)\left[3\nu_1^2\left(\neutral{\Delta r}^{*}_p+\neutral{\Delta r}'^{*}_p\right)^2-8\nu_3\right]
+\frac{1}{96}\nu_1^2\nZ^3J_3(\nZ)~.
\eeq

\renewcommand{\arraystretch}{1.4} 
\begin{table}[!htb]
  \centering
  \begin{tabular}{c|c}
  \hline
  \hline
  $S$ & Solution of $\partial_{\nX\nY}g+g=S$\\ \hline
  $0$ & $J_0(\nZ)$\\
  $J_0(\nZ)$ & $\nZ J_1(\nZ)/2$\\
  $(\nY-\nX)J_0(\nZ)$ & $(\nY-\nX)\nZ J_1(\nZ)/4$\\
  $(\nY-\nX)^2J_0(\nZ)$ & $\left[\nZ^2 J_2(\nZ) + 2(\nY-\nX)^2\nZ J_1(\nZ)\right]/12$\\
  $\nZ J_1(\nZ)$ & $\nZ^2J_2(\nZ)/4$\\
  $(\nY-\nX)\nZ J_1(\nZ)$ & $(\nY-\nX)\nZ^2 J_2(\nZ)/6$\\
  $(\nY-\nX)^2\nZ J_1(\nZ)$ & $\left[\nZ^3 J_3(\nZ) + 3(\nY-\nX)^2\nZ^2 J_2(\nZ)\right]/24$\\
  \hline
  \hline
  \end{tabular}
  \caption{Provision for the solution of the generalised Bessel equation $\partial_{XY}g+g=S$ with a source term $S$ written itself with a Bessel function \cite{barack2000}.}
  \label{TAB:bessel}
\end{table}

\paragraph{Reconstruction of the Green function ${G(x,x'_p)}$.}

%

Given the local behaviour of $g$ for large modes
\beq
g=g_0+g_1L^{-1}+g_2L^{-2}+\calo{L^{-3}}~,
\label{g:expand}
\eeq
we can reconstruct the function $\widehat{G}(x,x'_p)$ linked to $g$ through

\beq
\widehat{G}(x,x'_p)=2f(r_p)^{-1}g(x,x'_p)\cal{H}(u-u'_p)\cal{H}(v-v'_p)~,
\eeq
and itself connected to the Green function, Eqs. (\ref{link:psi:G},\ref{rwz:G}) through Eq. (\ref{green:reduced}), recalled herein

\beq
G=\left[\calhat{Q}(r'_p)-\calhat{F}(r'_p)\frac{d}{dr'_p}\right]\widehat{G}~,
\label{green:reduced2}
\eeq
with

\beq
\calhat{F}(r'_p)=-\kappa \Big[f^2(r'_p)L^{-4}+\calo{L^{-6}}\Big]~,
\label{F:expand}
\eeq
\beq
\calhat{Q}(r'_p)=\kappa\rho(r'_p)^2\left[L^{-2}+\left(\frac{9}4-\frac{4M}{r'_p}\right)L^{-4}+\calo{L^{-6}}\right]~.
\label{Q:expand}
\eeq

According to Eq. (\ref{green:reduced2}), the determination of $G$ implies the derivative of $\widehat{G}$ with respect to $r'_p$. This term necessarily involves the derivatives of $\neutral{\Delta r}'^*_p$, $\nZ$, and $\cal{H}_{u'_p}\cal{H}_{v'_p}$, listed here below
\beq
\frac{d\nDrpp}{dr'_p}=\frac{dr'^*_p}{dr'_p}\frac{d}{dr'^*_p}\nDrpp=L\rho(r_p)f(r'_p)^{-1}~,
\label{dDrppdrpp}
\eeq

\beq
\frac{d\nZ}{dr'_p}=L\rho(r_p)f(r'_p)^{-1}\frac{d}{dr'^*_p}\sqrt{(u-u_p)(v-v_p)}
=L\rho(r_p)f(r'_p)^{-1}\left(\frac{\nY-\nX}{\nZ}\right)
=L\rho(r_p)f(r'_p)^{-1}\left(\frac{\nDrp-\nDrpp}{\nZ}\right)~,
\label{dzdrpp}
\eeq

\beq
\frac{d}{dr'_p}\Big[\cal{H}(u'-u_p)\cal{H}(v-v'_p)\Big]=
f(r'_p)^{-1}\Big[\cal{H}_{v'_p}\delta(u-u'_p)-\cal{H}_{u'_p}\delta(v-v'_p)\Big]~.
\label{dHHdrpp}
\eeq

\begin{figure}[h!]
\centering
\includegraphics[width=0.7\linewidth]{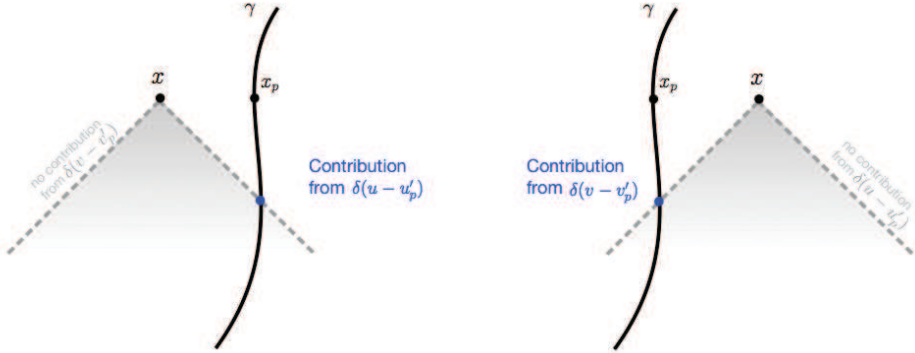}
\caption{{ The value of the derivative, Eq. (\ref{dHHdrpp}), depends on the limit, either taken on the right $r\to r_p^+$ or on the left hand-side $r\to r_p^-$ of the evaluation point. If $r\to r_p^-$ - left panel - the term involving $\delta(v-v'_p)$ in Eq. (\ref{dHHdrpp}) is null; instead, if $r\to r_p^+$ - right panel - the term involving $\delta(u-u'_p)$ in Eq. (\ref{dHHdrpp}) is null.}}
\label{fig04}
\end{figure}

Equation (\ref{dHHdrpp}) involves two non-vanishing terms of which the contribution depends on how the evaluation point is reached, from the right $r\to r_p^+$ or from the left $r\to r_p^-$, Fig. (\ref{fig04}). Therefore, for a simpler notation we adopt two additional neutral variables, displayed with the others in Tab. \ref{TAB:liste:neutral}

\renewcommand{\arraystretch}{1.4} 
\begin{table}[!htb]
  \centering
  {\small
  \begin{tabular}{cc}
  \hline
  \hline  
  Variable & Expression \\ \hline
  $\nDrp$ & $L(r^*-r^*_p)$\\
  $\nDrpp$ & $\rho(r_p)L(r'^*_p-r^*_p)$\\
  $\nX$ & $1/2\rho(r_p)L(u-u'_p)$\\
  $\nY$ & $1/2\rho(r_p)L(v-v'_p)$\\
  $\nZ$ & $\rho(r_p)L\sqrt{(u-u'_p)(v-v'_p)}$\\
  $\neutral{\omega}^+$ & $2\nX$\\
  $\neutral{\omega}^-$ & $2\nY$\\
  $\ntau$ & $-L\tau$ \\
  \hline
  \hline
  \end{tabular}
  \caption{List of standard and neutral variables used and their expressions as function of $L$.}
  \label{TAB:liste:neutral}
  }
\end{table}

\beq
\begin{aligned}
\neutral{\omega}^+&=2\nX\\&=L\rho(r_p)(u-u'_p)~,
\end{aligned}
\quad\text{and}\quad
\begin{aligned}
\neutral{\omega}^-&=2\nY\\&=L\rho(r_p)(v-v'_p)~.
\end{aligned}
\eeq
 { The above} change of variable gives
\beq
\begin{aligned}
\delta(\neutral{\omega}^+)
=\delta(u-u'_p)\left|\frac{d\neutral{\omega}^+}{du}\right|^{-1}_{u=u'_p}
=L^{-1}\rho(r_p)^{-1}\delta(u-u'_p)~,
\end{aligned}
\eeq
and
\beq
\begin{aligned}
\delta(\neutral{\omega}^-)
=\delta(v-v'_p)\left|\frac{d\neutral{\omega}^-}{dv}\right|^{-1}_{v=v'_p}
=L^{-1}\rho(r_p)^{-1}\delta(v-v'_p)~.
\end{aligned}
\eeq
Therefore,
\beq
\frac{d}{dr'_p}\Big[\cal{H}{u'_p}\cal{H}{v'_p}\Big]=\pm L\rho(r_p)f(r'_p)^{-1}\delta(\neutral{\omega}^\pm)~.
\label{dHHdrpp-bis}
\eeq
From Eqs. (\ref{green:causal},\ref{g:expand},\ref{Q:expand}), the first term of $G=\calhat{Q}(r'_p)\widehat{G}-\calhat{F}(r'_p)d\widehat{G}/dr'_p$ transforms into

\bea
\calhat{Q}(r'_p)\widehat{G}=& 2\calhat{Q}(r'_p)f(r'_p)^{-1}g(x,x'_p)\cal{H}(u-u'_p)\cal{H}(v-v'_p)\\
= &\left[\kappa\frac{f(r_p)}{2r_p}\left[L^{-2}+\left(\frac{9}4-\frac{4M}{r_p}\right)L^{-4}\right]+\mathcal{O}(L^{-6})\right]2f^{-1}(r_p)\Big[g_0+g_1L^{-1}+g_2L^{-2}+\mathcal{O}(L^{-3})\Big]\theta_{u_p}\theta_{v_p}\\
=& \frac{\kappa}{r'_p}\left\{g_0L^{-2}+g_1L^{-3}+\left[g_2+g_0\left(\frac{9}4-\frac{4M}{r'_p}\right)\right]L^{-4}
+\calo{L^{-5}}\right\}\cal{H}_{u'_p}\cal{H}_{v'_p}~.
\label{QG}
\eea

In the same way, for the second term, from Eqs. (\ref{green:causal},\ref{g:expand},\ref{dHHdrpp-bis})

\bea
\frac{d\widehat{G}}{dr'_p}=&\frac{d}{dr'_p}\Big[2f^{-1}(r'_p)g\cal{H}(u-u'_p)\cal{H}(v-v'_p)\Big]\\
=&\ub{-\frac{4M}{r'^2_p}f(r'_p)^{-2}g\cal{H}_{u'_p}\cal{H}_{v'_p}}_{d\widehat{G}_1(r'_p)}
+\ub{2f(r'_p)^{-1}\frac{dg}{dr'_p}\cal{H}_{u'_p}\cal{H}_{v'_p}}_{d\widehat{G}_2(r'_p)}
+\ub{2f(r'_p)^{-1}g\frac{d}{dr'_p}\cal{H}_{u'_p}\cal{H}_{v'_p}}_{d\widehat{G}_3(r'_p)}~,
\label{dGdrpp}
\eea
\begin{align}
&d\widehat{G}_1(r'_p)=\left[-\frac{4M}{r'^2_p}f^{-2}(r'_p)g_0+\mathcal{O}(L^{-1})\right]\cal{H}_{u'_p}\cal{H}_{v'_p}~,\label{dG1}\\
&\begin{aligned}
d\widehat{G}_2(r'_p)=&\Big[2f(r'_p)^{-1}\frac{dg_0}{dr'_p}+2f(r'_p)^{-1}\frac{dg_1}{dr'_p}L^{-1}
+\mathcal{O}(L^{-1})\Big]\cal{H}_{u'_p}\cal{H}_{v'_p}~,\label{dG2}
\end{aligned}\\
&
\begin{aligned}
d\widehat{G}^\pm_3(r'_p)=&\pm2L\rho(r_p)f(r'_p)^{-2}g_0\delta(\neutral{\omega}^\pm)
\pm2\rho(r_p)f(r'_p)^{-2}g_1\delta(\neutral{\omega}^\pm)+\mathcal{O}(L^{-1})~.
\end{aligned}
\label{dG3}
\end{align}

In Eq. (\ref{dG2}) the derivative of $\nZ$, computed with Eq. (\ref{dzdrpp}), and of $J_n(\nZ)$ are used. Useful properties are

\bea
&\frac{d}{d\nZ}\Big[\nZ^nJ_n(\nZ)\Big]=\nZ^nJ_{n-1}(\nZ)~,\\
&J_{-1}(\nZ)=-J_1(\nZ)~.
\eea

By multiplying Eq. (\ref{dGdrpp}) by Eq. (\ref{F:expand}), we get

\begin{align}
&\calhat{F}d\widehat{G}_1=\kappa\frac{4M}{r'^2_p}J_0(\nZ)L^{-4}+\calo{L^{-5}}~,\label{FdG1}\\
&\begin{aligned}
\calhat{F}d\widehat{G}_2=
\kappa\rho(r_p)\left(\nDrp-\nDrpp\right)\frac{J_1(\nZ)}{\nZ}L^{-3}+\frac{1}2\kappa\nu_1\rho(r_p)\left(\neutral{\Delta r}^{2*}_p-\neutral{\Delta r}'^{2*}_p\right)J_0(\nZ)L^{-4}+\frac{1}2\kappa\nu_1\rho(r_p)\nZ J_1(\nZ)L^{-4}+\calo{L^{-5}}~,
\end{aligned}
\label{FdG2}\\
&\begin{aligned}
\calhat{F}d\widehat{G}^\pm_3=\mp2\kappa\rho(r_p)J_0(\nZ)\delta(\neutral{\omega}^\pm)L^{-3}
\mp2\kappa\rho(r_p)g_1\delta(\neutral{\omega}^\pm)L^{-4}+\calo{L^{-5}}~.
\end{aligned}
\label{FdG3}
\end{align}

We inject Eqs. (\ref{QG},\ref{FdG1}-\ref{FdG3}) into Eq. (\ref{green:reduced})  to get $G(x,x'_p)$ in power series of $1/L$

\beq
G=G_0L^{-2}+G^\pm_1L^{-3}+G^\pm_2L^{-4}+\mathcal{O}(L^{-5})~,
\label{G:expand}
\eeq
where the coefficients $G_n$ depend on $r$ through $\nDrp$, and on $r'_p$ through $\nDrpp$
\begin{align}
&G_0=\frac{\kappa}{r'_p}J_0(\nZ)~,\label{G0}\\
&\begin{aligned}
G_1^\pm=&\frac{\kappa}{r_p}\Big[\frac{r_p}{r'_p}g_1-2f(r_p)^{1/2}\left(\nDrp-\nDrpp\right)\frac{J_1(\nZ)}{\nZ}
\pm2J_0(\nZ)\delta(\neutral{\omega}^\pm)\Big]~,
\end{aligned}
\label{G1}\\
&\begin{aligned}
G_2^\pm=&\frac{\kappa}{r_p}\Bigg\{\frac{r_p}{r'_p}g_2+\Bigg[\frac{1}4\left(\frac{r_p}{r'_p}\right)\left(9-\frac{32M}{r'_p}\right)
-\frac{1}2\nu_1f(r_p)^{1/2}\left(\neutral{\Delta r}^{2*}_p-\neutral{\Delta r}'^{2*}_p\right)\Bigg]J_0(\nZ)
-\frac{1}2\nu_1f(r_p)^{1/2}\nZ J_1(\nZ)\pm2g_1\delta(\neutral{\omega}^\pm)\Bigg\}~.\label{G2}
\end{aligned}
\end{align}

\paragraph{Expansion around $x'_p=x_p$.}

The behaviour of the $\psi$ wave-function for large values of $\ell$ is achieved by integrating Eq. (\ref{G:expand}) over the world line. According to Eq. (\ref{link:psi:G}), the integration of $G$, in proper time $\tau$, imposes first rendering the coefficients $G_n (x = x_p)$ explicitly function of $\tau$; put otherwise, expanding $G_n$ in powers of $\tau$ around the evaluation point $x'_p(\tau)=x_p$. We proceed as follows 
\begin{itemize}
\item The evaluation of Eq. (\ref{G:expand}) at $x=x_p$ for $r\to r_p$ and $\nDrp\to0$. Then, all $G_n$ coefficients will be only function of $r'_p$ and $r_p$.
\item All $r'_p$-dependent quantities are expanded in powers of $\tau$ around the point $r'_p=r_p$, that is to say around $\tau=0$ up to order $\tau^2$. This will lead us to introduce a neutral time variable $\ntau\propto L\tau$.
\item t constant $\ntau$, we find the expansion of $G$ in powers of $1/L$ such that
\beq
G=\widetilde{G}_0(\ntau)L^{-2}+\widetilde{G}^\pm_1(\ntau)L^{-3}+\widetilde{G}^\pm_2(\ntau)L^{-4}+\mathcal{O}(L^{-5})~,
\eeq
where the coefficients $\widetilde{G}_n(\ntau)$ explicitly depend upon $\tau$ and $L$ through $\ntau$.
\item The integration of $G$ to determine $\psi^{\ell\to\infty}(t,r_p)$ { involves} terms proportional to $\ntau^kJ_n(\ntau)$, with $k,n\in\mathbb{N}$. Caution is to be exercised, improper integrals arise.
\item The whole procedure is applicable to $\partial_r G$ to get $\partial_r\psi^{\ell\to\infty}(t,r_p)$.
\end{itemize}

So, first we take Eqs. (\ref{G0}-\ref{G2}), while requiring that $r\to r_p$ and $\nDrp\to0$

\beq
G_0=\frac{\kappa}{r'_p}J_0(\nZ)~,\label{G0rp}
\eeq
\beq
G_1^\pm=\frac{\kappa}{r_p}\left[\frac{r_p}{r'_p}g_1-2f(r_p)^{1/2}\nDrpp\frac{J_1(\nZ)}{\nZ}\pm2J_0(\nZ)\delta(\neutral{\omega}^\pm)\right]~,\label{G1rp}
\eeq
\beq
G_2^\pm=\frac{\kappa}{r_p}\Bigg\{\frac{r_p}{r'_p}\overline{g}_2+\Bigg[\frac{1}4\left(\frac{r_p}{r'_p}\right)\left(9-\frac{32M}{r'_p}\right)
-\frac{1}2\nu_1f(r_p)^{1/2}\neutral{\Delta r}'^{2*}_p\Bigg]J_0(\nZ)
-\frac{1}2\nu_1f(r_p)^{1/2}\nZ J_1(\nZ)\pm2\overline{g}_1\delta(\neutral{\omega}^\pm)\Bigg\}~,
\label{G2rp}
\eeq
with
{
\bea
\overline{g}_1=&-\frac{1}4f_1\nZ J_1(\nZ)\nDrpp~,\\
\overline{g}_2=&-\frac{1}6\nZ J_1\left[f_3\neutral{\Delta r}'^{*2}_p+3\nu_2\right]+\frac{1}{96}\nZ^2J_2(\nZ)\left[3\nu_1^2\neutral{\Delta r}'^{*2}_p-8f_3\right]+\frac{1}{96}f_1^2\nZ^3J_3(\nZ)~,
\eea
}
wherein all quantities $\neutral{\omega}^\pm$, $\nZ$, $\nDrpp$ are taken at $r=r_p$.
The quantities depending upon $r'_p$, and consequently on $\tau$, in Eqs. (\ref{G0rp}-\ref{G2rp}) are $\nDrpp,\quad  1/r'_p, \quad\nZ\quad$  and $\quad J_n(\nZ)$, named collectively $\cal Q$. 

Thus, an expansion around $r'_p=r_p$ corresponds to an expansion around $\tau=0$, since $r_p=r'_p(\tau=0)$. Let $\Gamma(\tau)$ be one of the quantities $\cal Q$, then the Taylor expansion can be written as

\beq
\Gamma=\Gamma(\tau=0)+\left.\frac{d\Gamma}{d\tau}\right|_{\tau=0}\tau+\frac{1}2\left.\frac{d^2\Gamma}{d\tau^2}\right|_{\tau=0}\tau^2+\frac{1}6\left.\frac{d^3\Gamma}{d\tau^3}\right|_{\tau=0}\tau^3+\cdots~,
\label{taylor:tau}
\eeq
where $\tau$ is a small entity ($\tau \rightarrow 0$) such that $\calo{\tau}\sim\calo{1/L}$. Introducing a neutral time variable

\beq
\ntau=-L\tau~, 
\eeq
for constant $\ntau$ we obtain the following expansion
\beq
\Gamma=\Gamma_0+\Gamma_1(\ntau)L^{-1}+\Gamma_2(\ntau)L^{-2}+\Gamma_3(\ntau)L^{-3}+\cdots~,
\label{taylor:ntau}
\eeq
where the coefficients $\Gamma_n(\ntau)$ depend on $\tau$ and $L$ only through $\ntau$. The coefficients $\Gamma_n(\ntau)$ are shown in Tab. (\ref{TAB:coef:ntau}) for each quantity $\cal Q$.
\renewcommand{\arraystretch}{2.4} 
\begin{widetext}
\begin{center}
\begin{table}[!htb]
  \begin{tabular}{c|c|c|c}
  \hline
  \hline
  $\Gamma$ & $\Gamma_0(\ntau)$ & $\Gamma_1(\ntau)$ & $\Gamma_2(\ntau)$ \\ \hline
  $\nDrpp$ & $-f(r_p)^{1/2}\dot{r}_p^*\ntau$ 
& ${\ds \frac{1}{2}}f(r_p)^{1/2}r_p\ddot{r}'^*_p\ntau^2$ 
& $-{\ds \frac{1}{6}}f(r_p)^{1/2}r_p^2\dot{\ddot{r}}'^{*}_p\ntau^3$\\
  $\ds \frac{1}{r'_p}$   
& $\ds \frac{1}{r_p}$    
& $\ds\frac{\dot{r}'_p}{r_p}\ntau$     
& ${\ds\frac{\dot{r}'^2_p}{r_p}-\frac{\ddot{r}'_p}{2\ntau^2}}$ \\
  $\nZ$      
& $-\dot{s}\ntau$    
& $\frac{1}2r_p\ddot{s}\ntau^2$                  
& $-{\ds\frac{1}{6}}r_p^2\dot{\ddot{s}}\ntau^3$ \\
  $J_n(\nZ)$ & $J_n(\ntau)$       
& ${\ds \frac{1}2r_p\ddot{s}\frac{dJ_n(\ntau)}{d\ntau}\ntau^2}$       
& ${\ds \left[-\frac{1}6r_p^2\dot{\ddot{s}}\frac{dJ_n(\ntau)}{d\ntau}\ntau^3+\frac{1}8r^2_p\ddot{s}^2\frac{d^2J_n(\ntau)}{d\ntau^2}\ntau^4\right]}$\\
  \hline
  \hline
  \end{tabular}
  \caption{Taylor coefficients $\Gamma=\Gamma_0+\Gamma_1L^{-1}+\Gamma_2L^{-2}+\calo{L^{-3}}$ when $\Gamma$ is one of the $\cal Q$ quantities.}
  \label{TAB:coef:ntau}
\end{table}
\end{center}
\end{widetext}

For completion of Tab. (\ref{TAB:coef:ntau}), here below the four-velocity, the four-acceleration and its derivative in the tortoise coordinate

\bea
\dot{r}^*_p=&f(r_p)^{-1}\dot{r}_p~,\\
\ddot{r}^*_p=&f(r_p)^{-2}\left(f(r_p)\ddot{r}_p-f'(r_p)\dot{r}^2_p\right)~,\\
\dot{\ddot{r}}^{*}_p=&f(r_p)^{-3}\Big[\left(2f'(r_p)^2-f''(r_p)f(r_p)\right)\dot{r}^3_p-f'(r_p)f(r_p)\dot{r}_p\ddot{r}_p+f(r_p)^2\dot{\ddot{r}}_p\Big]~,
\eea
where the primes indicate derivation with respect to $r$, while the point to $\tau$. Through the expression

\beq
\frac{ds^2}{d\tau^2}=-f(r'_p)\frac{du'_p}{d\tau}\frac{dv'_p}{d\tau}~,
\eeq
we are led to the normalisation relation $\dot{u}_p\dot{v}_p=f(r_p)^{-1}$. Taking then successive derivatives with respect to $\tau$ at point $r'_p=r_p$, we get $\ddot{u}_p\dot{v}_p+\dot{u}_p\ddot{v}_p=f'(r_p)^{-1}\dot{r}_p$ and $\dot{u}_p\dot{\ddot{v}}_p+2\ddot{u}_p\ddot{v}_p+\dot{\ddot{u}}_p\dot{v}_p=f''(r_p)^{-1}\dot{r}_p^2+f'(r_p)^{-1}\ddot{r}_p$. The following step is the derivation of $s$ with respect to $\tau$ at the { evaluation} point $\tau=0$

\bea
\dot{s}(\tau=0)=&-1~,\\
\ddot{s}(\tau=0)=&\frac{1}2f'(r_p)f^{-1}(r_p)\dot{r}_p~,\\ 
\dot{\ddot{s}}(\tau=0)=&\frac{1}{16f(r_p)^2}\Big[(8f''(r_p)f(r_p)-13f'(r_p)^2)\dot{r}^2_p+8f'(r_p)f(r_p)\ddot{r}\Big]+\frac{1}4f(r_p)\ddot{u}_p\ddot{v}_p~.
\eea

Finally, the derivatives of $J_n(\ntau)$ with respect to $\ntau$ for $n\geq0$

\bea
\frac{dJ_n(\ntau)}{d\ntau}&=J_{n-1}(\ntau)-\frac{n}\ntau J_{n}(\ntau)=\frac{n}\ntau J_{n}(\ntau)-J_{n+1}(\ntau)~,
\eea
\bea
\frac{d^2J_n(\ntau)}{d\ntau^2}&=\frac{1}{2^2}\Big[J_{n-2}(\ntau)-2J_{n}(\ntau)+J_{n+2}(\ntau)\Big]=J_n\left[\frac{n(n+1)}{\ntau^2}-1\right]+\frac{J_{n+1}}\ntau~.
\eea

\paragraph{Computation of $\psi^{\ell\to\infty}(x_p)$, $\partial_r\psi^{\ell\to\infty}(x_p)$, and $\partial_t^n\partial_r^m\psi^{\ell\to\infty}(x_p)$.}

Introducing expansions of $\cal Q$,  Tab. (\ref{TAB:coef:ntau}), in Eqs. (\ref{G0rp}-\ref{G2rp}), we finally express the coefficients of $G$ in function of the proper time

\beq
G=\widetilde{G}_0(\ntau)L^{-2}+\widetilde{G}^\pm_1(\ntau)L^{-3}+\widetilde{G}^\pm_2(\ntau)L^{-4}+\mathcal{O}(L^{-5})~,
\eeq
with
\begin{widetext}
\begin{align}
\widetilde{G}_0(\ntau)=&\frac{\kappa}{r_p}J_0(\ntau)\label{G0tau}~,\\
\widetilde{G}^\pm_1(\ntau)=&\pm\frac{4\kappa\delta(\neutral{\omega}^\pm)}{r_p}J_0(\ntau)-\frac{\kappa\nu_1f(r_p)^{1/2}\dot{r}^*_p\dot{s}}{4r_p}\ntau^2J_1(\ntau)-\frac{\kappa\ddot{s}}{2}\ntau^2J_1(\ntau)-\frac{\kappa f(r_p)\dot{r}^*_p}{r_p}\ntau J_2(\ntau)\label{G1tau}~,
\end{align}
\bea
\widetilde{G}^\pm_2(\ntau)=
\frac{\kappa}{96 r_p}& \left\{
- 12 \Big[64 M-18 r_p-4 f(r_p)\left(-2 M+2 f(r_p)r_p+f(r_p)^{1/2} \nu_1 r_p\right) \dot{r}_p^{*2}\ntau^2
\Big.\right.\\
&\left.\Big.
+ r_p^2 \ddot{s} \left(f(r_p)^{1/2} \nu_1 \dot{r}_p^* \dot{s}+r_p \ddot{s}\right)\ntau^4\Big] J_0(\ntau)
\right.\\
&\left.
+~r_p\ntau \Big[48 \nu_2 \dot{s}-24 f(r_p)^{3/2} \nu_1 \dot{r}_p^{*2} \dot{s}\ntau^2+16 f(r_p)\dot{r}_p^* (\nu_3 \dot{r}_p^* \dot{s}-3 r_p \ddot{s})\ntau^2
\Big.\right.\\
&\left.\Big.
+ 4r_p\ntau \left(4 r_p \dot{\ddot{s}}\ntau\mp 24 \delta(\neutral{\omega}^\pm) \ddot{s}+3 r_p \ddot{s}^2\ntau\right)
\Big.\right.\\
&\left.\Big.
+ 12 f(r_p)^{1/2} \nu_1 \left(4 \dot{s}\mp4 \delta(\neutral{\omega}^\pm) \dot{r}_p^* \dot{s}\ntau+r_p (\ddot{r}_p^* \dot{s}+\dot{r}_p^* (1+\dot{s}) \ddot{s})\ntau^2\right) J_1(\ntau)
\Big.\right.\\
&\left.\Big.
+ \left(-8 \nu_3 \dot{s}^2+48 f(r_p)r_p (\ddot{r}_p^*+2 \dot{r}_p^* \ddot{s})+3 f(r_p)\nu_1^2 \dot{r}_p^{*2} \dot{s}^2\ntau^2\right) \ntau J_2(\ntau)-\nu_1^2 \dot{s}^3\ntau^2 J_3(\ntau)\Big]
\right\}~.
\label{G2tau}
\eea
\end{widetext}

Thus, $\widetilde{G}^\pm_2$ is built from the contributions coming from the terms of $G^\pm_2$ in $\cal{O}(\ntau^0)$, of $G^\pm_1$ in $\cal{O}(\ntau^1)$, and of $G_0$ in $\cal{O}(\ntau^2)$. All quantities else than $\ntau$, involved in the formulation of $\widetilde{G}_n(\ntau)$, are evaluated at $r_p$. Returning to the definition given in Eq. (\ref{link:psi:G}), we can compute the integral of $G$ with respect to $\ntau$

\bea
&\psi^{\ell\to\infty}(x)=\int_{-\infty}^{0^+}G\big(x,x_p(\tau)\big)d\tau =\frac{r_p}{L}\int_{0^-}^{+\infty}\widetilde{G}\big(x,x_p(\hat{\tau})\big)d\hat{\tau}\\
&=\frac{r_p}{L}\int_{0^-}^{+\infty}\left(\widetilde{G}_0(\hat{\tau})L^{-2}+\widetilde{G}_1(\hat{\tau})L^{-3}+\widetilde{G}_2(\hat{\tau})L^{-4} \right.\left.+\calo{L^{-5}}\right)d\hat{\tau}\\
&=\ub{r_p\int_{0^-}^{+\infty}\widetilde{G}_0(\hat{\tau})L^{-3}d\hat{\tau}}_{\psi_\text{OL3}}+\ub{r_p\int_{0^-}^{+\infty}\widetilde{G}_1(\hat{\tau})L^{-4}d\hat{\tau}}_{\psi^\pm_\text{OL4}}+\calo{L^{-5}}~.\label{intG0G1}
\eea

Integrals in Eq. (\ref{intG0G1}) involve terms of the form

\beq
\int_{0^-}^{+\infty}\ntau^kJ_n(\ntau)d\ntau\quad k,n\in\mathbb{N}~,
\label{tauJ}
\eeq
which diverge for certain values of $k$ and $n$. Indeed, $J_n(\ntau)$ has an asymptotic behaviour of the form $J_n(\ntau)\approx\sqrt{2/\pi\ntau}\cos(\ntau-n\pi/2-\pi/4)$. Thus, for large positive values of $\ntau$, the integrand will be of the form $\sqrt{2/\pi}\ntau^m\cos(\ntau-n\pi/2-\pi/4)$ with $m=k-1/2$. To get a finite value from Eq. (\ref{tauJ}), we cancel the divergence through recasting the integral as \cite{barack2000} 

\beq
\int_{0^-}^{+\infty}\to\int_{0^-}^{+\widetilde{\infty}}~.
\eeq

The definition of the "tilde" integral is given by the limit of the same name, {\it i.e.} the "tilde limit"

\beq
\int_{0^-}^{+\widetilde{\infty}}\defeq\tildelim{\lambda\to+\infty}\int_{0^-}^{\lambda}~,
\eeq
where the limit, applied to any quantity $K$ depending on $\lambda$, is given by
\beq
\tildelim{\lambda\to+\infty}K(\lambda)=\lim_{\lambda\to+\infty}\left[K(\lambda)-\sum_jO_j(\lambda)\right]~.
\eeq

The terms $O_j(\lambda)$ have an oscillating form multiplied by a power law $O_j(\lambda)=a_j\lambda^{b_j}\cos(c_j\lambda+d_j)$ with $a_j,b_j,c_j,d_j\in\mathbb{R}$. The tilde limit appears as a standard limit, when we subtract all terms of type $O_j(\lambda)$ until the limit becomes finite. This method is very well detailed and clearly justified in Part III and Appendix A of \cite{barack2000}. When the integration is performed along the world line, the divergent terms for large $\ntau$ are ignored and written as oscillating term times a power of $\ntau$. Technically, the result of $\int_{0^-}^{+\widetilde{\infty}}\ntau^kJ_n(\ntau)d\ntau$ will be dependent of the relationship between $k$ and $n$. We propose an example where $0\leq k\leq n$ to show how the tilde limit acts concretely on the quantity to be regularised. 
Consider $\cal{I}^k_n(\lambda)$ the primitive of the function $\lambda^kJ_n(\lambda)$
\beq
\cal{I}^k_n(\lambda)\defeq\int\lambda^kJ_n(\lambda)d\lambda~.
\eeq

The integrand can be rewritten as
\beq
\lambda^kJ_n(\lambda)=\lambda^{k-n-1}\Big[\lambda^{n+1}J_n(\lambda)\Big]=\lambda^{k-n-1}\frac{d}{d\lambda}\Big[\lambda^{n+1}J_{n+1}(\lambda)\Big]~.
\eeq

Then, integration by parts leads to a recurrence relation on $\cal{I}^k_n(\lambda)$
\beq
\cal{I}^k_n(\lambda)=\lambda^kJ_{n+1}(\lambda)-(k-n-1)\cal{I}^{k-1}_{n+1}(\lambda) = \sum_{j=0}^{k-1}\left[\frac{(n-k-1+2j)!!}{(n-k-1)!!}\lambda^{k-j}J_{n+1+j}(\lambda)\right]+
\frac{(n+k-1)!!}{(n-k-1)!!}\cal{I}^0_{n+k}(\lambda)~.
\label{recurrenceIkn}
\eeq

Thus, using the tilde limit, the sum in Eq. (\ref{recurrenceIkn}) disappears because each term is of the form $O_j(\lambda)$ when $\lambda\to+\infty$. The term proportional to $\cal{I}^0_{n+k}(\lambda)$ is trivial since the standard integral $\int^{+\infty}_0J_n(\lambda)d\lambda=1\ \forall n\geq0$ is well defined and is finite. Accordingly,

\beq
\int^{+\widetilde{\infty}}_{0}\lambda^kJ_n(\lambda)d\lambda=\frac{(n+k-1)!!}{(n-k-1)!!}\quad\text{  { for} }0\leq k\leq n~.
\eeq

Following the same reasoning, the general expressions depending on the values of the integers $k$ and $n$ are

\begin{widetext}
\beq
\int_{0}^{+\widetilde{\infty}}\ntau^kJ_n(\ntau)d\ntau=
\begin{cases}
 (n+k-1)!!/(n-k-1)!!, & \text{if }0\leq k\leq n, \\
 (-1)^{(k-n)/2}(n+k-1)!!/(n-k-1)!!, & \text{if }k-n>0\text{ is even},\\
 0, & \text{if }k-n>0\text{ is odd}.
\end{cases}
\label{intbessel}
\eeq
\end{widetext}

Returning to the computation of $\psi^{\ell\to\infty}$, the evaluation of the integral in Eq. (\ref{intG0G1}) is done by replacing the standard by a tilde integral. The first term $\calo{L^{-3}}$ is obvious and gives

\beq
\psi_\text{OL3}=r_p\int_{0^-}^{+\widetilde{\infty}}\frac{\kappa}{r_p} J_0(\ntau)d\ntau=\kappa~.
\eeq
The term $\calo{L^{-4}}$ is written as
\beq
\psi^\pm_\text{OL4}=\int_{0^-}^{+\infty}\frac{\kappa}4\Big[\pm8\delta(\neutral{\omega}^\pm)J_0(\ntau)-\nu_1f(r_p)^{1/2}\dot{r}^*_p\dot{s}\ntau^2J_1(\ntau)
-2\ddot{s}\ntau^2J_1(\ntau)-4f(r_p)\dot{r}^*_p\ntau J_2(\ntau)\Big]d\ntau~,
\eeq
with $\delta(\neutral{\omega}^\pm)=\delta(\ntau)\left|{d\neutral{\omega}^\pm/d\ntau}\right|^{-1}=\left|f\dot{\omega}_\mp\right|\delta(\ntau)$, the second equality is found by using the normalisation condition. Applying Eqs. (\ref{intbessel}), the integral is simplified and can be computed

\begin{align}
\psi^\pm_\text{OL4}=&\pm2\kappa\int_{0^-}^{+\widetilde{\infty}}\left|f\dot{\omega}_\mp\right|\delta(\ntau)J_0(\ntau)d\ntau 
-\kappa f(r_p)\dot{r}_p^*\int_{0^-}^{+\widetilde{\infty}}\ntau J_2(\ntau)d\ntau = \pm2\kappa\left|f\dot{\omega}_\mp\right|_{\tau=0}J_0(0)-\kappa f\dot{r}^*_p\frac{2!!}{0!!}\nonumber\\
=&-2\kappa f\Big[\dot{r}^*_p\mp\dot{\omega}_\mp\Big] =\pm2\kappa f\dot{t} = \pm2\kappa \mE~.
\end{align}
{

Finally, we obtain the formulation of $\psi^\pm$ on the world line for large modes $\ell\to\infty$
\beq
\psi^{\pm\ell\to\infty}=\kappa\Big[L^{-3}\pm2\mE L^{-4}+\calo{L^{-5}}\Big]~.
\label{psi:linf}
\eeq

Then, by derivation of the Green function, we get for $\partial_r\psi^{\pm{ \ell}\to\infty}$

\beq
\partial_r\psi^{\pm\ell\to\infty}=\frac{\kappa}{r_p}f(r_p)^{-1}\Bigg[\mp \mE L^{-2}-\frac{3}2\mE^2L^{-3}\pm
\left(\frac{6M}{r_p}-\frac{9}4\right)\mE L^{-4}+\calo{L^{-5}}\Bigg]~.
\label{drpsi:linf}
\eeq

The next derivatives of $\psi^\pm$ are given by

\begin{widetext}
\begin{subequations}
\begin{align}
&\partial_r\psi^\pm=\frac{\kappa}{r_p}f(r_p)^{-1}\left[\mp \mE L^{-2}-\frac{3}2\mE^2L^{-3}\pm\left(\frac{6M}{r_p}-\frac{9}4\right)\mE L^{-4}+\mathcal{O}(L^{-5})\right]~,\\
&\partial^2_r\psi^\pm=\frac{\kappa}{r_p^2}f(r_p)^{-2}\Big[\mE^2L^{-1}\pm\left(2-\frac{3M}{r_p}\right)\mE L^{-2}+\mathcal{O}(L^{-3})\Big]~,\\
&\partial^3_r\psi^\pm=\frac{\kappa}{r_p^3}f(r_p)^{-3}\left\{\mp \mE^3+\mE^2\left[\frac{5}2\mE^2+\frac{9M}{r_p}-6\right]L^{-1}\mp3\mE\left[\frac{7M}{r_p}(\frac{M}{r_p}-1)+2\right]L^{-2}+\mathcal{O}(L^{-3})\right\}~,\\
&\partial_t\psi^\pm=\frac{\kappa}{r_p}\left[\pm \dot{r}_pL^{-2}+\frac{3}2\mE\dot{r}_pL^{-3}\mp\left(\frac{6M}{r_p}-\frac{9}4\right)\dot{r}_pL^{-4}+\mathcal{O}(L^{-5})\right]~,\\
&\partial^2_t\psi^\pm=\frac{\kappa}{r_p^2}\Big[\left(\mE^2-f(r_p)\right)L^{-1}\mp\frac{\mE}{2r_p}L^{-2}+\calo{L^{-3}}\Big]~,\\
&\partial^3_t\psi^\pm=\frac{\kappa}{r_p^3f(r_p)}\left\{\mp\mE\left(\mE^2-f(r_p)\right)+\frac{\mE^2\dot{r}_p}{2r_p}\Big[16M+5r_p(\mE^2-1)
\Big]L^{-1}\pm\frac{\mE\dot{r}_p}{r_p^2}\left(3M^2-2Mr_p\right)L^{-2}+\calo{L^{-3}}\right\}~,\\
&\partial_r\partial_t\psi^\pm=\frac{\kappa}{r_p^2f(r_p)}\Big[-\mE\dot{r}_pL^{-1}\pm\left(\frac{3M}{r_p}-1\right)\dot{r}_pL^{-2}+\mathcal{O}(L^{-3})\Big]~,\\
&\begin{aligned}
\partial_r\partial^2_t\psi^\pm=\frac{\kappa}{r_p^3f(r_p)}&\left\{\mp\mE\left(\mE^2-f(r_p)\right)+\frac{1}{2r_p^2}\Big[\left(5\mE^4-9\mE^2+4\right)r_p^2+\left(24M\mE^2-20M\right)r_p+6\Big]L^{-1}\right.\\
&\left.\ \pm\frac{\mE}{r_p^2}\left(Mr_p-3M^2\right)L^{-2}+\calo{L^{-3}}\right\}~,
\end{aligned}\\
&\begin{aligned}
\partial^2_r\partial_t\psi^\pm=\frac{\kappa}{r_p^3}f(r_p)^{-2}&\left\{\pm \mE^2\dot{r}_p-\mE\dot{r}_p\left[\frac{5}2\mE^2+\frac{9M}{r_p}-4\right]L^{-1}\pm\dot{r}_p\Big[\frac{3M}{r_p}(\frac{5M}{r_p}-4)+2\Big]L^{-2}+\mathcal{O}(L^{-3})\right\}~,
\end{aligned}
\end{align}
\label{Linf:psi}
\end{subequations}
\end{widetext}
and the asymptotic behaviour of the metric perturbation { functions} with respect to $\ell$ are given by 

\begin{align}
&H_1=-\kappa\left[\frac{\mE^2\dot{r}_p}{r_pf(r_p)^2}L^{-1}+\calo{L^{-2}}\right]~,\label{Linf:H1}\\
&\partial_tH_1^\pm=
\frac{\kappa}{r_p^2f(r_p)}\Big\{\Big.\mp\mE\left(\mE^2-f(r_p)\right)\left.+\frac{\kappa}{2r_p^2}\Big[(5\mE^4-7\mE^2+2)r_p^2+(18M\mE^2-10M)r_p+12M^2\Big]L^{-1}+\calo{L^{-2}}\right\}~,\\
&\partial_rH_1^\pm=
\frac{\kappa}{r_p^2f(r_p)^3}\left\{\mp\dot{r}_p\mE^3
-\frac{\kappa \mE^2\dot{r}_p}{2r_p^2f(r_p)}\Big[(5\mE^2-4)r_p^2+(8-5\mE^2)2Mr_p-16M^2\Big]L^{-1}
+\mathcal{O}(L^{-2})\right\}~,
\end{align}

\begin{align}
&H_2=
\kappa\left[\frac{1}{2r_pf(r_p)}\Big(2\mE^2-f(r_p)\Big)L^{-1}
+\mathcal{O}(L^{-2})\right]\label{Linf:H2}~,\\
&\partial_tH_2^\pm=
\frac{\kappa}{2r_p^2f(r_p)^2}\left\{\pm\mE\dot{r}_p\left(2\mE^2-f(r_p)\right)
-\frac{\mE^2\dot{r}_p}{2r_pf(r_p)}\Big[(10\mE^2-9)r_p+26M\Big]L^{-1}
+\mathcal{O}(L^{-2})\right\}~,\\
&\begin{aligned}
\partial_rH_2^\pm=&
\frac{\kappa}{2r_p^2f(r_p)^2}\Big\{\Big.\mp\mE\left(2\mE^2-f(r_p)\right)\left.+\frac{1}{2r_p^2f(r_p)}\Big[(10\mE^4-13\mE^2+2)r_p^2+(26M\mE^2-4)2Mr_p+8M^2\Big]L^{-1}
+\mathcal{O}(L^{-2})\right\}~.
\label{H2dr}
\end{aligned}
\end{align}

Equations (\ref{Linf:H1},\ref{Linf:H2}) confirm that also for $\ell\to\infty$, the perturbations are continuous at the position of the particle, see Sect. \ref{section:equation.of.motion}. The perturbations $K$, although not used in the computation of the SF for the radial fall

\begin{align}
&K^\pm=
\kappa\left[\frac{1}{2r_p}L^{-1}+\mathcal{O}(L^{-2})\right]~,\\
&\partial_tK=
\frac{\kappa\dot{r}_p}{2f(r_p)r_p^2}\left[\pm\mE
-\frac{\mE^2}{2}L^{-1}
+\mathcal{O}(L^{-2})\right]~,\\
&\partial_rK^\pm=
\frac{\kappa}{2f(r_p)r_p^2}\left[\mp\mE
\Big(\frac{\mE^2}2-f\Big)L^{-1}
+\mathcal{O}(L^{-2})\right]~,\\
&\begin{aligned}
&\partial_{tr}K^\pm=
\frac{\kappa\dot{r}_p}{2f(r_p)^2r_p^3}\Bigg\{-\mE^2L\pm\mE\Big[5M-2r_p\mE(1-\mE)\Big]
-\frac{\mE^2}{2fr_p}\Big(17M+4r_p\mE^2-11r_p\Big)L^{-1}
+\mathcal{O}(L^{-2})\Bigg\}~.
\label{Linf:perturbations}
\end{aligned}
\end{align}

We recall the relation between the $h^\text { ret}_\ab$ modes and the perturbation functions $H_1^\ell$ and $H_2^\ell$ 

\beq
h^{\text{ret}\ell}_\ab=\left(\begin{matrix}fH_2^\ell & H_1^\ell\\ H_1^\ell & f^{-1}H_2^\ell\end{matrix}\right)\sqrt{\frac{2\ell+1}{4\pi}}~.
\label{eq.hret.1}
\eeq

By putting Eqs. (\ref{Linf:H1}-\ref{H2dr}) in the expression of the retarded SF, Eq. (\ref{rappel:Fret}), we get

\beq
\begin{aligned}
F^{\alpha\ell}_\text{ret}=-\frac{m_0}{2f}\Bigg[f^\alpha_0\left(\frac{\partial H^\ell_2}{\partial t}-\frac{df}{dr}H^\ell_1\right) + f^\alpha_1\left(\frac{\partial H^\ell_1}{\partial t}-\frac{df}{dr}H^\ell_2\right) + f^\alpha_2\frac{\partial H^\ell_2}{\partial r} + f^\alpha_3\frac{\partial H^\ell_1}{\partial r}\Bigg]Y^{\ell 0}~.
\end{aligned}
\label{F:ret}
\eeq

Recalling now the definition of the $\ell$ independent regularisation parameters

\beq
\lim_{x\to x_p}F^{\alpha\ell}_{\text{sing}\pm}=F^{\alpha\ell\to\infty}_{\text{ret}\pm}(r_p)=A^\alpha_\pm L+B^\alpha+C^\alpha L^{-1}+\mathcal{O}(L^{-2})~, 
\label{mode:sum:one:mode}
\eeq
we obtain their explicit expression by equating each $L$ power of the right and left-hand sides of Eq.(\ref{mode:sum:one:mode}). The regularisation parameters for a radial geodesic in { an} SD black-hole in the RW gauge are given by

\beq
\begin{aligned}
&A^r_\pm=\mp\frac{m_0^2}{r^2_p}\mE~,\quad A^t_\pm=\mp\frac{m_0^2\dot{r}^2_p}{r^2_pf(r_p)}~,\\
&B^r=-\frac{m_0^2}{2r^2_p}\mE^2~,\quad B^t=-\frac{m_0^2\dot{r}_p}{2r^2_pf(r_p)}\mE~,\quad C^\alpha=0~.
\end{aligned}
\label{F:sing:RP}
\eeq
where $\dot{t}=\mE/f(r_p)$ and $\dot{r}_p=\sqrt{\mE^2-f(r_p)}$. The parameters are to be put into Eq.(\ref{mode:sum:Gideal}), noting that 
$D^\alpha = 0$ \cite{balo02,baetal02}.


}

\subsection{Non-radiative modes}\label{modes:l:zero}
In absence of a wave-equation for the non-radiative modes $\ell=0, 1$, it is necessary to identify an alternative way for evaluating their contribution. Incidentally, the contributions of the radiative and non-radiative modes, though they refer to different gauges have been summed in previous literature \cite{balo02}. 

\subsubsection{Zerilli gauge}

Zerilli \cite{ze70c} showed that the monopole $\ell=0$ expresses a variation of the mass parameter, while in radial fall the dipole $\ell = 1$ is associated to the { shift of the centre of mass and it may vanish with a proper gauge transformation}, see also Detweiler and Poisson \cite{depo04}. For the $\ell=0$ mode, it is possible to obtain an analytic solution for $F^{\alpha\ell=0}_\text{ret}$. 
With the gauge transformation $x^\alpha\to x^\alpha+\xi^\alpha_{\ell=0}$, we get 

\beq
\xi^\alpha_{\ell=0}=\Big(M_0(t,r),M_1(t,r),0,0\Big)Y^{00}~.
\eeq

The perturbations transform as 

\beq
h_\ab^{\text{(G')}}\to h_\ab^{\text{(G)}}+\nabla_\alpha\xi^{\ell=0}_\beta+\nabla_\beta\xi^{\ell=0}_\alpha~,
\eeq
and thus the components are related to the new gauge $(G')$ by, see Gleiser {\it et al.} \cite{gletal00}
\begin{align}
&H_0^{\ell=0\text{(G')}}=H_0^{\ell=0\text{(G)}}+2\frac{\partial}{\partial t}M_0^{\text{(G}\to\text{G')}}+\frac{2M}{r^2f}M_1^{\text{(G}\to\text{G')}}~,\\
&
\begin{aligned}
H_1^{\ell=0\text{(G')}}=&H_1^{\ell=0\text{(G)}}-f^{-1}\frac{\partial}{\partial t}M_1^{\text{(G}\to\text{G')}}+ f\frac{\partial}{\partial r}M_1^{\text{(G}\to\text{G')}}~,
\end{aligned}\\
&H_2^{\ell=0\text{(G')}}=H_2^{\ell=0\text{(G)}}-2\frac{\partial}{\partial r}M_1^{\text{(G}\to\text{G')}}+\frac{2M}{r^2f}M_1^{\text{(G}\to\text{G')}}~,\\
&K^{\ell=0\text{(G')}}=H^{\ell=0\text{(G)}}-\frac{2}{r}M_1^{\text{(G}\to\text{G')}}~.
\end{align}

{
The Zerilli (Z) gauge \cite{ze70c} implies that the two degrees of gauge freedom $M_0$ and $M_1$ must render $H_1^{\ell=0\text{(Z)}}=K^{\ell=0\text{(Z)}}=0$. 

 \beq
 F^{t\ell=0}_\text{ret} =  \frac{m_0}{4f^2}\Big[2f'f^2\dot{r}_p\dot{t}_p^3\Big(H_2^{\ell=0\text{(Z)}}+H_0^{\ell=0\text{(Z)}}\Big)- \Big(f^3\dot{t}_p^4+f\dot{r}_p^2\dot{t}^2-f^2\dot{t}_p^2+\dot{r}_p^2\Big)\frac{\partial}{\partial t}H_0^{\ell=0\text{(Z)}} -
  f\dot{r}_p\dot{t}\Big(f^2\dot{t}_p^2+\dot{r}_p^2-2f\Big)\frac{\partial}{\partial r}H_0^{\ell=0\text{(Z)}}
 \Big]~,
 \label{Ft:ret:G1}
 \eeq

 \beq
 F^{r\ell=0}_\text{ret}=\frac{m_0}{4f}\Big[
 2f'f(\dot{r}_p^2+f)\dot{t}_p^2\Big(H_2^{\ell=0\text{(Z)}}+H_0^{\ell=0\text{(Z)}}\Big) - 
 \Big(f^2\dot{r}_p^2\dot{t}_p^2 -f^3\dot{t}_p^2+\dot{r}_p^4+f\dot{r}_p^2\Big)\frac{\partial}{\partial r}H_0^{\ell=0\text{(Z)}} - \dot{r}_p\dot{t}_p\Big(f^2\dot{t}_p^2+\dot{r}_p^2+2f\Big)\frac{\partial}{\partial t}H_0^{\ell=0\text{(Z)}}
 \Big]~,
 \label{Fr:ret:G1}
 \eeq
 where

 \begin{align}
 &H_2^{\ell=0\text{(Z)}}=8\pi m_0\mE\frac{1}{rf}Y^{00\star}(0,0)\cal{H}(r-r_p)~,\\
 &\begin{aligned}
 H_0^{\ell=0\text{(Z)}}=&8\pi m_0\mE\Bigg[\frac{1}{rf}-\frac{1}{r_pf(r_p)}- 
\frac{1}{r_pf(r_p)^3}\left({\dot r}_p\right)^2\Bigg]Y^{00\star}(0,0)\cal{H}(r-r_p)~.
 \end{aligned}
 \end{align}

As noted in \cite{balo02}, the Z gauge leads to a pathological behaviour of $F^{\alpha\ell=0}_\text{self}$ approaching the horizon, see Fig. (\ref{fig08}). It is however possible to define another gauge condition.
 
 \subsubsection{The R gauge}

We thus made an other gauge choice, baptised as R. The two degrees of gauge freedom $M_0$ and $M_1$ may be chosen such that $H_0^{\ell=0\text{(R)}}=H_1^{\ell=0\text{(R)}}=H_2^{\ell=0\text{(R)}}=H^{\ell=0\text{(R)}}$ and $K^{\ell=0\text{(R)}}=0$. The obtained monopole solution for the retarded force and the self-acceleration (SA) are now compliant with the behaviour of $\ell\geq 2$ modes, see Figs. (\ref{fig08}, \ref{fig14}).}

 \beq
 \begin{aligned}
 F^{t\ell=0}_\text{ret}= &
 -\frac{m_0}{2{f}^{2}}
 \Big( f\dot{t}_p +\dot{r}_p \Big)
 \Big[\Big.\left( {f}^{2}{\dot{t}_p }^{3}+f\dot{r}_p {\dot{t}_p }^{2}-f\dot{t}_p +\dot{r}_p \right)\frac{\partial}{\partial t}H^{\ell=0\text{(R)}}\\
 &+f\dot{r}_p \left( f{\dot{t}_p }^{2}+\dot{r}_p \dot{t}_p -2\right)\frac{\partial}{\partial r}H^{\ell=0\text{(R)}}
 -\Big. f' \left( {f}^{2}{\dot{t}_p }^{3}+f\dot{r}_p {\dot{t}_p }^{2}-f\dot{t}_p +\dot{r}_p \right) H^{\ell=0\text{(R)}}
 \Big]~,
 \end{aligned}
 \label{Ft:ret:G2}
 \eeq
 \beq
 \begin{aligned}
 F^{r\ell=0}_\text{ret}= & -\frac{m_0}{2f}\Big( f\dot{t}_p +\dot{r}_p \Big) 
 \Big[\Big.\dot{t}_p \left( f\dot{r}_p \dot{t}_p +{\dot{r}_p }^{2}+2f\right)\frac{\partial}{\partial t}H^{\ell=0\text{(R)}} \\
 &+\left( f{\dot{r}_p }^{2}\dot{t}_p -{f}^{2}\dot{t}_p +{\dot{r}_p }^{3}+f\dot{r}_p \right) \frac{\partial}{\partial r}H^{\ell=0\text{(R)}} -f' \dot{t}_p \left( f\dot{r}_p \dot{t}_p +{\dot{r}_p }^{2}+2f\right) H^{\ell=0\text{(R)}}
 \Big]~,
 \end{aligned}
 \label{Fr:ret:G2}
 \eeq
 with
 \beq
 H^{\ell=0\text{(R)}}=8\pi m_0\mE\frac{1}{rf}Y^{*00}(0,0)\cal{H}(r-r_p(t))~.
 \eeq

\section{Numerical approach, performance and code validation}

\subsection{{ Computation of the perturbations, and the gravitational SF}\label{numersubsect}}

%

We can test the robustness and validity of our code by comparing the numerical results to the outcomes of 
Eqs. (\ref{psi:linf}-\ref{Linf:psi},\ref{Linf:H1}-\ref{H2dr}), 
knowing the analytic asymptotic behaviour for large $\ell$. For the evaluation of the fields on the worldline, we use an interpolation method described in App. \ref{Interpolation}. 

\begin{turnpage}
    \begin{figure}[h!]
    \centering
    \includegraphics[width=1.0\linewidth]{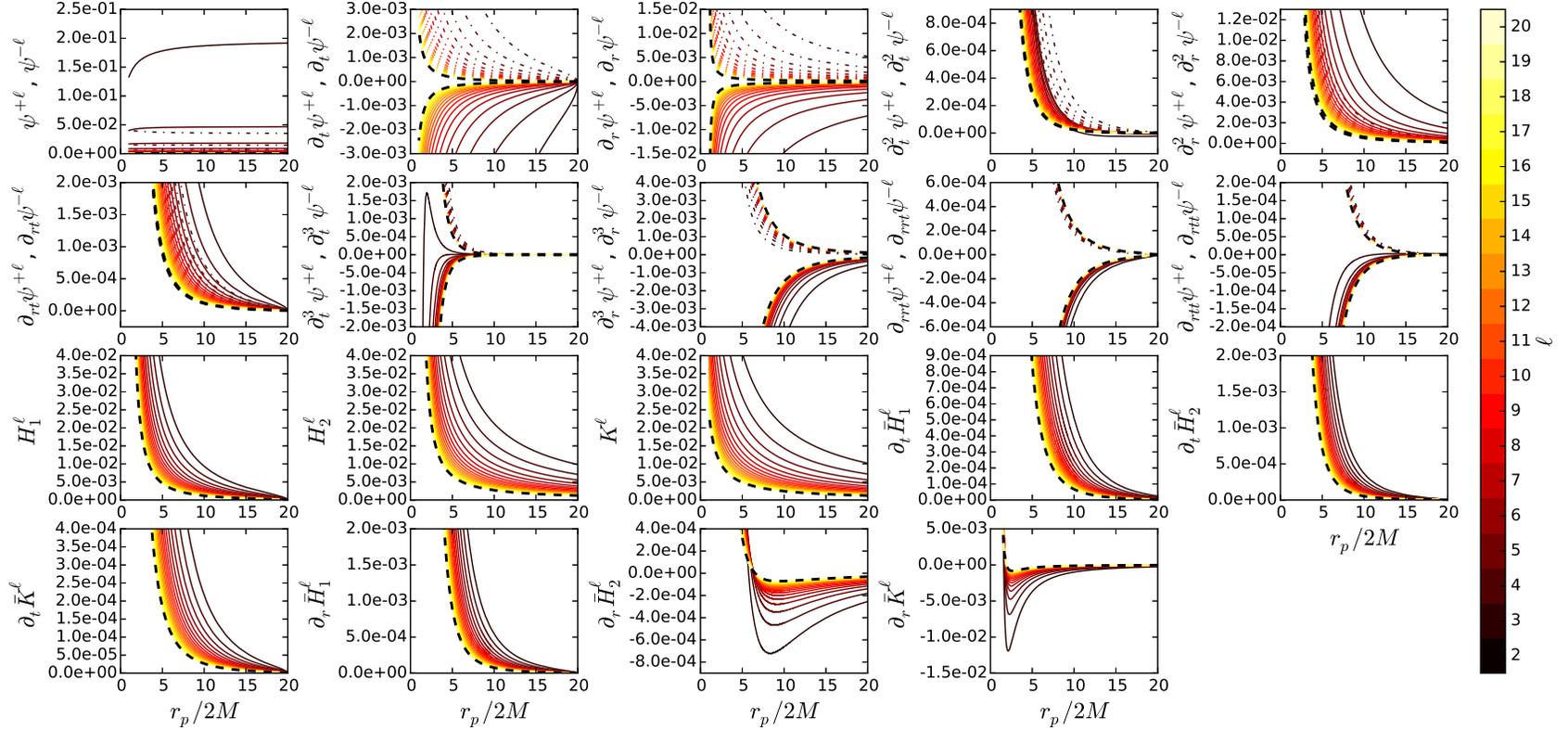}
    \caption{Radial fall of a particle at rest from  $r_0/2M=20$. The wave-function and its derivatives up to third order, $H^\ell_1$ and $H^\ell_2$ and their first derivatives are shown at the position of the particle. Each quantity is given for $2 \leq \ell \leq 20$ (colour palette). We plot in black the asymptotic behaviour given by Eqs. (\ref{psi:linf}-\ref{Linf:psi},\ref{Linf:H1}-\ref{H2dr}). The dashed curves are related to $r\to r_p^-$ (superscript "-") and the solid curves to the part $r\to r_p^+$ (superscript "+"). 
     The values are in SI units of $2M/m_0\kappa^{-1}$.}
    \label{fig05}
    \end{figure}
\end{turnpage}

Figure (\ref{fig05}) shows the quantities (the wave-function and its derivatives up to third order, $H^\ell_1$ and $H^\ell_2$ and their first derivatives) that the code is able to extract at the position of the particle during its fall from an initial rest position at $r_0/2M=20$. Each quantity is given for $2 \leq \ell \leq 20$. We plot in black the asymptotic behaviour given by Eqs. (\ref{psi:linf}-\ref{Linf:psi},\ref{Linf:H1}-\ref{H2dr}). The dashed curves are related to the side $r\to r_p^-$ (superscript "-") and the solid curves to $r\to r_p^+$ (superscript "+"). 
The values are in SI units of $2M/m_0\kappa^{-1}$. 

\begin{turnpage}
    \begin{figure}[h!]
    \centering
    \includegraphics[width=1.0\linewidth]{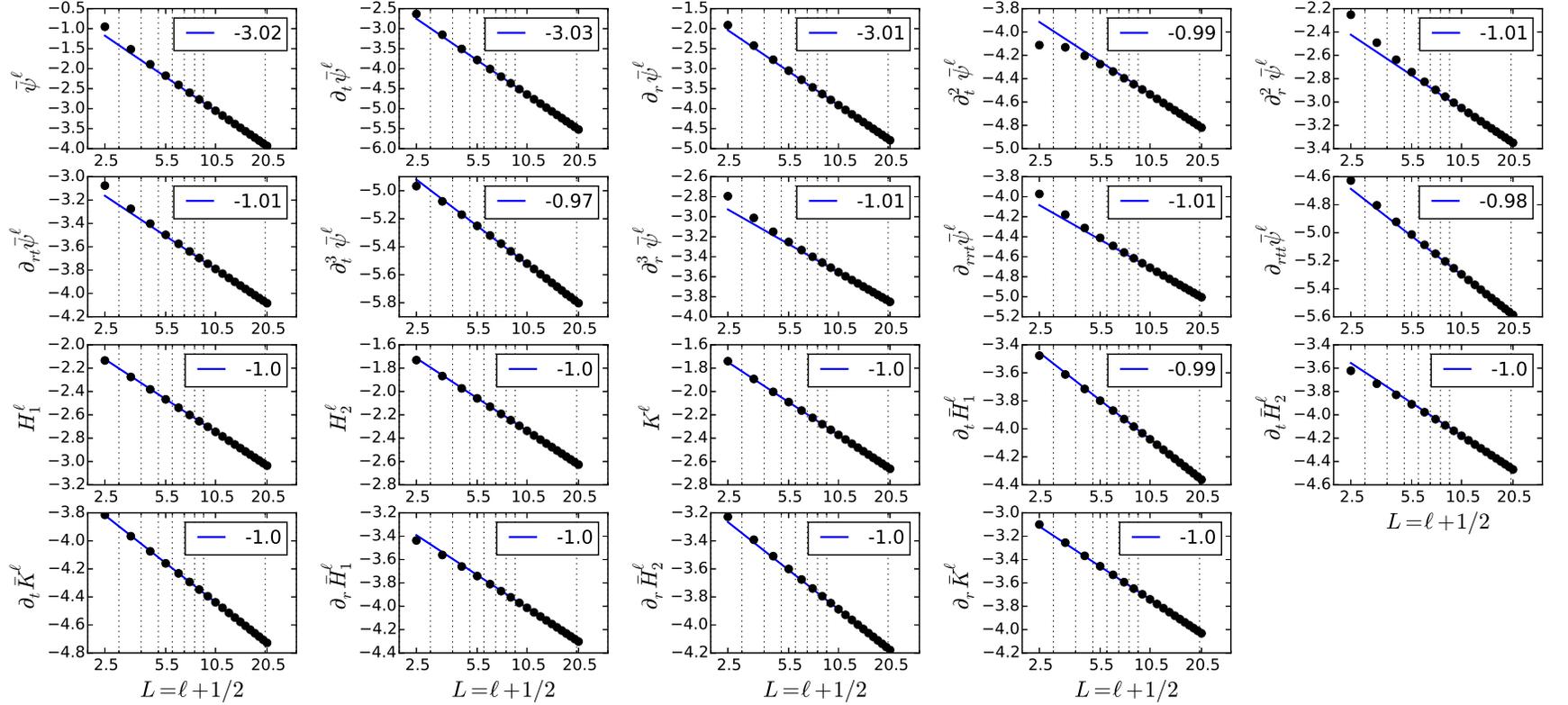}
    \caption{Radial fall of a particle at rest from  $r_0/2M=20$. The quantities displayed in Fig. (\ref{fig05}) are shown {\it vis \'a vis} the $L$ mode for $r_p/2M\approx 10$. On the vertical axis, the $\log_{10}$ of the averaged quantities in $2M/m_0\kappa^{-1}$ units; the average of 
$\psi^{\ell\pm}$ provides $\bar{\psi}^\ell(r_p)=1/2\left[\psi^{\ell+}(r_p)+\psi^{\ell-}(r_p)\right]$. The slope of each straight line corresponds to the leading order in Eqs. (\ref{psi:linf}-\ref{Linf:psi},\ref{Linf:H1}-\ref{H2dr}). The agreement is found for different $r_p$ values.}
    \label{fig06}
    \end{figure}
\end{turnpage}

In Fig. (\ref{fig06}), we check the asymptotic behaviour of the modes with $\ell$. We observe example, at fixed $r_p$, $H^{\ell\to\infty}_{1,2}\propto \kappa L^{-1}$. The straight line formed by the points $\log_{10}|H_{1,2}|$ in terms of $L$ has a slope $-1$.

\begin{figure}[h!]
\centering
\includegraphics[width=0.5\linewidth]{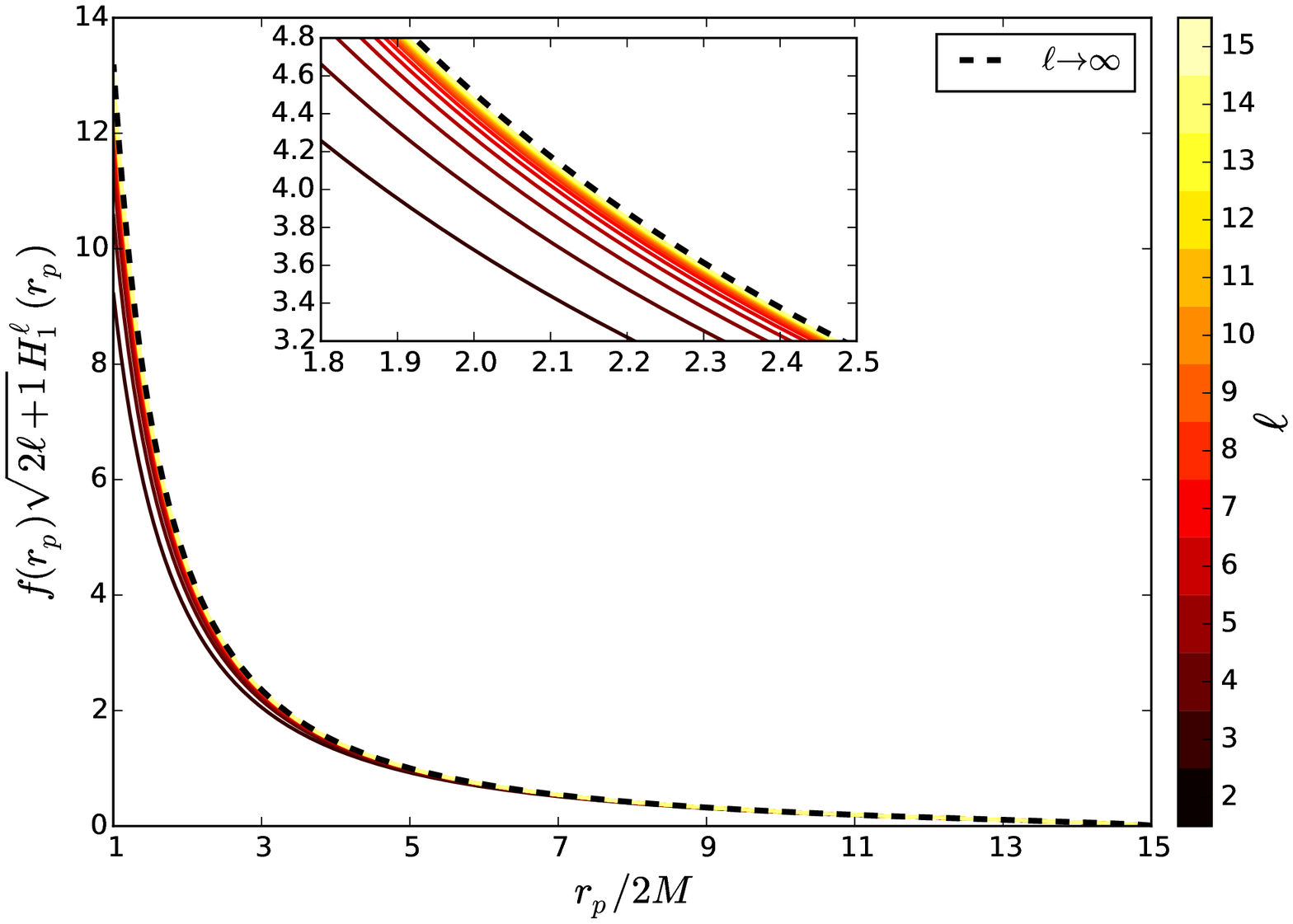}\\
\includegraphics[width=0.5\linewidth]{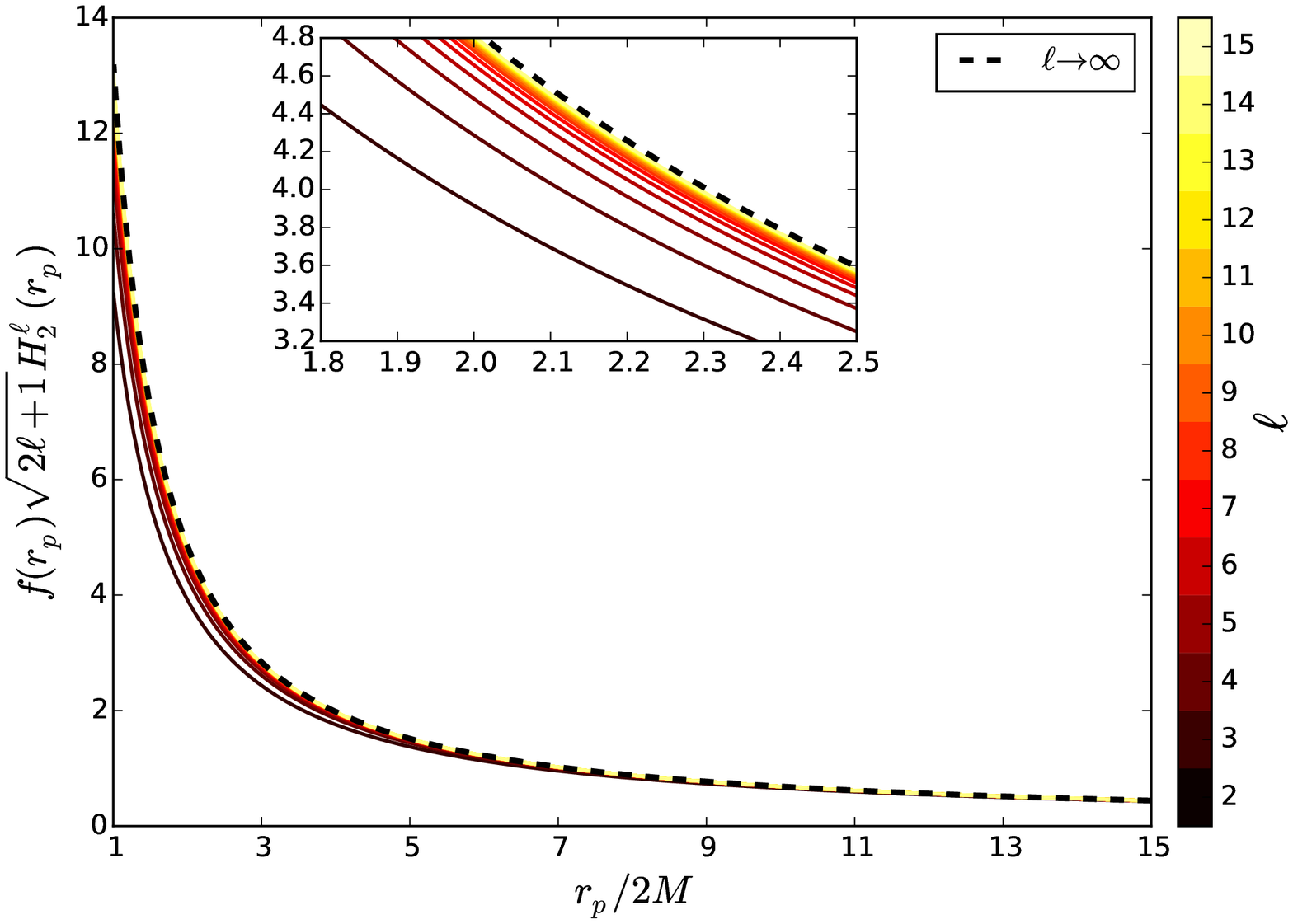}
\caption{Radial fall of a particle at rest from  $r_0/2M=15$. The perturbation functions $H_1^\ell$ (upper panel) and $H_2^\ell$ (lower panel) are computed at the particle position $r_p(t)$ for the modes $2 \leq \ell \leq 20$ (colour palette). The asymptotic behaviour of $H_1^{\ell\to\infty}$ and $H_2^{\ell\to\infty}$, Eqs. (\ref{Linf:H1},\ref{Linf:H2}) traced in black, is confirmed.}
\label{fig07}
\end{figure}

Figure (\ref{fig07}) displays the perturbation functions of the retarded field for a fall from $r_0/2M=15$ for  
$2 \leq \ell \leq 20$. For the modes $\ell>8$, the behaviour tends to $H_{1,2}^{\ell\to\infty}$ as expressed by Eqs. (\ref{Linf:H1},\ref{Linf:H2}). The divergent feature of the series is due to the infinite sum of finite contributions. The standard theorem by Courant \cite{cohi53} states that on the 2-sphere of constant $t$ and $r$, a function must be at least $\cal{C}^2$ for the uniform and absolute convergence of its expansion in spherical harmonics. This condition is clearly not satisfied in the case of the radial perturbation tensor which is  $\cal{C}^0$.

For the computation of the retarded force mode by mode, we use Eq. (\ref{F:ret}). The latter may 
provide also $F^{\alpha\ell}_{\text{ret}\pm}$, where the $\pm$ sign indicates one of the { particle worldline sides. }
Since the $\ell$-modes of the retarded field $h^{\text{ret}\ell}_\ab$ are continuous at the position of the particle in the RW gauge,  their derivatives have a jump and that is why the sign $\pm$ is needed. 
Indeed, the value of $F^{\alpha\ell}_{\text{ret}\pm}(t,r_p)$ depends on the direction in which the derivatives are taken through the limit $r\to r_p(t)$. In the following, we will consider the average of each mode only

\beq
F^{\alpha\ell}_\text{ret}=\frac{1}2\Big(F^{\alpha\ell}_{\text{ret}+}+F^{\alpha\ell}_{\text{ret}-}\Big)~.
\label{average}
\eeq

\begin{figure}[h!]
\centering
\includegraphics[width=0.5\linewidth]{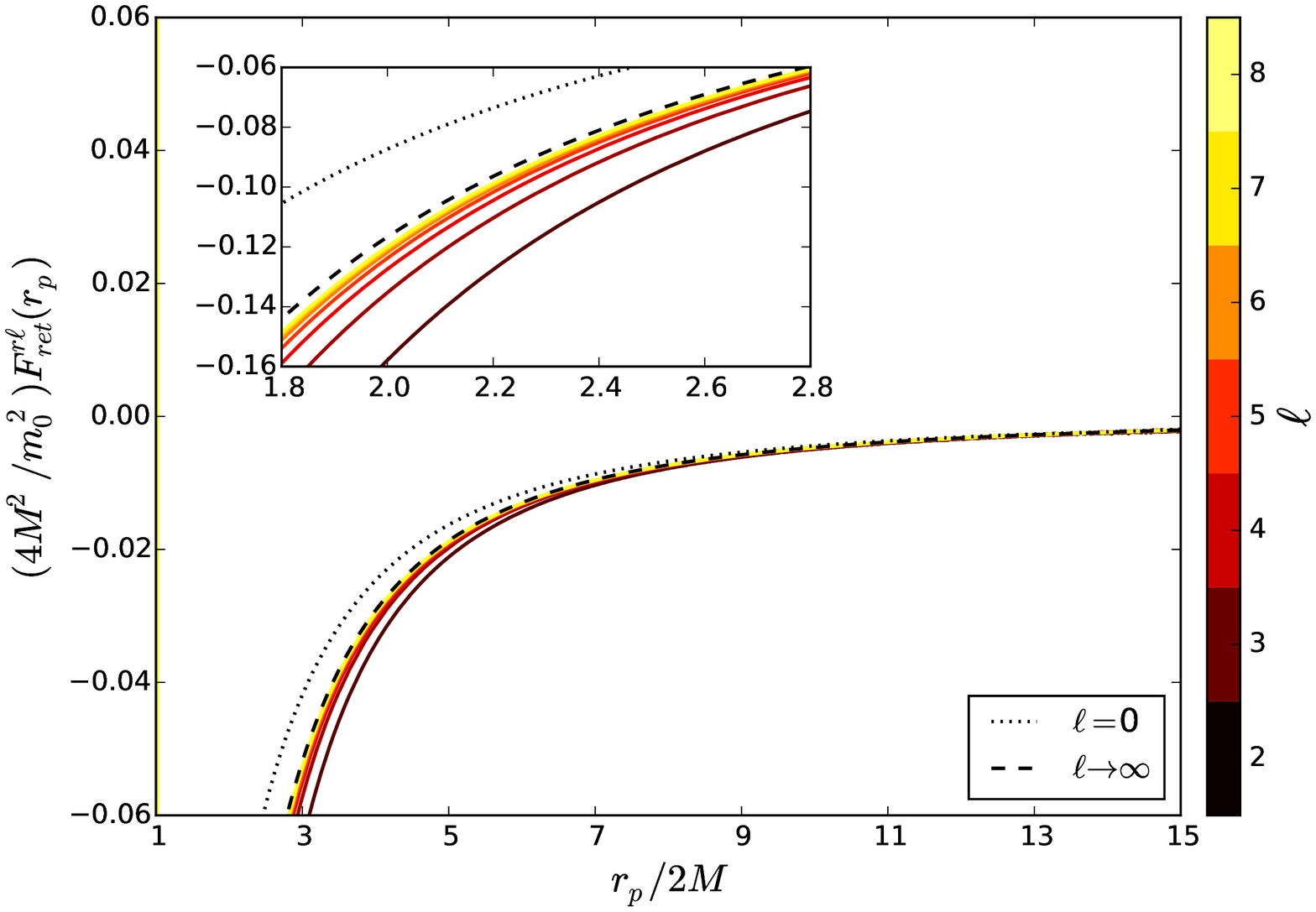}\\
\includegraphics[width=0.5\linewidth]{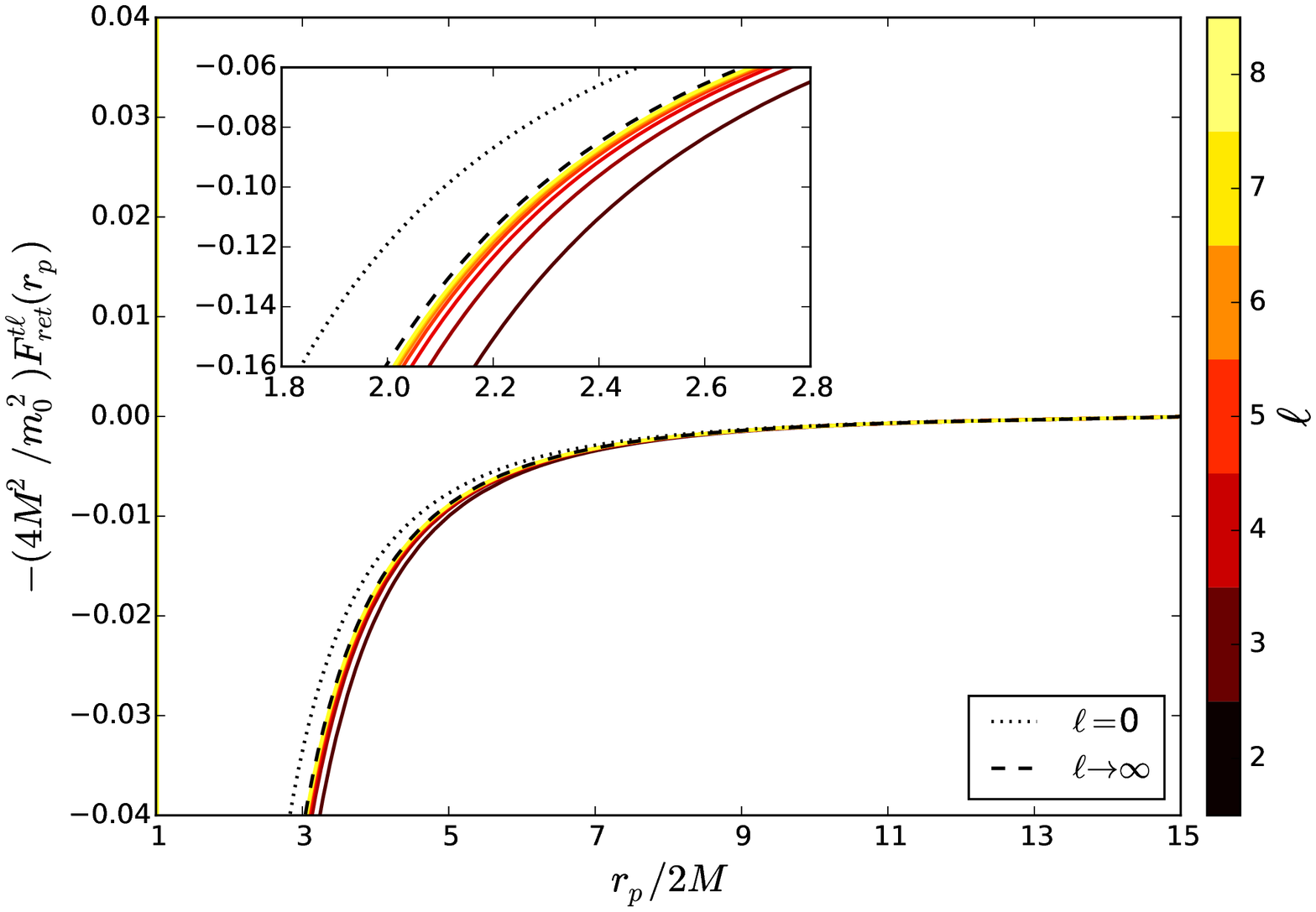}
\caption{Radial fall of a particle at rest from  $r_0/2M=15$. The average retarded force $F^{\alpha\ell}_\text{ret}$ is computed at the particle position $r_p(t)$ for the modes $2 \leq \ell \leq 20$ (colour palette). The asymptotic behaviour of $F^{\alpha\ell\to\infty}_\text{ret}=B^\alpha$, Eq. (\ref{F:sing:RP}) traced in black, is confirmed. The mode $\ell=0$ for the $Z$ and $R$ gauges is shown by the  dot-dash and dash lines, respectively. For the former, we note the divergent behaviour at the horizon.}
\label{fig08}
\end{figure}

Figure (\ref{fig08}) shows the eight first modes of the retarded force both for $r$ and $t$ { components} as { a} function of the particle position $r_p(t)$ for a fall from $r_0/2M=15$. For the $\ell=0$ mode, two curves are plotted for the $Z$ and $R$ gauges. The former shows a divergent behaviour as expected.     
The black solid line refers to the parameter $B^\alpha$ which describes the asymptotic form of $F^{\alpha\ell}_\text{ret}$ when $\ell\to\infty$, since because $A^\alpha_+=-A^\alpha_-$, Eq. (\ref{F:sing:RP}). The divergent feature of the series appears again due to the infinite sum of finite modal contributions. The modes tend to $B^\alpha$ when $\ell\to\infty$. 

\begin{figure}[h!]
\centering
\includegraphics[width=0.6\linewidth]{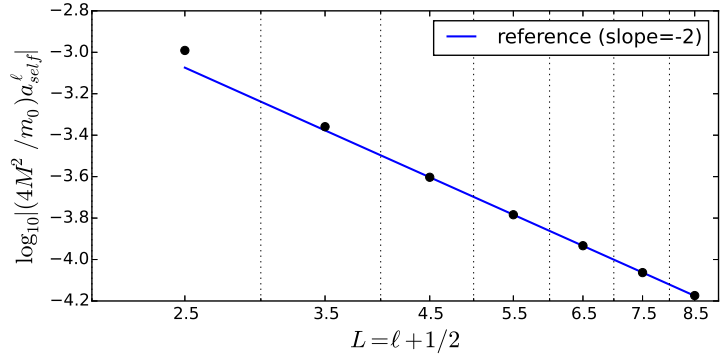}
\caption{Speed of the convergence of the series $a_\text{self}=\sum_\ell a^\ell_\text{ret}-B_a$.}
\label{fig09}
\end{figure}

The value of the SF does not depend on the sign "$\pm$" shown in Eq. \ref{F:ret}). Thus, the average $F^{\alpha\ell}_{\text{ret}\pm}$ is taken for regularisation. Given Eq. (\ref{average}), 
and considering the regularisation parameters obtained, Eq. (\ref{F:sing:RP}), we have

\beq
F^\alpha_\text{self}=\sum_{\ell=0}^\infty\Big[ F^{\alpha\ell}_\text{ret}-B^\alpha\Big]~,
\label{mode:sum:average}
\eeq
where the superscript (G) of Eq. (\ref{mode:sum:Gideal}), in our case (RW), has been removed. This expression ensures the $L^{-2}$ convergence, Fig. (\ref{fig09}).

\clearpage
\clearpage
\clearpage

\begin{figure}[h!]
\centering
\includegraphics[width=0.6\linewidth]{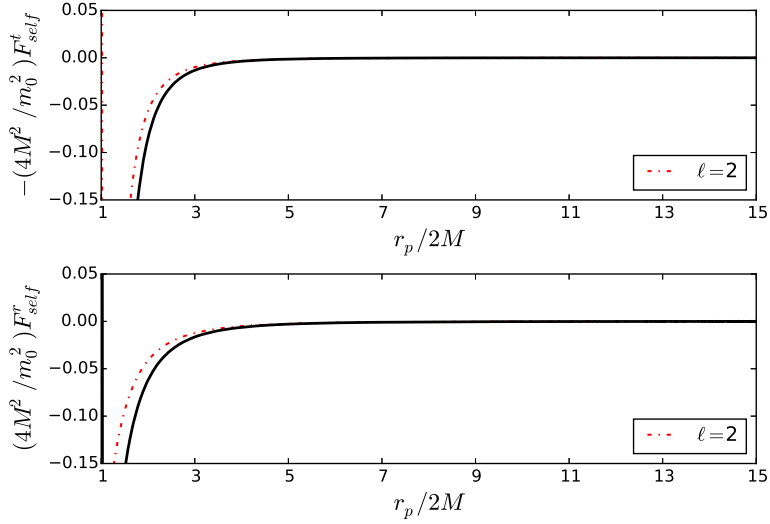}
\caption{Radial fall of a particle at rest from  $r_0/2M=15$. After the regularisation, Fig. (\ref{fig08}), the modes are summed together with the $\ell=0$ mode and the analytic contribution of the $\ell >8$ modes. The curves in black represent the SF, Eq.  (\ref{serie:FP}), for the time $F^t_\text{self}$ (upper curve) and radial $F^r_\text{self}$ (lower curve) components. For comparison, the quadrupole mode is also traced (red curve), thereby showing the relevance of the modes $\ell >2$ for the SF computation.}
\label{fig10}
\end{figure}

\begin{figure}[h!]
\centering 
(a)
\includegraphics[width=0.46\linewidth]{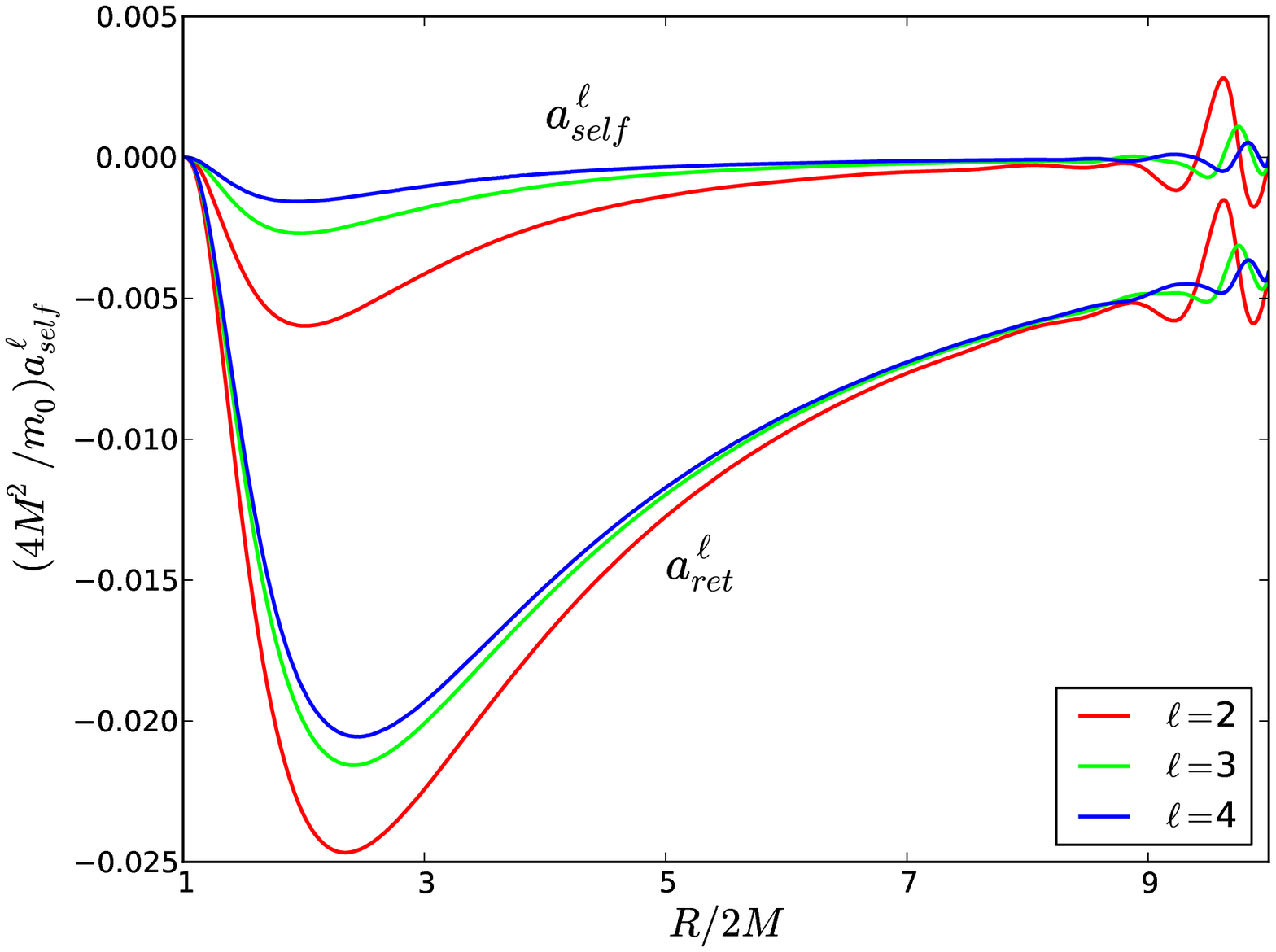}
(b)
\includegraphics[width=0.46\linewidth]{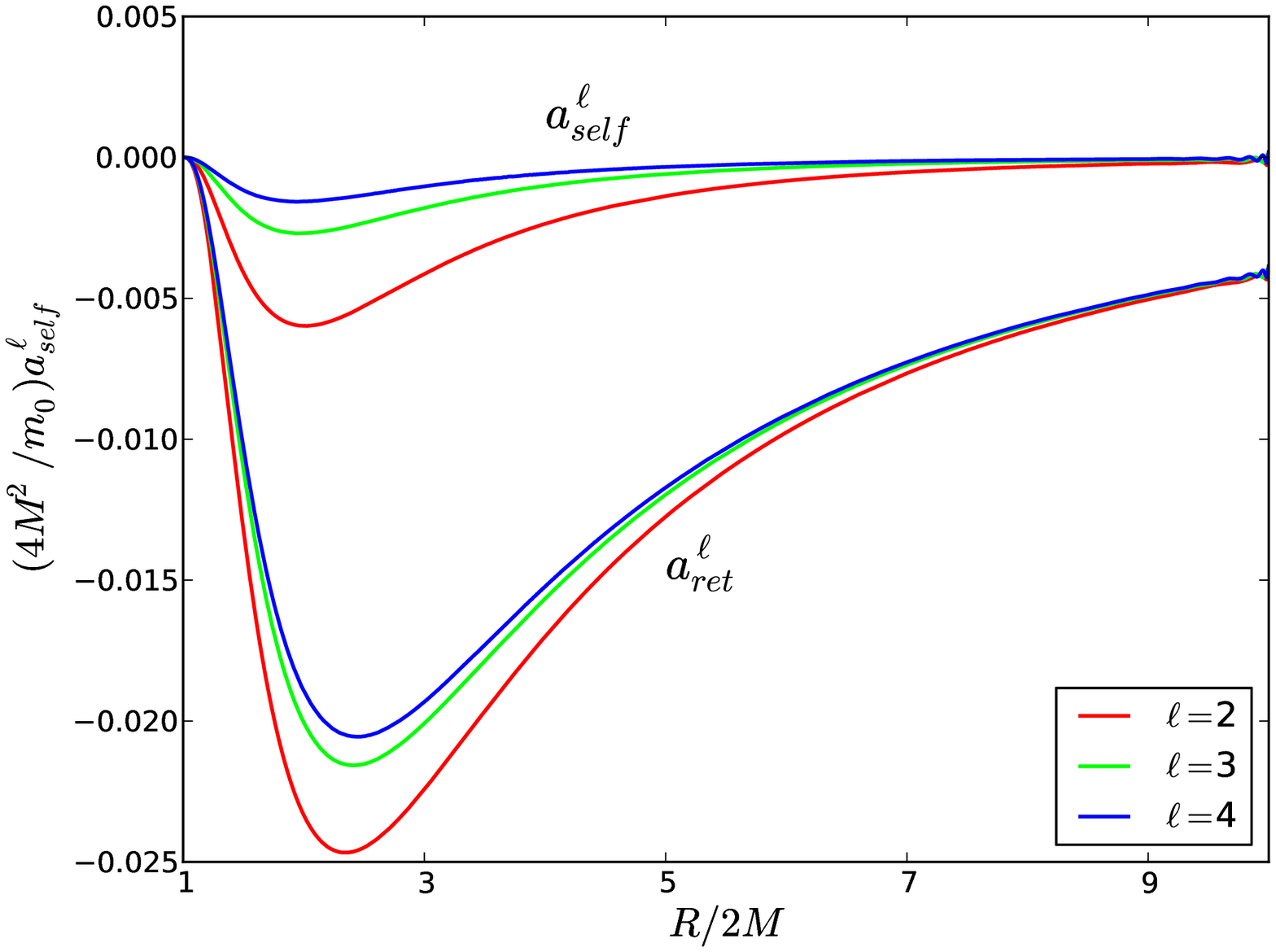}
\hspace{4cm}(c)
\includegraphics[width=0.46\linewidth]{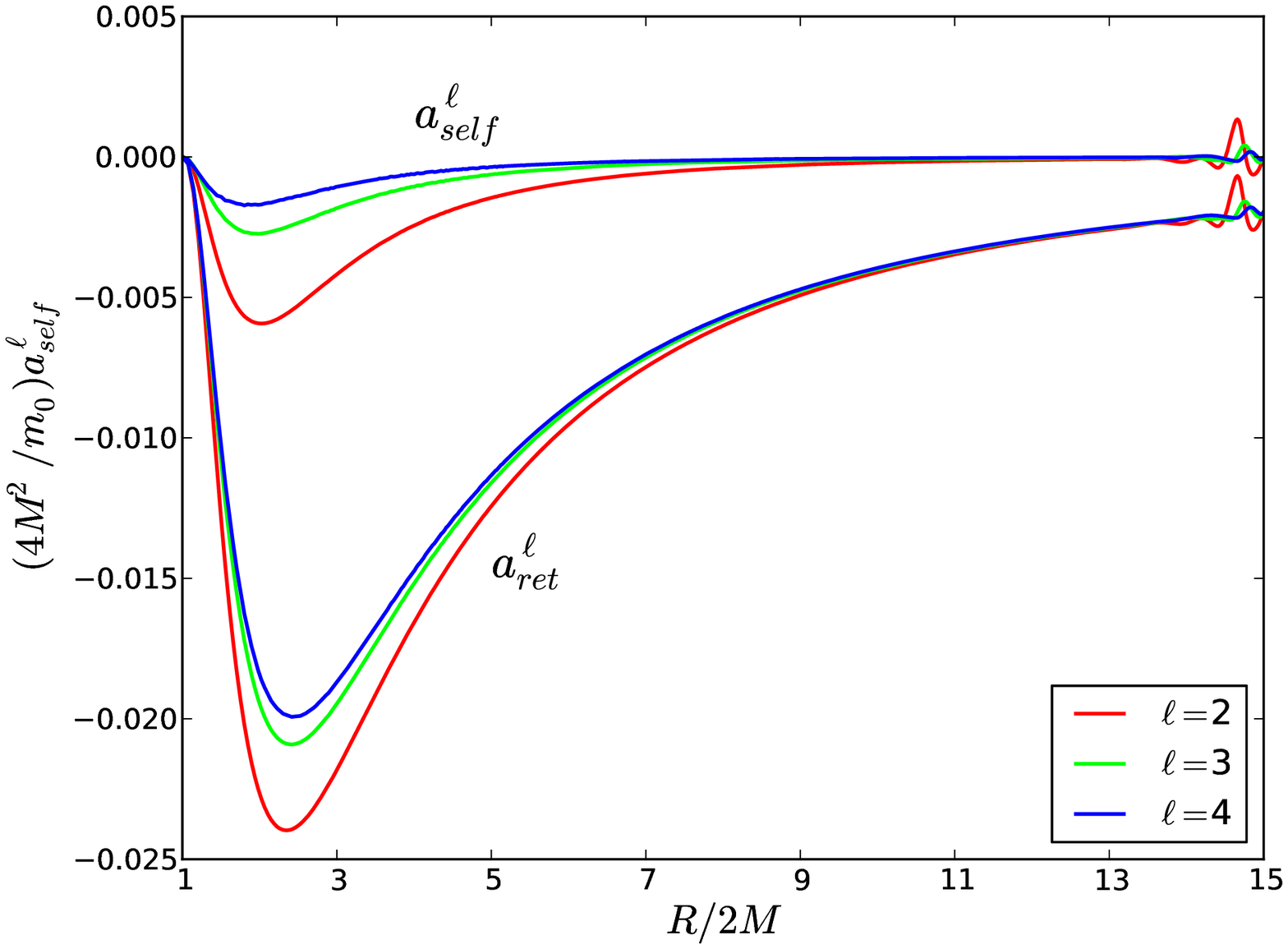}
(d)
\includegraphics[width=0.46\linewidth]{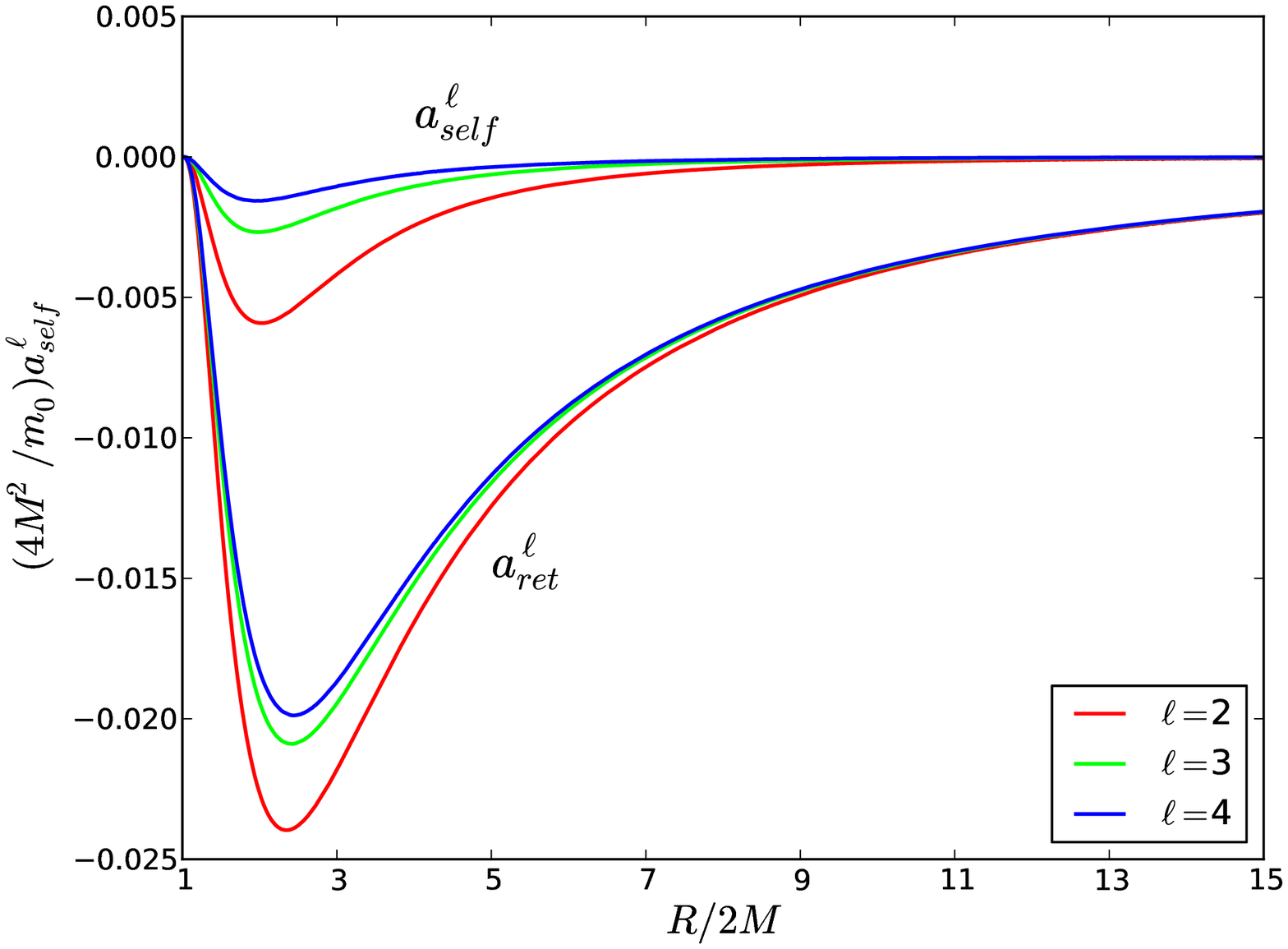}
\hspace{4cm}
\caption{The Brill-Lindquist \cite{brilllindquist1963} initial conditions imposed at $r=r_0$ induce non-physical oscillations polluting the first stage of  the fall. This corresponds to the graphs (a) and (c), respectively associated to a fall from $r_0/2M=10$ and $r_0/2M=15$. In order to exploit the first stage of the fall, we use symmetric trajectories ($m$ is thrown up vertically) and then consider only the portion of the data corresponding to the fall into the black hole (the non-physical oscillations of the upward part vanish at infinity). This scenario is displayed in the graphs (b) and (d), respectively associated to a fall from $r_0/2M=10$ and $r_0/2M=15$.}
\label{fig11} 
\end{figure}

We fix a value $\ell=\ell_\text{max}$ corresponding to an acceptable threshold error. The contribution of higher modes than $\ell_\text{max}$ is computed analytically using the asymptotic behaviour with respect to $L$, Eqs. (\ref{Ft:self:Linf},\ref{Fr:self:Linf})

\beq
F^\alpha_\text{self}=F^{\alpha\ell=0}_\text{self}+\ub{\sum_{\ell=2}^{\ell_\text{max}}F^{\alpha\ell}_\text{self}}_\text{numerical}\ \ +\ub{\sum_{\ell=\ell_\text{max}+1}^{\infty}F^{\alpha\ell\to\infty}_\text{self}}_\text{analytic}~.
\label{serie:FP}
\eeq

Figure (\ref{fig10}) shows the SF computed from the modes of the retarded force plotted in Fig. (\ref{fig08}), for $\ell_\text{max}=8$. The case corresponds to $r_0/2M=15$ but  the { general} behaviour of the components of the SF remains the same regardless the value of $r_0$. Indeed, the radial component is always oriented toward the black hole which suggests a positive work of the force during the fall (attractive nature) and therefore the energy $\mE$ parameter increases \cite{spri14}.

Figure (\ref{fig10}) can be compared to Fig. (3) of \cite{balo02}: qualitatively the behaviour $F^\alpha_\text{self}$ is consistent but unlike \cite{balo02} our curves do not suffer of the non-physical oscillations that pollute the first stages of the fall. We use a symmetric trajectory ($m$ is thrown up vertically) to overcome this problem, { which makes} the first stage of the fall exploitable for  our analysis, Sect. \ref{section:conditions:initiales}.

The attractive nature of the SF in RW, H and all smoothly related gauges is compliant with the findings by Barack and Lousto in \cite{balo02}, but at odds with those by Lousto alone in \cite{lo00,lo01}, where the SF is repulsive for some modes and attractive for others. This is largely discussed in \cite{spri14}. 

The modes beyond $\ell_\text{max}$ are approached analytically by the quantity $F^{\alpha\ell\to\infty}_\text{self}$ which corresponds to the contribution $\calo{L^{-2}}$ contained in Eq. (\ref{mode:sum:one:mode}). This term is computed by following the procedure in Sect. \ref{regmsp} but keeping the higher order of development of the Green function. We obtain

\begin{align}
&F^{t\ell\to\infty}_\text{self}=\frac{15}{16}\frac{m_0^2\mE}{r_p^2f(r_p)}\dott{r}_p\left(2r_p\ddott{r}_p-\dott{r}_p^2\right)L^{-2}+\calo{L^{-4}}~,
\label{Ft:self:Linf}\\
&F^{r\ell\to\infty}_\text{self}=-\frac{15}{16}\frac{m_0^2\mE^2}{r_p^2}\left(\mE^2+\frac{4M}{r_p}-1\right)L^{-2}+\calo{L^{-4}}~,
\label{Fr:self:Linf}
\end{align}

where `$\circ$' is a full derivation operator with respect to ccordinate time. The derivation of Eqs. (\ref{Ft:self:Linf},\ref{Fr:self:Linf}) is obtained with a similar computation appeared in Sect. \ref{regmsp}, but for a higher order. This is the first independent confirmation of Eqs. (6a,6b) 
in \cite{balo02}, for which derivation the reader was reminded to an accompanying paper, that finally was never published. 

\subsection{Initial conditions}\label{section:conditions:initiales}
The Brill-Lindquist \cite{brilllindquist1963} initial conditions generate quasi-normal modes and induce non-physical oscillations that pollute the first stage as shown in Figs. (\ref{fig11}a,\ref{fig11}c) and in \cite{balo02}. We circumvent the nuisance by adopting a symmetric trajectory ($m_0$ is thrown up vertically) and consider only the portion for which $\dott{r}_p\leq0$, Figs. (\ref{fig11}b,\ref{fig11}d).

\begin{figure}[h!]
\centering
\includegraphics[width=0.5\linewidth]{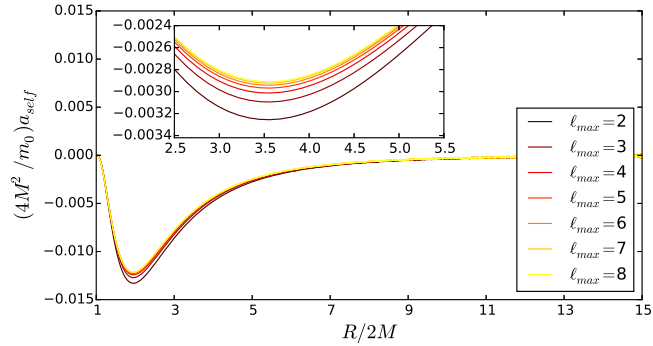}
\caption{{$a_\text{self}$ for different values of the truncation parameter  $\ell_\text{max}$ appearing in the series in Eq. (\ref{somme:a}). }}
\label{fig12}
\end{figure}

\subsection{Sensitivity to $\ell_\text{max}$}\label{sensibilite:lmax}
The truncation of the series in Eq. (\ref{somme:a}) depends on $\ell_\text{max}$, that is the highest mode to be computed numerically. Obviously, { the} larger { is $\ell_\text{max}$}, and more $a_\text{self}$ tends to its exact value, Fig. 
(\ref{fig12}). However, to avoid the burden of an heavy numerical computation or conversely a large error on  $a_\text{self}$, we pick $\ell_\text{max}$ such that its contribution to the truncated series  is less than $0.1\%$. This contribution is quantified by the term $\Delta\nLun{a_\text{self}}^{\ell_\text{max}}$ corresponding to the relative error between the 
$\text{L}^1$-norm of the truncated series at $\ell_\text{max}-1$ and the truncated series at $\ell_\text{max}$

\beq
\Delta\nLun{a_\text{self}}^{\ell_\text{max}}=\frac{\nLun{a_\text{self}}^{\ell_\text{max}-1}-\nLun{a_\text{self}}^{\ell_\text{max}}}{\nLun{a_\text{self}}^{\ell_\text{max}}}~,
\eeq
where the $L^1$-norm is given by the integral through the whole history of the particle position $R$ on the background metric

\beq
\nLun{a_\text{self}}^{\ell_\text{max}}=\int|a_\text{self}|dR~.
\eeq

Table \ref{table:L2self:de:lmax} gives value of $a_\text{self}$ with respect to the truncation parameter $\ell_\text{max}$. $\nLun{a_\text{self}}^{\ell_\text{max}}$ converges toward a finite value such that the criterion
\beq
\Delta\nLun{a_\text{self}}^{\ell_\text{max}}\leq0.1\%
\eeq
is satisfied for $\ell_\text{max}=8$.

\setlength{\tabcolsep}{0.03\linewidth} 
\begin{table}[h!]
  \centering
  \begin{tabular}{ccc}
  \hline
  \hline
  $\ell_\text{max}$ & $\nLun{a_\text{self}}^{\ell_\text{max}}$ & $\Delta\nLun{a_\text{self}}^{\ell_\text{max}}$\\
  \hline
  $2$ & $0.03248$ & $ - $\\
  $3$ & $0.03042$ & $ 6.8\% $\\
  $4$ & $0.02960$ & $ 2.7\% $\\
  $5$ & $0.02922$ & $ 1.3\% $\\
  $6$ & $0.02904$ & $ 0.6\% $\\
  $7$ & $0.02896$ & $ 0.2\% $\\
  $8$ & $0.02893$ & $ 0.1\% $\\
  \hline
  \hline
  \end{tabular}
  \caption{{ Estimate of  $\ell_\text{max}$ for a given accuracy. }}
  \label{table:L2self:de:lmax}
\end{table}

\subsection{Sensitivity to $h$}
The grid step parameter $h$ must be chosen carefully { to reach the desired accuracy without useless extra computation}. We follow the same reasoning with $\ell_\text{max}$ considering
\beq
\Delta\nLun{a_\text{self}}^{h_k}=\frac{\nLun{a_\text{self}}^{h_{k-1}}-\nLun{a_\text{self}}^{h_k}}{\nLun{a_\text{self}}^{h_k}}~,
\eeq
where $\nLun{a_\text{self}}^{h_k}$ corresponds to self-acceleration computed with $\ell_\text{max}=8$ with an integration step $h/2M=h_k$ such that $h_0=0.01$, $h_1=0.005$, $h_2=0.0025$, $h_3=0.001$. It is found that the criterion
\beq
\Delta\nLun{a_\text{self}}^{h}\leq0.1\%
\eeq
is satisfied for $h/2M=0.001$.

\subsection{Asymptotic $\ell$-behaviour}\label{section:comportement:asymptotique:en:L}
In section \ref{section:GSF:RW}, we observed that the code assured the correct asymptotic behaviour of quantities for large  $\ell$, Fig. (\ref{fig06}). We   confirm that the regularisation technique works and the regularisation parameters are computed correctly, through the asymptotic behaviour of the self-quantities (acceleration and force), with respect to $L$.

\begin{figure}[h!]
\centering
\includegraphics[width=0.5\linewidth]{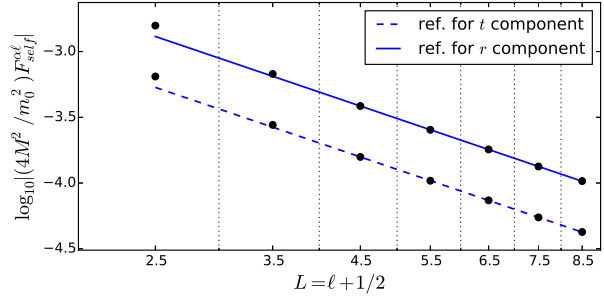}
\caption{Speed of convergence of the series $F^\alpha_\text{self}=\sum_\ell F^{\alpha\ell}_\text{ret}-B^\alpha$ for both components of the SF $t$ (continuous line) and $r$ (dashed line).}
\label{fig13}
\end{figure}

Figure (\ref{fig09}) exhibits the values of $a^\ell_\text{self}$ in terms of $L$. The slope of the line indicates the rate of the convergence of the series $a_\text{self}=\sum_\ell a^\ell_\text{ret}-B^\alpha_a$, that is $\calo{L^{-2}}$. We recall that $a^\ell_\text{ret}$ is an average $a^\ell_\text{ret}=1/2(a^\ell_\text{ret+}+a^\ell_\text{ret-})$. A good behaviour has been found for several values of $R\in[2M,r_0]$ and of $r_0\geq10$. 
Figure (\ref{fig13}) also displays a good exhibit of values for the two components of the SF.

{
\section{Equations of motion}\label{section:equation.of.motion}

The geodesic equation of motion of a test particle in the SD spacetime is
\beq
u^\alpha\nabla_\beta u^\beta=0~.
\label{eq:geod:1}
\eeq

In radial fall, the angular momentum is zero ($\mL=0$), and without loss of generality, the azimuthal angle is chosen to be null too ($\theta=0$). In coordinate time the geodesic equation is given by \cite{sp11}
 \beq
 \frac{d^2r_p}{dt^2}=a_0(r_p,\dott{r}_p)
 =-\left(\Gamma^r_{\ \ab}-\dott{r}_p\Gamma^t_{\ \ab}\right)\dott{x}_p^\alpha\dott{x}_p^\beta
 =-\frac{1}2f(r_p)f'(r_p)\left[1-\frac{3}{f(r_p)^2}\left(\frac{dr_p}{dt}\right)^2\right]
 =f(r_p)f'(r_p)\left[1-\frac{3}{2}\frac{f(r_p)}{\mE^2}\right]~.
 \label{eq:geod:radial:t:coo}
\eeq 

Instead, the perturbed motion is seen as an accelerated motion in the background SD spacetime. The worldline $x^\alpha_p(\tau)$ is not geodesic anymore. The equation of motion changes into

\beq
m_0u^\alpha\nabla_\beta u^\beta=F^\alpha_\text{self}~,
\label{eq:geod:compact:2}
\eeq
where $F^\alpha_\text{self}\sim\calo{m_0/M}$ is the SF computed in the RW gauge. Equation (\ref{eq:geod:compact:2}) can be rewritten in coordinate time

\beq
\frac{d^2r_p}{dt^2}=a_0(r_p,\dott{r}_p)+a_\text{self}(r_p,\dott{r}_p)~.
\label{ode}
\eeq

The trajectory $r_p(t)$ does not have the same meaning as in Eq. (\ref{eq:geod:radial:t:coo}), where it described a geodesic in the background spacetime. The acceleration term $a_0$ is now taken on the perturbed path $r_p$, and the velocity is $\dott{r}_p=dr_p/dt$. The $a^\ell_\text{self}$ term is 

\beq
a_\text{self}=\frac{f(r_p)^2}{m_0\mE^2}\left[F^r_\text{self}-\dott{r}^2_pF^t_\text{self}\right]~.
\label{aminchia}
\eeq
}
\subsection{Pragmatic approach}
In the pragmatic approach \cite{lo00,lo01,spao04,spri14}, we consider Eq. (\ref{ode}) in its linearised version at first order around the reference geodesic $X^\alpha(\tau)$ which is the solution of

\beq
\frac{d^2X^\alpha}{d\tau^2}+\Gamma^{\alpha}_{\beta\gamma}(X^\alpha)\frac{dX^\beta}{d\tau}\frac{dX^\gamma}{d\tau}=0~,
\eeq 
where $\Gamma^{\alpha}_{\beta\gamma}(X^\alpha)$ is the affine connection associated to the background metric. The perturbed trajectory labelled by the coordinates $x^\alpha_p(\tau)=(t,r_p)$ is the solution of Eq. (\ref{eq:geod:compact:2}), developed as

\beq
\frac{d^2x_p^\alpha}{d\tau^2}+\Gamma^{\alpha}_{\beta\gamma}(x_p^\alpha)\frac{dx_p^\beta}{d\tau}\frac{dx_p^\gamma}{d\tau}=\frac{F^\alpha_\text{self}}{m_0}~.
\label{eq:geod:expanded:2}
\eeq

It differs from the reference geodesic by $\Delta X^\alpha\propto m_0/M$ such that
\begin{align}
x^\alpha_p=X^\alpha+\Delta X^\alpha~,
\label{pradec1}
\\
\dot{x}^\alpha_p=\dot{X}^\alpha+\Delta \dot{X}^\alpha~,
\label{pradec2}
\\
\ddot{x}^\alpha_p=\ddot{X}^\alpha+\Delta \ddot{X}^\alpha~,
\label{pradec3}
\end{align}
having supposed that the perturbed motion remains close to the geodesic. By injecting Eqs. (\ref{pradec1}-\ref{pradec3}) into Eq. (\ref{eq:geod:expanded:2}), we have 

\beq
\frac{d^2}{d\tau^2}\Big(X^\alpha+\Delta X^\alpha\Big)
+\Big(\Gamma^{\alpha}_{\beta\gamma}+\partial_\delta\Gamma^{\alpha}_{\beta\gamma}\Delta X^\delta\Big)
\frac{d}{d\tau}\Big(X^\beta+\Delta X^\beta\Big)\frac{d}{d\tau}\Big(X^\gamma+\Delta X^\gamma\Big)=\frac{F^\alpha_\text{self}}{m_0}~.
\label{pragma:tau}
\eeq

For $d/d\tau=(dt/d\tau) d/dt$, and $d^2/d\tau^2=(d^2t/d\tau^2)d/dt+\left(dt/d\tau\right)^2d^2/dt^2$, Eq. (\ref{pragma:tau}) is expanded to first order in coordinate time 

\beq
\frac{d^2t}{d\tau^2}\left(\frac{d\tau}{dt}\right)^2(\dott{X}^\alpha+\Delta\dott{X}^\alpha)
+\ddott{X}^\alpha
+\Delta\ddott{X}^\alpha
+\Gamma^{\alpha}_{\beta\gamma}\dott{X}^\beta \dott{X}^\gamma
+2\dott{X}^\gamma\Gamma^{\alpha}_{\beta\gamma}\dott{X}^\beta
+\Delta X^\delta\partial_\delta\Gamma^{\alpha}_{\beta\gamma}\dott{X}^\beta \dott{X}^\gamma
=\frac{F^\alpha_\text{self}}{m_0}\left(\frac{d\tau}{dt}\right)^2~.
\label{pragma:tcoo}
\eeq

Assuming $\Delta X^\alpha=\sigma^\alpha\Delta R$ with $\sigma^\alpha\defeq(0,1)$ we find for the time and radial components

\beq
\frac{d^2t}{d\tau^2}\left(\frac{d\tau}{dt}\right)^2=
\frac{F^t_\text{self}}{m_0}\left(\frac{d^2\tau}{dt^2}\right)^2
-\Gamma^{t}_{\beta\gamma}\dott{X}^\beta \dott{X}^\gamma
-2\Gamma^{t}_{\beta\gamma}\dott{X}^\beta \sigma^\gamma\Delta\dott{R}
-\partial_r\Gamma^{t}_{\beta\gamma}\dott{X}^\beta \dott{X}^\gamma\Delta R~,
\label{pragma:comp:t}
\eeq

\beq
\frac{d^2t}{d\tau^2}\left(\frac{d\tau}{dt}\right)^2\frac{d}{dt}\Big(R+\Delta R\Big)
+\ddott{R}+\Delta\ddott{R}
+\Gamma^{r}_{\beta\gamma}\dott{X}^\beta \dott{X}^\gamma+
2\Gamma^{r}_{\beta\gamma}\dott{X}^\beta \sigma^\gamma\Delta\dott{R}
+\partial_r\Gamma^{r}_{\beta\gamma}\dott{X}^\beta \dott{X}^\gamma\Delta R
=\frac{F^r_\text{self}}{m_0}\left(\frac{d\tau}{dt}\right)^2~.
\label{pragma:comp:r}
\eeq

By injecting  Eq. (\ref{pragma:comp:t}) into Eq. (\ref{pragma:comp:r}), we find at first order

\bea
&\ddott{R}+\Delta\ddott{R}
+\Big(\Gamma^{r}_{\beta\gamma}-\Gamma^{t}_{\beta\gamma}\dott{R}\Big)\dott{X}^\beta \dott{X}^\gamma+
\Delta R\Big(\partial_r\Gamma^{r}_{\beta\gamma}-\partial_r\Gamma^{t}_{\beta\gamma}\dott{R}\Big)\dott{X}^\beta \dott{X}^\gamma+\\
&\Delta\dott{R}\Big(2\Gamma^{r}_{\beta\gamma}\sigma^\gamma-2\Gamma^{t}_{\beta\gamma}\sigma^\gamma\dott{R}-\Gamma^{t}_{\beta\gamma}\dott{X}^\gamma\Big)\dott{X}^\beta=\frac{\dot{t}^2}{m_0}\left[F^r_\text{self}(R,\dott{R})-\dott{R}F^t_\text{self}(R,\dott{R})\right]~,
\label{pragma:explicit}
\eea
where $\dot{t}=dt/d\tau=\mE/f$. Equation (\ref{pragma:explicit}) is presented as

\beq
\frac{d^2r_p}{dt^2}=a_0(R,\dott{R})+a_1(R,\dott{R})\Delta R + a_2(R,\dott{R})\Delta\dott{R} + 
a_\text{self}(R,\dott{R})+\mathcal{O}(\Delta R^2,\Delta\dott{R}^2)~,
\label{ode_lin}
\eeq
with 
\begin{align}
&
\begin{aligned}
a_0(R,\dott{R})&=-\Big(\Gamma^{r}_{\beta\gamma}-\Gamma^{t}_{\beta\gamma}\dott{R}\Big)\dott{X}^\beta \dott{X}^\gamma 
=-\frac{1}2ff'\left[1-\frac{3}{f^2}\dott{R}^2\right]~,
\label{a0:geod}
\end{aligned}\\
&\nonumber\\
&
\begin{aligned}
a_1(R,\dott{R})&=-\Big(\partial_r\Gamma^{r}_{\beta\gamma}-\partial_r\Gamma^{t}_{\beta\gamma}\dott{R}\Big)\dott{X}^\beta \dott{X}^\gamma~
=\frac{2M}{R^3}\left[1-\frac{3M}R-3\left(1-\frac{M}R\right)f^{-2}\dott{R}^2\right]~,
\label{a1:geod}
\end{aligned}\\
&\nonumber\\
&
\begin{aligned}
a_2(R,\dott{R})&=-\Big(2\Gamma^{r}_{\beta\gamma}\sigma^\gamma-2\Gamma^{t}_{\beta\gamma}\sigma^\gamma\dott{R}-\Gamma^{t}_{\beta\gamma}\dott{X}^\gamma\Big)\dott{X}^\beta~
=\frac{6M}{R^2}f^{-1}\dott{R}~,
\label{a2:geod}
\end{aligned}\\
&\nonumber\\
&a_\text{self}(R,\dott{R})=\frac{f^2}{m_0\mE^2}\left[F^r_\text{self}(R,\dott{R})-\dott{R}F^t_\text{self}(R,\dott{R})\right]~.
\label{aself:geod}
\end{align}

Given $\ddott{R}=a_0(R,\dott{R})$, at first order we get through the Taylor expansion of $\ddott{r}_p(r_p,\dott{r}_p)$ (considered as a function of two variables) around the point $(R,\dott{R})$ the $\Delta$ variation

\beq
\Delta\ddott{R}\approx a_1(R,\dott{R})\Delta R + a_2(R,\dott{R})\Delta\dott{R} + a_\text{self}(R,\dott{R})~.
\label{edo}
\eeq

The expression of $a_1(R,\dott{R})$ differs from \cite{lo00} by $4M/R^3(1+\dott{R}/f^2)$, as already pointed out 
in \cite{spao04,sp11,aosp11}. 
The variation of the acceleration is given by the $a^\ell_\text{self}$ computed along the reference geodesic, while the terms $a_1(R,\dott{R})\Delta R + a_2(R,\dott{R})\Delta\dott{R}$ represent the background geodesic deviation.
Indeed, by recasting Eq. \ref{pragma:tau} at first order we find

\beq
2\Gamma^\alpha_{\beta\gamma}\Delta\dot{X}^\beta\dot{X}^\gamma+\partial_\delta\Gamma^\alpha_{\beta\gamma}\dot{X}^\beta\Delta X^\delta\dot{X}^\gamma=-R_{\beta\gamma\delta}^{\ \ \ \ \alpha}\dot{X}^\beta\Delta X^\gamma\dot{X}^\delta~,
\eeq
where the right hand-side term appears in the rigourous derivation of the perturbation equation at first order in the H gauge
\cite{grwa08,grwa11}

\beq
\frac{D^2\Delta X^\alpha}{d\tau^2}=-R_{\beta\gamma\delta}^{\ \ \ \ \alpha}\dot{X}^\beta\Delta X^\gamma\dot{X}^\delta+F^{\alpha\text{(H)}}_\text{self}~.
\eeq

\begin{figure}[h!]
\centering
\includegraphics[width=0.5\linewidth]{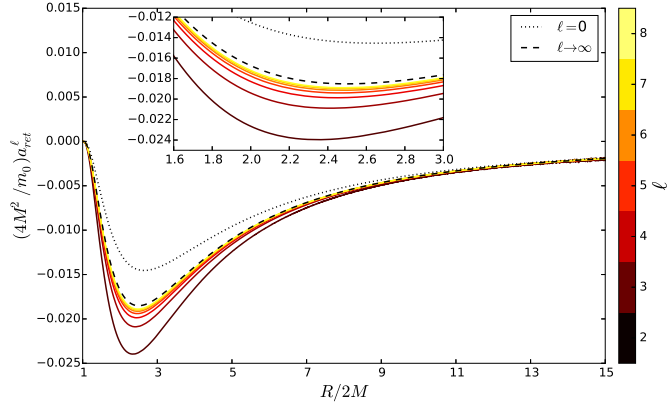}
\caption{For the radial fall of a particle, initially at rest, from  $r_0/2M=15$, the modes of the acceleration term (colour palette) computed from the modes of retarded force, Fig. (\ref{fig08}), are shown. The continuous curve represents the analytical behaviour for large $\ell$, Eq. (\ref{Ba}). The dashed curve corresponds to the mode $\ell=0$.}
\label{fig14}
\end{figure}

{
Fig. (\ref{fig14}) provides the first modes of $a_\text{ret}$,  the acceleration term constructed from retarded force computed on the reference geodesic $X^\alpha$
\beq
a^\ell_\text{ret}(R,\dott{R})=\frac{f^2}{m_0\mE^2}\Big[F^{r\ell}_\text{ret}-\dott{R}F^{t\ell}_\text{ret}\Big]~,
\label{retarded:acceleration}
\eeq
for $\ell\geq0$. The sum over all modes { needs} to be regularised. Thus from Eq. (\ref{mode:sum:average}) and Eq. (\ref{retarded:acceleration}) we have
\beq
\sum_{\ell=0}^\infty a^\ell_\text{self}=\sum_{\ell=0}^\infty\Big[a^\ell_\text{ret}-B_a\Big]~,
\label{serie:aself}
\eeq
where the modes of $a^\ell_\text{ret}$ tend to an asymptotic value $B_a(R)$ obtained analytically from 
\beq
B_a=\frac{f^2}{m_0\mE^2}\Big[B^r-\dott{R}B^t\Big]=-\frac{m_0\mE}{2R^2}\left[\frac{f}{\mE}\right]^3~.
\label{Ba}
\eeq
}

\begin{figure}[h!]
\centering
\includegraphics[width=0.5\linewidth]{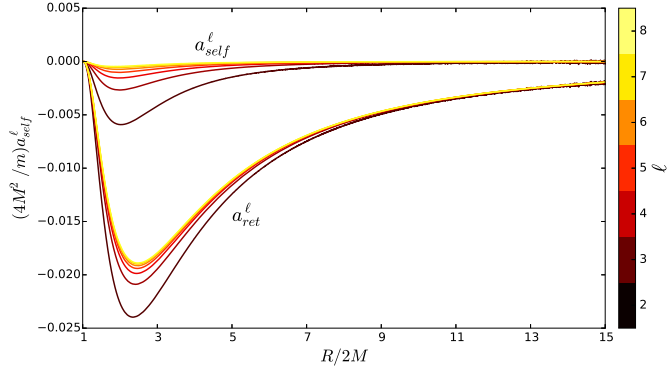}
\caption{We compare the first modes of the non-regularised acceleration term $a_\text{ret}$ (upper part of the colour palette) and the modes of the regularised acceleration term $a_\text{self}$ (upper part of the colour palette). After regularisation, $a^\ell_\text{self}$ well satisfies the convergence criterion, Eq. (\ref{conv:criterion}), validating the method and the formulation of $B_a$ \cite{spri14}.}
\label{fig15}
\end{figure}

\begin{figure}[h!]
\centering
\includegraphics[width=0.5\linewidth]{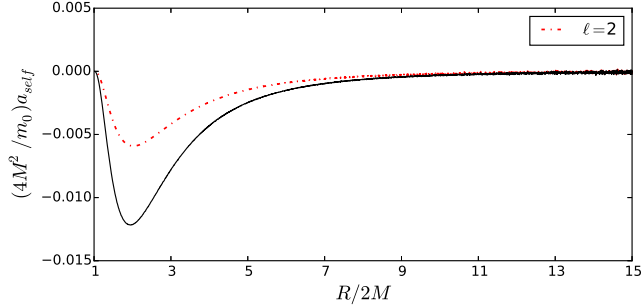}
\caption{{ After adding all modes $a^\ell_\text{self}$ numerically computed up to $\ell_\text{max}=8$, we add the non-radiative mode $\ell=0$ mode and the analytically computed higher modes for $\ell>8$. We get the $a^\ell_\text{self}$ (solid curve) for a falling particle from $r_0/2M=15$. We plot the mode $\ell=2$ (dot-dash curve) which is roughly $\sim55\%$ of the total $a_\text{self}$ (integrated over $R$).}}
\label{fig16}
\end{figure}

In Fig. (\ref{fig15}) $a^\ell_\text{self}$ is also plotted for $\ell=2$ to $\ell=8$ and compared to modes $a^\ell_\text{ret}$ (before regularisation). The required criterion of convergence of the series, Eq.  (\ref{serie:aself}) namely

\beq
\text{if }\sum_{\ell=0}^\infty a^\ell_\text{self}\text{ converge,}\quad\text{ then }\quad a^\ell_\text{self}\xrightarrow{\ell\to\infty}0~.
\label{conv:criterion}
\eeq

is satisfied; thereby it ensures the regularisation of the self-acceleration, and validates the formulation of $B_a$ \footnote{Curves $a^\ell_\text{self}$ in Fig. (\ref{fig15}) disagree with Fig. (2a) in \cite{lo01}. The first modes of the $a^\ell_\text{self}$ named $C^\ell_\text{ren}$ do not appear to satisfy the convergence criterion of Eq. (\ref{conv:criterion}). Therefore, the curves in Fig. (2b) \cite{lo01} will not be consistent with ours \cite{spri14}.}. The convergence speed is displayed in Fig. (\ref{fig09}), and it is discussed in Sect. 
\ref{section:comportement:asymptotique:en:L}. It appears that the $a^\ell_\text{self}$ is dominated by the lowest modes, the quadrupole mode $\ell=2$ representing itself $\sim55\%$ of total, Fig. (\ref{fig16}).

{
The total SF is computed by summing all numerical modes up to $\ell=\ell_\text{max}$ plus an additive part for higher modes contribution
\beq
a_\text{self}=a^{\ell=0}_\text{self}+\ub{\sum_{\ell=2}^{\ell_\text{max}}a_\text{self}^\ell}_\text{numeric}+\ub{\sum_{\ell=\ell_\text{max}+1}^{+\infty}a_\text{self}^{\ell\to\infty}}_\text{analytic}~,
\label{somme:a}
\eeq
where $a_\text{self}^{\ell\to\infty}=f^2m_0^{-1}\mE^{-2}\Big[F_\text{self}^{r\ell\to\infty}-\dott{R}F_\text{self}^{t\ell\to\infty}\Big]$ is the analytic term from Eqs. (\ref{Ft:self:Linf},\ref{Fr:self:Linf}). It allows to take into account the contribution of the modes $\ell>\ell_\text{max}$

\beq
a_\text{self}^{\ell\to\infty}=\mathcal{A}^\infty_\text{self}L^{-2}+\calo{L^{-4}}~,
\eeq
with
\beq
\mathcal{A}^\infty_\text{self}=-\frac{15}{16}m_0\frac{f^2}{R^3\mE^2}\Bigg[\mE^2\left(4M+R(\mE^2-1)\right)+
R\dott{R}\left(2R\ddott{R}-\dott{R}^2\right)\Bigg]~.
\eeq

Therefore the analytic part of the sum, Eq. (\ref{somme:a}), is approximated by
\beq
\sum_{\ell=\ell_\text{max}+1}^{+\infty}{a_\text{self}^{\ell\to\infty}}\approx\mathcal{A}^\infty_\text{self}\sum_{\ell=\ell_\text{max}+1}^{+\infty}(\ell+1/2)^{-2}\\
\approx\mathcal{A}^\infty_\text{self}\ \zeta(2,\ell_\text{max}+1)~,
\eeq
where $\zeta$ is the Riemann-Hurwitz function defined by \cite{ri59,hu82}, see App. (\ref{rihureg})

\beq
\zeta(s,n)=\sum_{\ell=0}^\infty(\ell+n)^{-s}~.
\eeq

If $\ell_\text{max}=8$, we have

\beq
\zeta(2,9)=\frac{\pi^2}{6}-\frac{1077749}{705600}\approx 0.117512~.
\eeq
}

Figure (\ref{fig16}) displays $a^\ell_\text{self}$ for all $\ell$ contributions as computed in Eq. (\ref{somme:a}) for $\ell_\text{max}=8$, and $a^2_\text{self}$ both for $r_0/2M=15$. The choice in the order of truncation of the series $\ell_\text{max}=8$ admits a relative error less than 0.1\% and is justified in Sect. (\ref{sensibilite:lmax}).

In \cite{spri14}, we present our analysis on the impact of the SF on the motion of the particle. Herein we summarise the main findings and produce some new insights. For a particle supposedly released from $15 r_{\rm g}$, we have identified four zones according to the sign of $\Delta R$, $\Delta {\dott R}$, $\Delta {\ddott R}$, for $r_0 = 15r_g$, where $r_g$ is the SD black hole radius.  

\setlength{\tabcolsep}{0.0100\linewidth} 
\begin{table*}
\caption{The four zones according to the sign of the deviation from the nominal position, velocity and acceleration for $r_0 = 15r_g$, where $r_g$ the SD black hole radius, see reference \cite{spri14}.}

\begin{tabular}{|l|c|c|c|}
\hline
Zone                                 & $\Delta R$ & $\Delta {\dott R}$ & $\Delta {\ddott R}$ \\ \hline
I~~~ $r_0 - 3.5~r_{\rm g}$           &     -      &         -         &         -              \\ \hline
II~~~$3.5~r_{\rm g} - 2.2~r_{\rm g}$ &     -      &         -         &         +              \\ \hline
III~~$2.2~r_{\rm g} - 1.2~r_{\rm g}$ &     -      &         +         &         +              \\ \hline
IV~~$1.2~r_{\rm g} - r_{\rm g}$      &     -      &         +         &         -              \\ \hline
\end{tabular}
\label{fourroses}
\end{table*}

\begin{figure}[h!]
\centering
\includegraphics[width=0.6\linewidth]{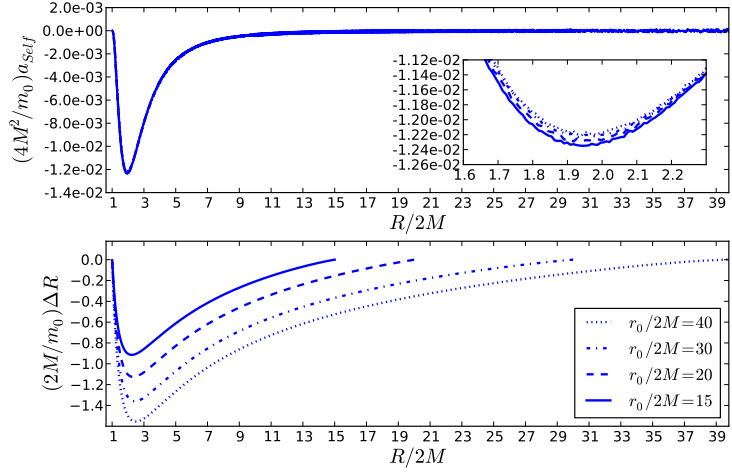}
\caption{For different values of $r_0/2M$, in the upper panel, $a_\text{self}$, and in the lower panel the $\Delta R$ deviation are plotted as function of $R$. The behaviour {\it vis \`a vis} $r_0$ differs between these two quantities: the amplitude of $a_\text{self}$ diminishes for larger $r_0$, conversely to $\Delta R$ that increases. We interpret this as a manifestation of the effect of the geodesic deviation term or of the longer time in which the particle radiates.} 
\label{fig17}
\end{figure}

\begin{enumerate}
  \item {$a_\text{self}$ is strictly negative in $R\in[2M,r_0]$, and tends to a finite value at the horizon . This behaviour is independent of the initial position $r_0$. It reaches its maximum amplitude, in absolute value, when the particle approaches the maximun of the Zerilli potential, $R\approx 3.1M$. After, the derivative changes sign, and $a_\text{self}$ tends to zero at the horizon, compatibly with the findings of an external observer.}
\item {As expected, the amplitude of the orbital deviation $\Delta R$ is of the order of the mass ratio. It reaches its maximum amplitude, in absolute value, when the particle approaches the maximun of the Zerilli potential, $R\approx 3.1M$. The $\Delta R$ term  has the same sign of $a_\text{self}$, that is to say, the particle will reach faster the black  hole { horizon} premises than the geodetic motion (obviously the horizon will never be reached).}
\item{Table \ref{fourroses} describes the four different zones the particle passes through.    In zone I ($3.5~r_{\rm g} < r \leq r_0 = 15~r_{\rm g}$), the particle falls faster than in a background geodesic. Approaching the potential, it radiates more and it undergoes a breaking phase: in zone II ($2.2~r_{\rm g} < r < 3.5~r_{\rm g}$), the acceleration deviation $\Delta {\ddot r}$ becomes positive, but the velocity deviation $\Delta {\dot r}$ remains negative; in zone III ($1.2~r_{\rm g} < r < 2.2~r_{\rm g}$), the breaking is stronger and even the velocity deviation turns positive. Finally, in zone IV ($r_{\rm g} < r < 1.2~r_{\rm g}$), the acceleration deviation reappears negative, but not sufficiently to render the velocity deviation again negative. The particle tends to acquire the geodesic behaviour at the horizon where indeed $\Delta R=\Delta\dott{R}=\Delta\ddott{R}=0$). }
\item{In zones I and II, $\Delta \dott{R}<0$, the particle increases its velocity relatively to the geodesic motion. Instead, in zones III and IV, $\Delta \dott{R}>0$, the particle loses velocity relatively to the geodesic motion.}
\item{$\Delta\ddott{R}>0$ { in} zones II, III as opposed to $a_\text{self}$ which is negative. The absolute amplitude of the former is much larger and it is due to the relevant role of the background geodesic deviation that counteracts the effects of the self-acceleration. Nevertheless, we refrain from attributing a repulsive behaviour to the SF due to the constant negative sign of $\Delta R$.}
\item{In Fig. (\ref{fig17}), we solve Eq. (\ref{edo}) for different values of $r_0/2M$. The amplitude of $a_\text{self}$ diminishes for larger $r_0$, conversely to $\Delta R$ that increases.  We interpret this as a manifestation of the effect of the geodesic deviation term or of the longer time in which the particle radiates.} 
\item{The computation for different values of $r_0/2M=15,20,30,40$ shows that $|\Delta\dott{R}|_\text{max}$ increases linearly with $r_0$. In fact, the position in which a test particle reaches the maximum value is  $6Mr_0/(r_0+4M)$, which remains in the strong field area where Zerilli potential is high and wherein the effect of the SF on the motion is 
the most important.}
\item {The self-quantities (acceleration and force) are dominated by the mode  $\ell=2$ that represents more than $50\%$ of total.}
\end{enumerate}

\subsection{Orbital evolution}

The pragmatic approach builds a perturbed trajectory from $a_\text{self}$ computed on the reference geodesic under the constraint that $\Delta R\sim\calo{m_0/M}$. When considering a fall from a very far initial position we may doubt about the applicability of Eq. (\ref{edo}), and instead,  consider a 2\textsuperscript{nd} order development to ensure accuracy. But the arguments in \cite{grwa08,grwa11} lead to conclude that at sufficiently late times a second order perturbative development will fail too, and instead it is preferable to correct the motion iteratively with a first order scheme. 

\begin{figure}[h!]
\centering
\includegraphics[width=0.6\linewidth]{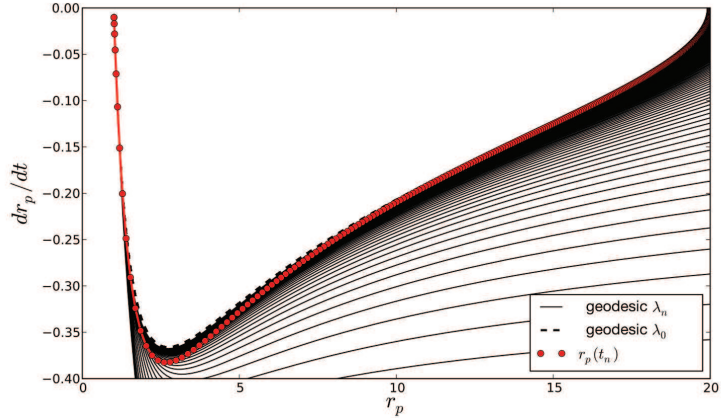}
\caption{Illustration of the osculating method. During evolution, each point of the perturbed trajectory (red dots) given by the equation of motion Eq. (\ref{ode}), is approximated by a geodesic (black solid curves) tangent to the perturbed path and passing through this point. The dashed curve corresponds to the geodesic followed by a test particle initially coinciding with the perturbed trajectory at $r_0/2M=20$.}
\label{fig18}
\end{figure}

Strict self-consistency implies that the applied SF at some instant is what arises from the actual field at that same instant. 
This has been done  for a scalar charged particle around an SD black hole \cite{divewade12}, and never for a massive particle. 
In other works \cite{waakbagasa12,labu12}, the applied SF is what would have resulted if the particle
were moving along the geodesic that only instantaneously matches the true orbit. Herein, we adopt the latter acception. 
Our approach in orbital evolution (in the RW gauge) consists thus in computing the total acceleration through self-consistent (osculating) geodesic stretches of orbits. The  self-consistent method would require solving Eq. (\ref{ode}), and therefore the evaluation of $a_\text{self}$ on the trajectory taken by the particle. But the regularisation parameters are evaluated onto a geodesic. This renders quite natural to choose an osculating method, Fig.(\ref{fig18}).

We don't compute the trajectories for large values of $r_0$, but remain within the values previously analysed, as we wish to develop an algorithm handling a self-consistent computation. 

A geodesic $\lambda$ describes the trajectory $Z^\lambda(t)=(R^\lambda(t),\dott{R}^\lambda(t))$ in the phase space. The osculating method finds, at each point of the perturbed trajectory $z_p(t_n)=(r_p(t_n),\dott{r}_p(t_n))$ at time $t_n$, a geodesic $\lambda_n$ which passes through $z_p(t_n)$. Thus the value of the $a^\ell_\text{self}$ on the perturbed trajectory is given by the computed value on the osculating geodesic. The osculating method therefore assumes that

\beq
a_\text{self}(r_p(t_n),\dott{r}_p(t_n))\approx a_\text{self}(R^{\lambda_n}(t_n),\dott{R}^{\lambda_n}(t_n))~.
\eeq

\begin{figure}[h!]
\centering
\includegraphics[width=0.5\linewidth]{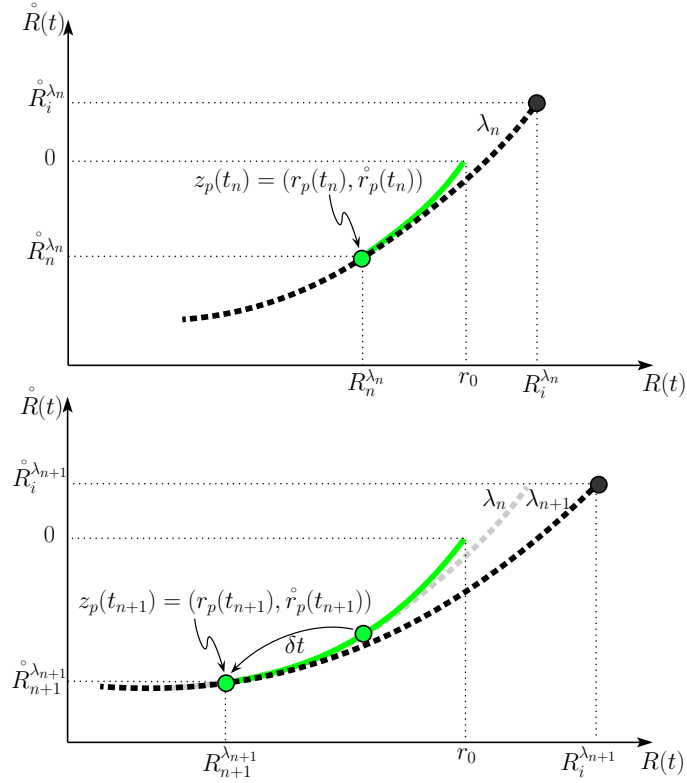}
\caption{The osculating method consists in identifying each point of the trajectory $z_p(t_n)=(r_p(t_n),\dottb{r}_p(t_n))$ (green dots and curves) and at time $t_n$, a geodesic $\lambda_n$ (black and grey curves in the upper and lower panels, respectively) passing through 
$z_p(t_n)$. Using $a_\text{self}$ computed at $Z^{\lambda_n}_n$ point, Eq. (\ref{ode}) indicates a new point $z_p(t_{n+1})$ at time 
$t_{n+1}$. Therein, a new geodesic $\lambda_{n+1}$ is searched again, such that $z_p(t_{n+1})=Z^{\lambda_{n+1}}_{n+1}$ (black curve in the lower panel).}
\label{fig19}
\end{figure}

In a discretised version, adapted to numerical processing, we consider a family of geodesics $\lambda_n$ that we will find at every time step $t_n$. We introduce a series of notations, Fig. (\ref{fig19})

\begin{itemize}
\item $\lambda_n$: the geodesic passing at time $t_n$ through the $r_p(t_n)$ point with the velocity $\dott{r}_p(t_n)$.
\item $t_n=t_0+n\delta t$, where $t_0$ is the instant when $r_p(t_0)=r_0$ and $\delta t=k h,\ k\in\mathbb{N}$.
\item $z_p(t_n)=(r_p,\dott{r}_p)(t_n)$: the point of the perturbed trajectory in phase space at time $t_n$.
\item $Z^{\lambda_n}_n=Z^{\lambda_n}(t_n)=(R^{\lambda_n}_n,\dott{R}^{\lambda_n}_n)$: the point of the geodesic $\lambda_n$ in phase space at time $t_n$.
\item $R^{\lambda_n}_n=R^{\lambda_n}(t_n)$: the position at time $t_n$ on the geodesic $\lambda_n$.
\item $\dott{R}^{\lambda_n}_n=\dott{R}^{\lambda_n}(t_n)$: the velocity at time $t_n$ on the geodesic $\lambda_n$.
\item $a_\text{self}(t_n)=a_\text{self}\left(R^{\lambda_n}_n,\dott{R}^{\lambda_n}_n\right)$: the self-acceleration computed on the geodesic $\lambda_n$ at point $Z^{\lambda_n}_n$ { and} time $t_n$.
\item $\mE_n$: the energy associated with the geodesic $\lambda_n$. It is directly given by the coordinates of the point $z_p(t_n)=Z^{\lambda_n}_n$\beq
\mE_n=\sqrt{\frac{f(R^{\lambda_n}_n)^3}{f(R^{\lambda_n}_n)^2-(\dott{R}^{\lambda_n}_n)^2}}~.
\label{E:lambdan}
\eeq
\item $Z^{\lambda_n}_i=(R^{\lambda_n}_i,\dott{R}^{\lambda_n}_i)$, where $R^{\lambda_n}_i$ and $\dott{R}^{\lambda_n}_i$ are the initial position and velocity required for the geodesic $\lambda_n$ to reach the point $z_p(t_n)$ at time $t_n$. The initial velocity is linked to the initial position and the energy via
\beq
\dott{R}^{\lambda_n}_i=\pm\frac{f(R^{\lambda_n}_i)}{\mE_n}\sqrt{\mE_n^2-f(R^{\lambda_n}_i)}~,
\label{vi:lambdan}
\eeq
where "$\pm$" is the sign of the initial velocity.
\end{itemize}

Knowing $a_\text{self}(t_n)$ at each time step $t_n$, we solve numerically Eq. (\ref{ode}), starting from  $r_p(t_0)=r_0$ and $\dott{r}_p(t_0)=0$. The diagram (\ref{fig20}) shows the conceptual flow of the algorithm. The main steps are  :

\begin{figure}[h!]
\centering
\includegraphics[width=0.7\linewidth]{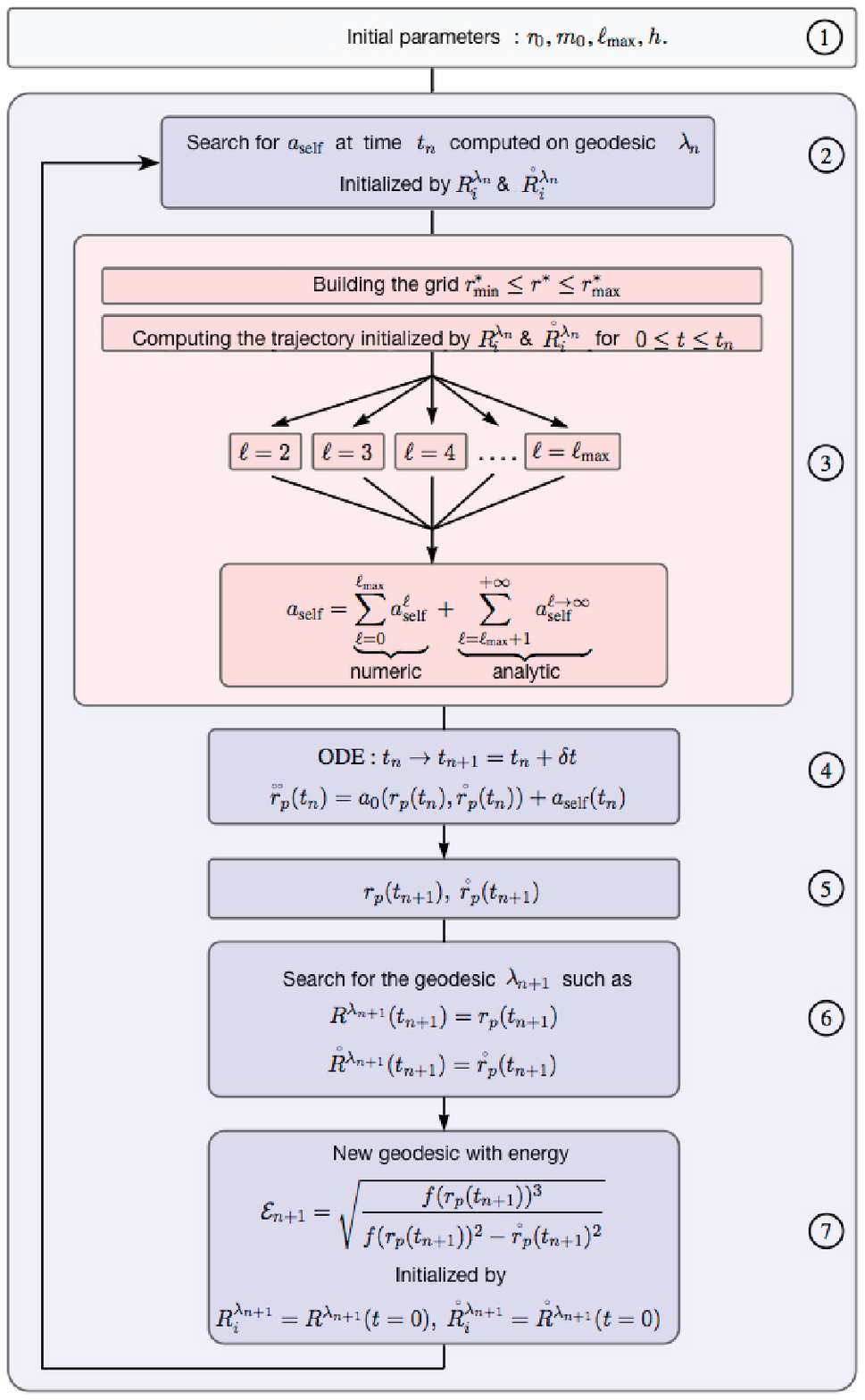}
\caption{Algorithm for the osculating method. Frame 3 is relative to the computation of $a_\text{self}$ for a given point $z_p(t_n)$ at a given instant $t_n$. The other frames correspond to the iterative procedure determining the single geodesic passing through the following point $z_p(t_{n+1})$. }
\label{fig20}
\end{figure}

\begin{enumerate}
\item Initialisation of the numerical parameters, $m_0$, $h$ $\ell_\text{max}$, and $r_0$.
\item Loop resolution of the ordinary differential equation (\ref{ode}) where at each step $t_n$, the quantity $a_\text{self}$ is computed on the osculating geodesic $\lambda_n$ { at} $r_p(t_n)$.
\item Discretisation of the grid defined by its boundaries  $r^*_\text{min}$ and $r^*_\text{max}$; generation of the trajectory $\lambda_n$ passing through the numerical domain; computation of $a_\text{self}$ at time $t_n$ for different modes $\ell$ to $\ell=\ell_\text{max}$. For the optimisation of the computation time, this multi-modal operation is distributed on multiple processors. The sum is then performed over all modes ($\ell=0$ included) to which the contribution of higher modes $\ell_\text{max}$ are added analitically.
\item Iterative solution of the equation of motion by Euler's method.
\item The new position $z_p(t_{n+1})=\left(r_p(t_{n+1}),\dott{r}_p(t_{n+1})\right)$ is obtained in phase space.
\item The new geodesic $\lambda_{n+1}$ which passes through the point $z_p(t_{n+1})$ at time $t_{n+1}$ is searched through a modified Newton method. The output parameter is the initial position $R^{\lambda_{n+1}}_i=R^{\lambda_{n+1}}(t=0)$ of $\lambda_{n+1}$.
\item The new geodesic $\lambda_{n+1}$ is characterised by the initial position $R^{\lambda_{n+1}}_i$ from which the particle releases the energy $\mE_{n+1}$ given by Eq. (\ref{E:lambdan}). The latter determines the initial speed $\dott{R}^{\lambda_{n+1}}_i=\dott{R}^{\lambda_{n+1}}(t=0)$ given by Eq. (\ref{vi:lambdan}).
\end{enumerate}

The search of new geodesics $\lambda_n$ requires to fix the only free parameter $R^{\lambda_n}_i$, by scanning in a range of initial positions $R^{\lambda_k}_i$, and then evaluating $Z^k(t_n)$ points to be compared to the targeted $z_p(t_n)$. A Newton's method assures that the quantity $\left|r_p(t_n)-R^{\lambda_k}(t_n)\right|$ is below an arbitrary value $10^{-6}$, and thereby that the geodesic starting at $R^{\lambda_k}_i$ is the right geodesic. 

\begin{figure}[h!]
\centering
\includegraphics[width=0.7\linewidth]{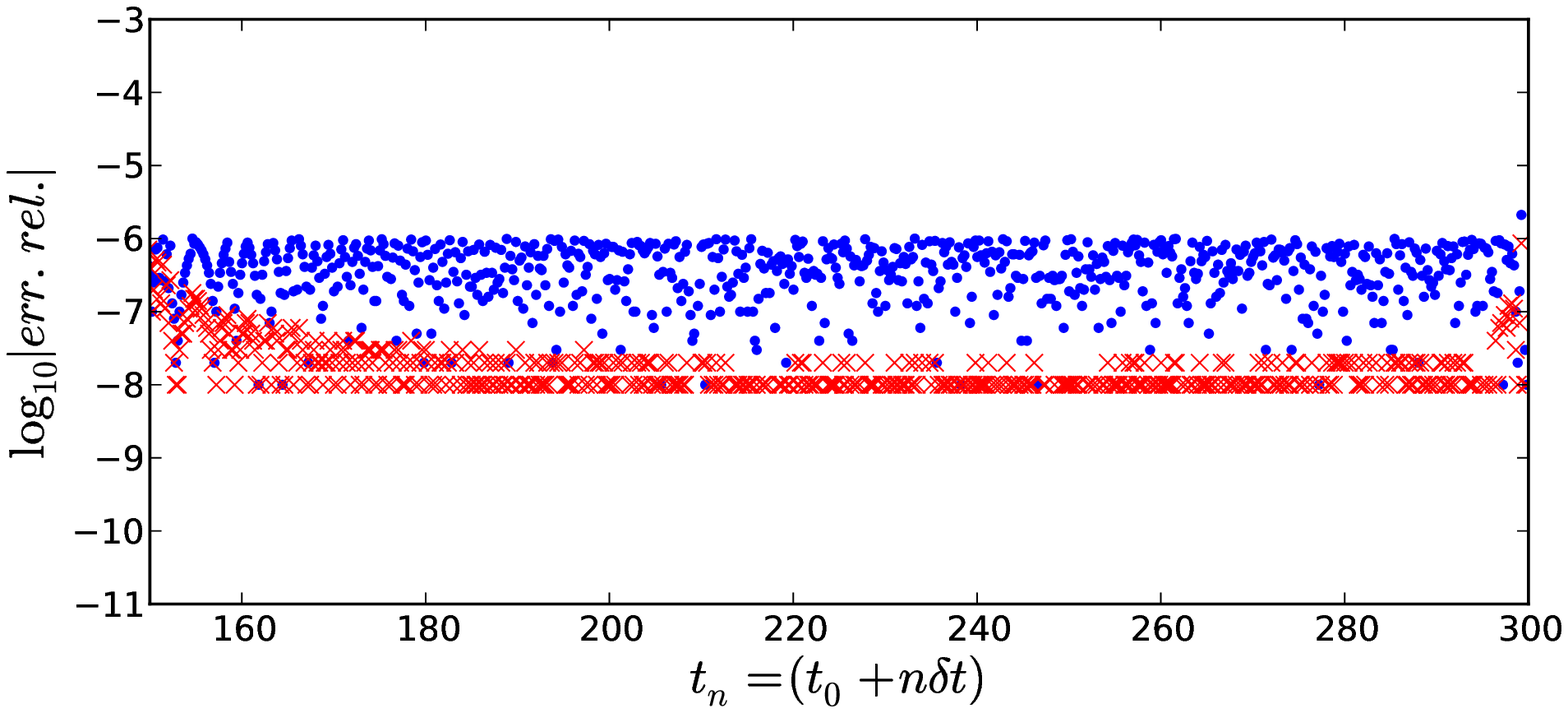}
\caption{The relative error between the $z_p(t)$ point of the perturbed trajectory and the $Z(t)$ point of the geodesic passing through 
$z_p(t)$ which is determined by the algorithm. The blue dots correspond to $\log_{10}\left|r_p(t_n)-R^{\lambda_n}(t_n)\right|$ and the red crosses to $\log_{10}\left|\dottb{r}_p(t_n)-\dottb{R}^{\lambda_n}(t_n)\right|$. }
\label{fig21}
\end{figure}

In Fig. (\ref{fig21}), it is shown the relative error between the $z_p(t)$ point of the perturbed trajectory and the $Z(t)$ point of the geodesic passing through $z_p(t)$ which is determined by the algorithm.

\beq
\left|r_p(t_n)-R^{\lambda_k}(t_n)\right|\leq10^{-6}\ \Rightarrow\ \lambda_n=\lambda_k~,
\eeq

In order to compare the pragmatic analysis to the osculating one, we introduce the following quantities:

\begin{align}
&\Delta r^\text{prag} \defeq \Delta R~,\\
&\Delta \dott{r}^\text{prag} \defeq \Delta \dott{R}~,
\end{align}
where $\Delta R(t)$ is a solution of Eq. (\ref{ode_lin}). In the same way we define a deviation term in the osculating formalism
\begin{align}
&\Delta r^\text{osc} \defeq r_p-R^{\lambda_0}~,\\
&\Delta \dott{r}^\text{osc} \defeq \dott{r}_p-\dott{R}^{\lambda_0}~,
\end{align}
where $r_p$ is the perturbed trajectory built from the osculating algorithm, and $R^{\lambda_0}(t)$ is the { first} reference geodesic passing through the initial point $z_p(t_0)=(r_0,0)$. Explicitly, $\Delta r^\text{prag}$ is the 1\textsuperscript{st} order deviation with respect to the geodesic motion for which $a_\text{self}$ is computed along the reference geodesic. Then, $\Delta r^\text{osc}$ is the deviation from the geodesic motion, wherein $a_\text{self}$ is provided at each point of the perturbed trajectory $r^\text{osc}_p$ by its value computed on the osculating geodesic.

\begin{figure}[h!]
\centering
\includegraphics[width=0.8\linewidth]{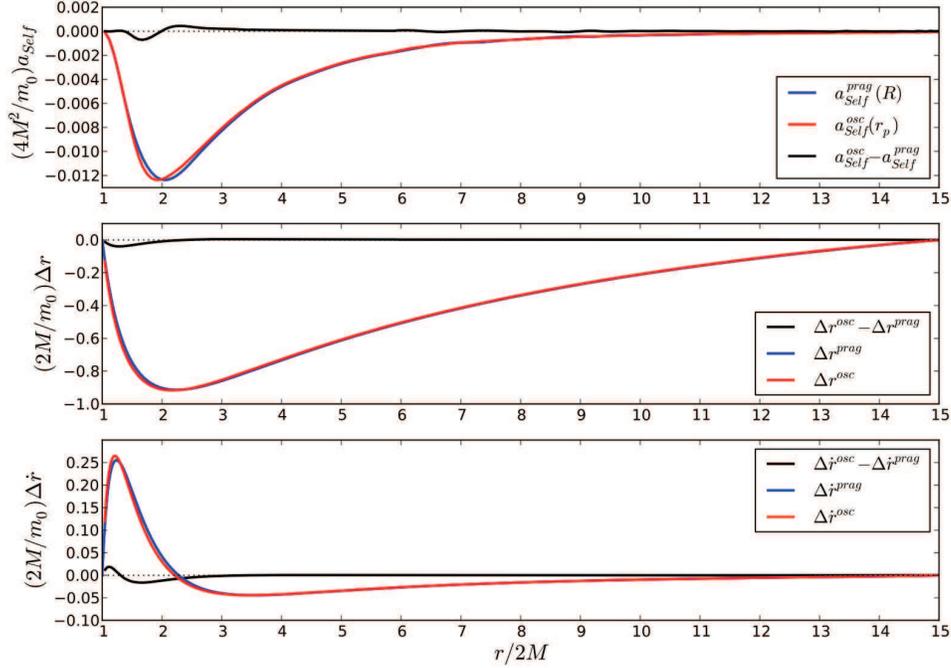}
\caption{Comparison between the pragmatic solution, Fig. (\ref{fig16}), and the osculating solution for $r_0/2M=15$ \cite{spri14}.}
\label{fig22}
\end{figure}

In Fig. (\ref{fig22}), we choose $m_0=10^{-5}$ and an initial position $r_0/2M=15$. Comparison is made between the solution built by the pragmatic method, and that from the osculating algorithm (red curves). 
At the top of the graph, we compare the amplitude of $a^\text{prag}_\text{self}\defeq a_\text{self}(R)$ previously given in Fig. (\ref{fig16}) to the amplitude of $a^\text{osc}_\text{self}$ given by the values of $a_\text{self}$ taken on the osculating geodesics. The absolute difference between these two quantities has a maximum relative amplitude of approximately $3\%$. The notable difference is localised in the strong field region ($R\lesssim3$); the minimum of $a^\text{osc}_\text{self}$, $\Delta r^\text{osc}$ and $\Delta \dott{r}^\text{osc}$ are shifted toward the horizon with respect to $a^\text{prag}_\text{self}$. For $R\gtrsim3$ all curves are identical respectively.

The perturbed trajectory coming from the osculating algorithm is consistent with the pragmatic approach (there is a correction of about $3\%$), thereby confirming our code. The correction of few percent remains valid for different values of $m_0$. We have noted that the osculating analysis shifts slightly the four zones towards the horizon.

\subsection{Perturbed wave-forms}

We wish here to evaluate the effect of the SF on the wave-form (WF) and the energy radiated to infinity. For the computation of the perturbed trajectory, the osculating algorithm will be used as described above. For the generation of the WFs we will use the code developed in Part I.

\begin{figure}[h!]
\centering
\includegraphics[width=0.7\linewidth]{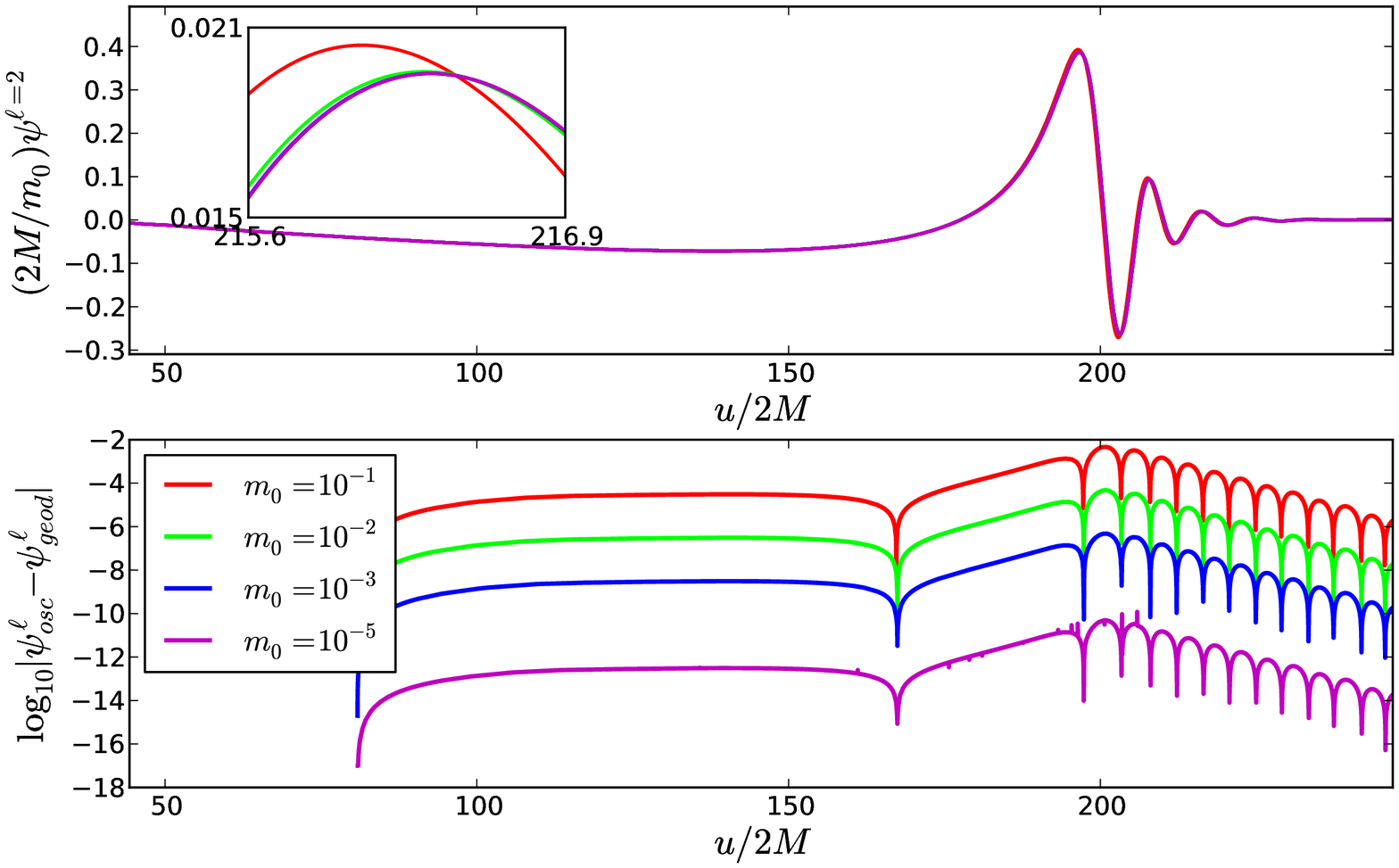}
\caption{Perturbed WFs (top of the graph) as a function of the retarded time $u=t-r^*_\text{obs}$ (with $r^*_\text{obs}=500M$) for the quadrupole mode seen at infinity for a particle falling from $r_0/2M=15$. The computation is done for different values of $m_0$. At the bottom of the graph is plotted the absolute difference in $\log_{10}$ scale between the perturbed WFs and the WFs generated when the particle follows a geodetic motion.}
\label{fig23}
\end{figure}

\begin{figure}[h!]
\centering
\includegraphics[width=0.7\linewidth]{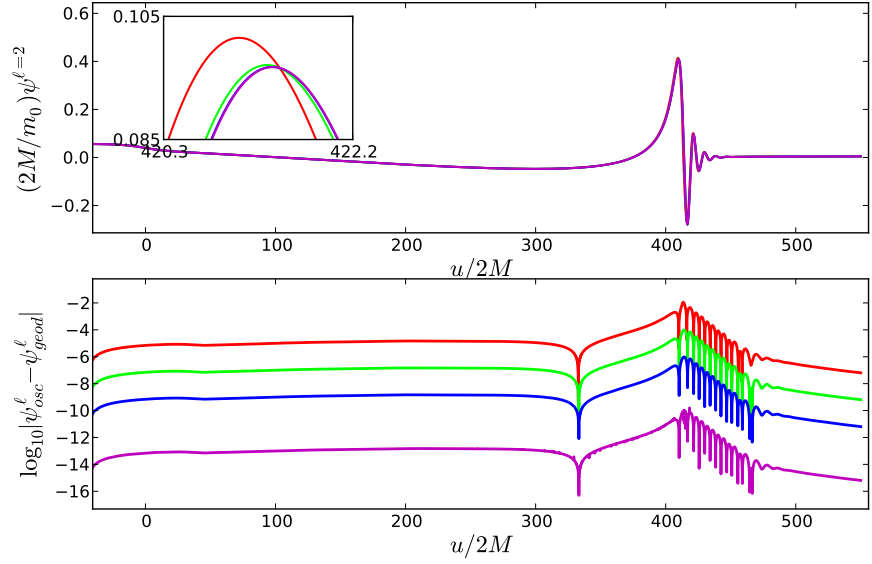}
\caption{Perturbed WFs (top of the graph) as a function of the retarded time $u=t-r^*_\text{obs}$ (with $r^*_\text{obs}=500M$) for the quadrupole mode seen at infinity for a particle falling from $r_0/2M=40$. The computation is done for different values of $m_0$. At the bottom of the graph is plotted the absolute difference in $\log_{10}$ scale between the perturbed WFs and the geodesic WFs.}
\label{fig24}
\end{figure}

Figures (\ref{fig23},\ref{fig24}) show the perturbed WFs for a particle falling from $r_0/2M=15$ and $r_0/2M=40$, respectively. The WFs are superimposed in the top of the graph for different values of $m_0$. At the bottom of the graph we plot the absolute difference between the perturbed WFs and the geodesic WFs. The area where the difference is maximal ($t\in[175,250]$) corresponds to the motion in the strong field area where the particle reaches the horizon and then produce quasi-normal modes.

Knowing the perturbed WFs, the associated radiated energy  can be computed. Table \ref{table:E:FOP} lists energy values for the two trajectories  ($r_0/2M=15$ and $r_0/2M=40$) for the modes $\ell=2$ to $\ell=5$. In each case, we compute the energy difference $\delta E_\ell$ with respect to the energy radiated from a particle following a geodesic. This computation is performed for three different values of $m_0$. Note that $\delta E_\ell$ is much smaller when the mass ratio $m_0/M$ is small . This is explined by $a_\text{self}\propto m_0$. Moreover, for the same value of $m_0$, $\delta E_\ell$ seems more important for $r_0/2M=40$ than for $r_0/2M=15$ since the gravitational radiation effect occurs for a longer time for $r_0/2M=40$. Likewise, the relative difference in energy increase when $\ell$ is large. However, for the sum of the modes, from $\ell=2$ to $\ell=5$, the difference in total energy exceeds $1\%$ for $m_0=10^{-2}$ and remains well below $1\%$ for masses $m_0=10^{-3}$ or $m_0=10^{-5}$.

Thus, for a typical EMRI  system ($m_0<10^{-5} { M}$),  where the compact star is in free fall, the energy variation is very negligible compared to the criterion  that we set on the computation of $E$. For larger values of $m_0$, orbital adjustment becomes significant and can be taken into account. In all cases, the difference in energy is positive, which is consistent with the fact that the system loses energy carried by GW until the distant observer.

\begin{widetext}

\setlength{\tabcolsep}{0.005\linewidth} 
\begin{table}[!htb]
  \centering
  {\footnotesize

  \begin{tabular}{cccccc|cccccc}
  \hline
  \hline
  $r_0/2M$ & $m_0$ & $\ell$  & $E_\ell$ & $\delta E_\ell$ & $\delta E_\ell/E_\ell$ & $r_0/2M$ & $m_0$ & $\ell$  & $E_\ell$ & $\delta E_\ell$ & $\delta E_\ell/E_\ell$\\\hline
  15&$10^{-2}$ & $2$ & $1.65301.10^{-2}$&$8.2.10^{-5}$  & $0.50\%$  &40&$10^{-2}$ & $2$ & $1.85187.10^{-2}$ &$8.5.10^{-5}$ & $0.45\%$\\
  &           & $3$ & $1.95258.10^{-3}$ &$1.67.10^{-5}$ & $0.85\%$ &&            & $3$ & $2.12044.10^{-3}$ &$1.87.10^{-5}$ & $0.88\%$\\
  &           & $4$ & $2.60437.10^{-4}$ &$3.14.10^{-6}$ & $1.21\%$ &&            & $4$ & $2.85501.10^{-4}$ &$3.52.10^{-6}$ & $1.23\%$\\
  &           & $5$ & $3.68305.10^{-5}$ &$5.81.10^{-7}$ & $1.58\%$ &&            & $5$ & $4.06409.10^{-5}$ &$6.54.10^{-7}$ & $1.61\%$\\
  &$10^{-3}$  & $2$ & $1.64562.10^{-2}$ &$8.10^{-6}$    & $0.05\%$ &&$10^{-3}$   & $2$ & $1.85187.10^{-2}$ &$8.5.10^{-6}$  & $0.05\%$\\
  &           & $3$ & $1.93750.10^{-3}$ &$1.66.10^{-6}$ & $0.09\%$ &&            & $3$ & $2.10344.10^{-3}$ &$1.77.10^{-6}$ & $0.09\%$\\
  &           & $4$ & $2.57592.10^{-4}$ &$2.98.10^{-7}$ & $0.12\%$ &&            & $4$ & $2.82311.10^{-4}$ &$3.34.10^{-7}$ & $0.12\%$\\
  &           & $5$ & $3.63100.10^{-5}$ &$6.08.10^{-8}$ & $0.17\%$ &&            & $5$ & $4.00538.10^{-5}$ &$6.66.10^{-8}$ & $0.17\%$\\
  &$10^{-5}$  & $2$ & $1.64483.10^{-2}$ &$8.1.10^{-8}$  & $4.5.10^{-4}\%$    &&$10^{-5}$   & $2$ & $1.85102.10^{-2}$ &$8.6.10^{-7}$ & $4.6.10^{-4}\%$\\
  &           & $3$ & $1.93586.10^{-3}$ &$2.10^{-8}$    & $1.10^{-3}\%$&&            & $3$ & $2.10169.10^{-3}$ &$2.5.10^{-8}$ & $1.2.10^{-3}\%$\\
  &           & $4$ & $2.57296.10^{-4}$ &$2.7.10^{-9}$  & $1.10^{-4}\%$  &&            & $4$ & $2.81975.10^{-4}$ &$5.10^{-9}$ & $1.7.10^{-3}\%$\\
  &           & $5$ & $3.62498.10^{-5}$ &$6.6.10^{-10}$ & $1.8.10^{-4}\%$  &&            & $5$ & $3.99876.10^{-5}$ &$3.6.10^{-10}$ & $9.10^{-4}\%$\\
  \hline
  \hline
  \end{tabular}
  \caption{Radiated energy for a falling particle starting from $r_0/2M=15$ and $r_0/2M=40$ on a perturbed trajectory. Energy is given mode by mode (in units of $(2M/m_0^2)$) and compared to the radiated energy for a geodesic motion via $\delta E_\ell$. The computation is done for 3 values of $m_0$}
  \label{table:E:FOP}
  }
\end{table}
\end{widetext}

\section{Conclusions}

In this second part of the work, we exploited our code for the determination of effect of the gravitational self-force on the motion of the particle and on the wave-forms. We have computed the regularisation parameters exclusively in the Regge-Wheeler gauge. The procedure has been described in detail since never appeared in the literature. The perturbation tensor components and the retarded gravitational self-force were then computed numerically and regularised. 

The gravitational self-force was found with less than 0.1\% error. The equation of motion was solved using two approaches: 
pragmatic and self-consistent (osculating). 

The convergence of the two methods results validates our indirect integration method. We confirm our previous findings stating that in Regge-Wheeler and harmonic gauges, the self-force induces an additional push on the particle towards the black hole, conversely to previous results. This is emphasised when the self-consistent approach is used. 
  
The latter improves the orbital accuracy by a factor of few percents. For the computation of the radiated energy and the display of the wave-forms, we have shown feeble but existing differences between geodesic and non-geodesics orbits. The correction factor could be important for intermediate mass ratios. 
 
\section*{Acknowledgements}

We thank C. F. Sopuerta (Barcelona), L. Blanchet (Paris) and L. Burko (Huntsville) for the comments received. P. Ritter's thesis, on which this paper is partly based, received an honorable mention by the Gravitational Wave International (GWIC) and Stefano Braccini Thesis Prize Committee in 2013.  

\appendix

\section{{Riemann-Hurwitz regularisation}\label{rihureg}}

The Riemann-Hurwitz $\zeta$ function \cite{ri59,hu82} is formally defined for complex arguments $n$ with $\mathbb{R}(n) > 1 $ and $m$ with $\mathbb{R}(m) > 0$ by 
\beq
 \zeta(n,m)=\sum_{\ell=0}^\infty(\ell+m)^{-n}~.
 \label{riemann:hurwitz}
 \eeq
 
This series is absolutely convergent for the given values of $n$ and $m$. The $\zeta$ function has been adopted for regularisation in \cite{lo00,lo01,spao04}. From the behaviour of the metric coefficients, we consider that $h_\ab^\text{ret}$ can be decomposed in two pieces. The first piece, noted $h_\ab^{\text{reg}\ell}$, tends quickly towards zero when $\ell\to\infty$, ensuring the convergence of the sum. The second piece $h_\ab^{\infty}$ generates the limit behaviour when $\ell\to\infty$, observed in Fig. (\ref{fig07}), and responsible for the divergence of the sum. For example, for the $tt$ component, we can write   

 \beq
  H_2(t,r)=\sum_{\ell=0}^\infty H_2^{\ell}Y^{\ell0}
=\sum_{\ell=0}^\infty H_2^{\text{reg}\ell}Y^{\ell0}+\sum_{\ell=0}^\infty (2\ell+1)^{-\beta}H_2^{\infty}Y^{\ell0}~,
 \eeq
  where $\beta$ is a parameter to be determined numerically to ensure the limit behaviour of $H_2^{\infty}$ when $\ell\to\infty$. 
  So, when regularising  the field at the particle position, we will have 

 \beq
 H_2^\text{ret}(t,r_p)=\sum_{\ell=0}^\infty H_2^{\text{reg}\ell}\sqrt{\frac{2\ell+1}{4\pi}}+2^{1/2-\beta}\frac{H_2^{\infty}}{4\pi}\zeta(\beta-1/2,1/2)~.
\label{h2rh}
 \eeq

Numerically, we get  $\beta=1/2$ and the analytical extension of the $\zeta$ function gives $\zeta(0,1/2)=0$. Thus, for this  regularisation, it is sufficient to subtract from each $\ell$ mode the asymptotic value, i.e. the highest mode  computed numerically $H_2^{\infty}=H_2^{\ell_\text{max}}$ such that 
 
\beq
 H_2^{\text{reg}}(t,r_p)=\sum_{\ell=0}^{\ell_\text{max}}\Big[H_2^{\text{ret}\ell}(t,r_p)-H_2^{\ell_\text{max}}(t,r_p)\Big]Y^{\ell0}~.
 \eeq
 
Referring to the Mode-Sum formalism, we have $D^\alpha\propto\zeta(0,1/2)=0$, where $D^\alpha$ is the residual parameter linked to regularisation of the field  $H_2$. The difference between the two regularisation approaches can be represented by the quantity $H_2^{\ell_\text{max}}-H_2^{\ell\to\infty}$ which is equal to zero when  $\ell_\text{max}$ is sufficiently large. Thus, at least in the case of a radial orbit, a correlation could be done between $\zeta$ and Mode-Sum regularisations.    

\section{Numerical extraction of the field on the worldline}
\label{Interpolation}

\begin{figure}[h!]
\centering
\includegraphics[width=0.5\linewidth]{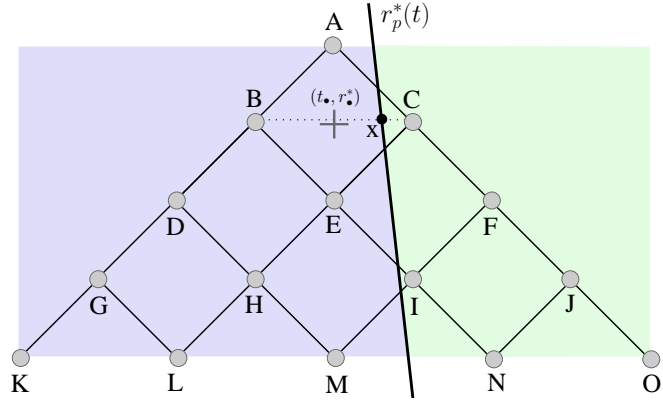}
\caption{Stencil for the interpolation of $\psi$ at the particle position $(t_\text{x},r^*_\text{x})$. We consider fifteen collocation points and fifteen jump relations to define the thirty coefficients $p^\pm_{(n,m)}$, where $\ n+m\leq4$, of the interpolated polynomial at the cell central point $(t_\bullet,r^*_\bullet)$, Eq. (\ref{formule:interp:psi}).}
\label{fig25}
\end{figure}

We approximate $\psi$ at the position of the particle $(t_\text{x},r^*_\text{x})$ with a polynomial $P(t,r^*)$ of fourth order { centred at $(t_\bullet,r^*_\bullet)$}

\beq
\psi(t,r^*)\approx P(t,r^*)=\sum_{n+m\leq4}\frac{p_{(n,m)}}{n!m!}\left(r^*-r^*_\bullet\right)^n\left(t-t_\bullet\right)^m~.
\eeq

The stencil contains fifteen points in the past light cone of the point $A$ (included), {\it i.e.} $\triangle=\{A,B,C,D,E,F,G,H,I,J,K,L,M,N,O\}$, Fig. (\ref{fig25}). Being $\psi$ discontinuous at $r_p$, we approach $\psi^+(t,r^*)$ and $\psi^-(t,r^*)$ with two interpolation polynomials $P^+(t,r^*)$ and $P^-(t,r^*)$

\beq
\psi^\pm(t,r^*)\approx P^\pm(t,r^*)=\sum_{n+m\leq4}\frac{p^\pm_{(n,m)}}{n!m!}\left(r^*-r^*_\bullet\right)^n\left(t-t_\bullet\right)^m~.
\label{formule:interp:psi}
\eeq

At $(t_\text{x},r^*_\text{x})$, $P^\pm(t_\text{x},r^*_\text{x})\approx\psi^\pm(t_\text{x},r^*_p(t_\text{x}))$. The thirty interpolation coefficients $p^\pm_{(n,m)}$ are uniquely determined by fifteen relations at the collocation points
\beq
P^\pm(t_i,r^*_i)=\psi^\pm(t_i,r^*_i)\quad\forall\ i\in\triangle~,
\eeq
and fifteen jump relations
\beq
\begin{aligned}
&\partial^n_{r^*}\partial^m_tP^+(t_\text{x},r_\text{x})-\partial^n_{r^*}\partial^m_tP^-(t_\text{x},r_\text{x})=\jump{\partial^n_{r^*}\partial^m_t\psi}_\text{x}\\
&\forall\ n,m\ |\ n+m\leq4~.
\end{aligned}
\eeq

For the errors, if the function $\psi$ is computed with a fourth  order interpolation scheme, then Eq. (\ref{formule:interp:psi}) at most provides an accuracy of order one for the perturbations $H^\ell_1$ and $H^\ell_2$, due to the third derivatives of the $\psi$ function.


\section{Jump conditions: radial orbits} \label{ANNEXE_JC_RAD}

We list here the explicit forms of the jump conditions of the $\psi^\ell$ function and its derivatives until $4$\textsuperscript{th} order for radial orbits localised by $R(t)$. The coordinate time derivative of $R$ is $\dott{R} = f(R)/\mE\sqrt{\mE^2-f(R)}$. 

\begin{widetext}
\subsubsection*{0th order}
\beq
\jump{\psi^\ell}  = 
\frac{\kappa \mE R}{(\lambda +1) (3 M+\lambda   R)}~,
\label{jump-psi} 
\eeq

\subsubsection*{1st order}
\beq
\jump{\partial_t\psi^\ell}  =  
-\frac{\kappa \mE R \dott{R}}{(2 M - R) (3
   M+\lambda R)}~,
\label{jump-psit} 
\eeq

\beq
\jump{\partial_r\psi^\ell} = 
\frac{\kappa \mE \left[6 M^2+3 M \lambda  R+\lambda 
   (\lambda +1) R^2\right]}{(\lambda +1) (2 M-R) (3 M+\lambda  R)^2}~,
\label{jump-psir}
\eeq

\subsubsection*{2nd order}
\beq
 \jump{\partial^2_r\psi^\ell} = 
-\frac{\kappa \mE \left[3 M^3 (5 \lambda -3)+6 M^2 \lambda (\lambda -3)
    R+3 M \lambda ^2
(\lambda -1)    R^2-2 \lambda ^2 (\lambda +1)
   R^3\right]}{(\lambda +1) (2M-R)^2 (3 M+\lambda  R)^3}~,
\label{jump-psirr}
\eeq

\beq
\jump{\partial_t\partial_r\psi^\ell} = 
\frac{\kappa \mE \left(3 M^2+3 M \lambda 
   R - \lambda  R^2\right)\dott{R}}{(2M - R)^2 (3 M+\lambda  R)^2}~,
\label{jump-psitr} 
\eeq

\beq
\jump{\partial^2_t\psi^\ell} = -\frac{\kappa\,\mE\,M}{{R}^{2}\,\left( 3\,M+R\,\lambda\right) }~,
\label{jump-psitt} 
\eeq

\subsubsection*{3rd order}
\bea
\jump{\partial^3_r\psi^\ell}=&
\frac{\kappa \mE}{R\left(  \lambda+1\right) {\left( 2M-R\right) }^{3}{\left( 3M+R \lambda\right) }^{4}}\bigg[81\left( \lambda+1\right) {M}^{5}+9R\left( 19\lambda^{2}+18{\mE}^{2}\lambda + \right.\nonumber\\
&\left. 3\lambda+18{\mE}^{2}\right) {M}^{4}+9R^{2}\lambda\left(7\lambda^{2}+24{\mE}^{2}\lambda-14\lambda
+24{\mE}^{2}+3\right) {M}^{3}+3R^{3}\lambda^{2}\left( 7 \lambda ^2\right.\nonumber\\
&\left.+36{\mE}^{2}\lambda-11\lambda+36{\mE}^{2}+18\right){M}^{2}+3R^{4}\lambda^{3}\left( 8{\mE}^{2}\lambda-7\lambda+8{\mE}^{2}-1\right) M + \nonumber\\
& 2R^{5}\lambda^{3}\left( \lambda+1\right) \left( {\mE}^{2}\lambda+3\right) \bigg]~,
\label{jump-psirrr}
\eea

\bea
\jump{\partial_t\partial^2_r\psi^\ell}= & 
\frac{-\kappa \mE\dott{R}}{R{\left( 2M-R\right) }^{3}{\left( 3M+R \lambda\right) }^{3}}\bigg[27{M}^{4}+6R\left( 5\lambda+9{\mE}^{2}-3\right) {M}^{3}+3R^{2}\lambda \left( 5\lambda + \right.\nonumber\\
&\left. 18{\mE}^{2}-6\right) {M}^{2}+6R^{3}\lambda^{2}\left( 3{\mE}^{2}-2\right) M+2R^{4}\lambda^{2}\left( {\mE}^{2}\lambda+1\right) \bigg]~,
\eea

\bea
\jump{\partial^2_t\partial_r\psi^\ell}=&
\frac{\kappa \mE}{R^{3}\left( 2M-R\right) {\left( 3M+R \lambda\right) }^2}\bigg[39{M}^{3}+9R\left( 3\lambda+2{\mE}^{2}-2\right) {M}^{2}+R^{2}\lambda\left( 4\lambda + \right. \nonumber\\
& \left. 12{\mE}^{2}-13\right) M+2R^{3}\lambda^{2}\left( \mE^2-1\right) \bigg]~,
\eea

\beq
\jump{\partial^3_t\psi^\ell}=\frac{-\kappa \mE\dott{R}}{R^{3}\left(2M-R\right)\left( 3M+R \lambda\right)}\bigg[9{M}^{2}+2R\left( 2\lambda+3{\mE}^{2}-2\right) M+2R^{2}\lambda \left( \mE^2-1\right)\bigg]~,
\eeq

\subsubsection*{4th order} 
\bea
\jump{\partial^4_r\psi^\ell}=& \frac{-3\kappa \mE}{R^2\left(  \lambda+1\right) {\left( 2M-R\right) }^{4}{\left( 3M+R \lambda\right) }^{5}}\bigg[567\left( \lambda+1\right) {M}^{7}+162R\left( \lambda+1\right) \left( 6\lambda\right.\nonumber\\
&\left.+16{\mE}^{2}-5\right) {M}^{6}+6R^{2}\left( 139\lambda^{3}+738{\mE}^{2}\lambda^{2}-123\lambda^{2}+162{\mE}^{4}\lambda+441{\mE}^{2}\lambda\right.\nonumber\\
&\left.-171\lambda+162{\mE}^{4}-297{\mE}^{2}+27\right) {M}^{5}+12R^{3}\lambda\left( 21\lambda^{3}+252{\mE}^{2}\lambda^{2}-85\lambda^{2}+\right.\nonumber\\
&\left.135{\mE}^{4}\lambda-24\lambda+135{\mE}^{4}-252{\mE}^{2}+18\right) {M}^{4}+3R^{4}\lambda^{2}\left( 21\lambda^{3}+344{\mE}^{2}\lambda^{2}-\right.\nonumber\\
&\left.95\lambda^{2}+360{\mE}^{4}\lambda-340{\mE}^{2}\lambda+100\lambda+360{\mE}^{4}-684{\mE}^{2}+24\right) {M}^{3}+2R^{5}\lambda^{3}\cdot\nonumber\\
&\left( 88{\mE}^{2}\lambda^{2}-47\lambda^{2}+180{\mE}^{4}\lambda-260{\mE}^{2}\lambda+25\lambda+180{\mE}^{4}-348{\mE}^{2}-24\right) {M}^{2}\nonumber\\
&+2R^{6}\lambda^{4}\left( 6{\mE}^{2}\lambda^{2}+30{\mE}^{4}\lambda-53{\mE}^{2}\lambda+23\lambda+30{\mE}^{4}-59{\mE}^{2}+11\right) M+\nonumber\\
&4R^{7}\lambda^{4}\left( \lambda+1\right) \left( {\mE}^{4}\lambda-2{\mE}^{2}\lambda-2\right)\bigg]~,
\eea

\bea
\jump{\partial_t\partial^3_r\psi^\ell}=&\frac{3\kappa \mE\dott{R}}{R^2{\left( 2M-R\right) }^{4}{\left( 3M+R \lambda\right) }^{4}}\bigg[135{M}^{6}+27R\left( 7\lambda+32{\mE}^{2}-6\right) {M}^{5}+3R^{2}\cdot\nonumber\\
&\left( 35\lambda^{2}+396{\mE}^{2}\lambda-75\lambda+108{\mE}^{4}-144{\mE}^{2}+18\right) {M}^{4}+R^{3}\lambda\left( 35\lambda^{2}+\right.\nonumber\\
&\left.612{\mE}^{2}\lambda-120\lambda+432{\mE}^{4}-594{\mE}^{2}+72\right) {M}^{3}+R^{4}\lambda^{2}\left( 140{\mE}^{2}\lambda-45\lambda+\right.\nonumber\\
&\left.216{\mE}^{4}-306{\mE}^{2}+36\right) {M}^{2}+2R^{5}\lambda^{3}\left( 6{\mE}^{2}\lambda+24{\mE}^{4}-35{\mE}^{2}+9\right) M+\nonumber\\
&2R^{6}\lambda^{3}\left( 2{\mE}^{4}\lambda-3{\mE}^{2}\lambda-1\right)\bigg]~,
\eea

\bea
\jump{\partial^2_t\partial^2_r\psi^\ell}=&\frac{-\kappa \mE}{R^{4}{\left( 2M-R\right) }^2{\left( 3M+R \lambda\right) }^{3}}\bigg[1431{M}^{5}+6R\left( 251\lambda+234{\mE}^{2}-210\right) {M}^{4}+\nonumber\\
&9R^{2}\left( 59\lambda^{2}+160{\mE}^{2}\lambda-148\lambda+36{\mE}^{4}-66{\mE}^{2}+30\right) {M}^{3}+6R^{3}\lambda\left( 10\lambda^{2}+\right.\nonumber\\
&\left.82{\mE}^{2}\lambda-79\lambda+54{\mE}^{4}-102{\mE}^{2}+48\right) {M}^{2}+2R^{4}\lambda^{2}\left( 28{\mE}^{2}\lambda-27\lambda+54{\mE}^{4}\right.\nonumber\\
&\left.-105{\mE}^{2}+52\right) M+12R^{5}\lambda^{3}{\left( \mE^2-1\right) }^{2}\bigg]~,
\eea

\bea
\jump{\partial^3_t\partial_r\psi^\ell}=&\frac{\kappa \mE\dott{R}}{R^{4}\left( 2M-R\right)^2{\left( 3M+R \lambda\right) }^2}\bigg[243{M}^{4}+3R\left( 61\lambda+132{\mE}^{2}-64\right) {M}^{3}+3R^{2}\cdot\nonumber\\
&\left( 12\lambda^{2}+92{\mE}^{2}\lambda-49\lambda+36{\mE}^{4}-48{\mE}^{2}+12\right) {M}^{2}+2R^{3}\lambda\left( 24{\mE}^{2}\lambda-15\lambda+\right.\nonumber\\
&\left.36{\mE}^{4}-51{\mE}^{2}+14\right) M+6R^{4}\lambda^{2}\left( \mE^2-1\right) \left( 2{\mE}^{2}-1\right) \bigg]~,
\eea

\bea
\jump{\partial^4_t\psi^\ell}=&\frac{-\kappa \mE}{{R^{6}\left( 3M+R \lambda\right) }}\bigg[189{M}^{3}+2R\left( 36\lambda+84{\mE}^{2}-77\right) {M}^{2}+6R^{2}\left( \mE^2-1\right)\left( 10\lambda+\right.\nonumber\\
&\left.6{\mE}^{2}-5\right) M+12R^{3}\lambda{\left( \mE^2-1\right) }^{2}\bigg]~.
\eea
\end{widetext}

\bibliography{references_spallicci_150913}
\end{document}